%% file: paper.tex
\documentclass[conference,a4paper]{IEEEtran}

\input{packages}

\input{macros}

\begin{document}

\title{Adversary-Augmented Simulation for Fairness Evaluation and Defense in Hyperledger Fabric}

\author{
\IEEEauthorblockN{
Erwan Mahe \orcidlink{0000-0002-5322-4337},~~
Rouwaida Abdallah \orcidlink{0009-0007-9803-3511},~~
Sara Tucci-Piergiovanni \orcidlink{0000-0001-9738-9021}
}
\IEEEauthorblockA{\textit{Université Paris Saclay, CEA LIST} \\
Palaiseau, France}
\and
\IEEEauthorblockN{Pierre-Yves Piriou \orcidlink{0000-0001-6061-7270}}
\IEEEauthorblockA{\textit{EDF Lab, Dpt. PRISME} \\
Palaiseau, France}
}

\maketitle

\begin{abstract}
This paper presents an adversary model and a simulation framework specifically tailored for analyzing attacks on distributed systems composed of multiple distributed protocols, with a focus on assessing the security of blockchain networks. Our model classifies and constrains adversarial actions based on the assumptions of the target protocols—defined by failure models, communication models, and the fault tolerance thresholds of Byzantine Fault Tolerant (BFT) protocols. The goal is to study not only the intended effects of adversarial strategies but also their unintended side effects on critical system properties.
We apply this framework to analyze fairness properties in a Hyperledger Fabric (HF) blockchain network. Our focus is on novel fairness attacks that involve coordinated adversarial actions across various HF services. Simulations show that even a constrained adversary can violate fairness with respect to specific clients ({\em client fairness}) and impact related guarantees ({\em order fairness}), which relate the reception order of transactions to their final order in the blockchain.
This paper significantly extends our previous work by introducing and evaluating a mitigation mechanism specifically designed to counter transaction reordering attacks. We implement and integrate this defense into our simulation environment, demonstrating its effectiveness under diverse conditions.
\end{abstract}


\input{1_intro}

\input{2_0_prelims}

\input{3_0_adv_model}

\input{4_0_usecase}

\input{5_0_attacks}

\input{6_0_sim_attacks}

\input{7_0_mitig}

\input{8_0_sims}

\input{10_conc}

\bibliographystyle{IEEEtran}
\bibliography{
biblio/adversary_models,
biblio/attacks_in_real_life,
biblio/biblio_formal,
biblio/biblio_max_gitlab,
biblio/blind,
biblio/blockchain_simulators,
biblio/clock,
biblio/consensus,
biblio/distr_comp_models,
biblio/fairness,
biblio/hyperledger,
biblio/multiagent,
biblio/others,
biblio/own,
biblio/voting
}

\end{document}

%% file: packages.tex
\usepackage{graphicx}
\usepackage{textcomp}
\usepackage{xcolor}
\def\BibTeX{{\rm B\kern-.05em{\sc i\kern-.025em b}\kern-.08em
    T\kern-.1667em\lower.7ex\hbox{E}\kern-.125emX}}
    
\usepackage{amsmath}
\usepackage{amsthm}
\usepackage{amssymb}

\usepackage{subcaption}
\usepackage{wrapfig}
\usepackage{multirow}
\usepackage{makecell}
\usepackage{url}
\usepackage{orcidlink}

\usepackage{fontawesome}
\usepackage{filecontents}
\usepackage{pgfplots}
\newsavebox{\imagebox}

\usepackage{soul}
\setstcolor{red}

\usepackage{diagbox}
\usepackage{bussproofs}

\usepackage{tikz}
\usetikzlibrary{
positioning,
arrows,
automata,
shapes,
hobby,
backgrounds,
calc,
trees,
fit,
shapes.multipart,
fadings,
through,
patterns,
decorations.pathreplacing,
patterns,
shapes.geometric}
\usepackage{tikz-cd}
\pgfdeclarelayer{bg}
\pgfsetlayers{bg,main}

\usepackage{algorithm2e}
\SetKwComment{Comment}{\tiny/* }{\tiny */}
\SetKw{KwCustomType}{Type}
\SetKw{KwCustomInput}{Input}

%% file: macros.tex
\newcommand{\advactInject}{$\mathtt{inject}$}
\newcommand{\advactDelay}{$\mathtt{delay}$}
\newcommand{\advactSkip}{$\mathtt{skip}$}
\newcommand{\advactStop}{$\mathtt{stop}$}
\newcommand{\advactReveal}{$\mathtt{reveal}$}
\newcommand{\advactListen}{$\mathtt{listen}$}
\newcommand{\advactSend}{$\mathtt{send}$}

\definecolor{darkspringgreen}{rgb}{0.09, 0.45, 0.27}

%% file: 1_intro.tex
\section{Introduction\label{sec:intro}}

Distributed Systems (DS), by virtue of their decentralized nature, complexity and scale, present a unique set of security challenges. 
While decentralization might favor fault tolerance, it also introduces vulnerabilities.
Indeed, in addition of providing a greater surface of attack, most DS require specific global properties to hold (e.g., for blockchains a common property is the coherence of the replicated state, expressed as the absence of forks).
Each sub-system, each connection linking them, and even properties of communication protocols in use are potential targets for malicious entities. 
How then can we ensure the security of DS?

Cybersecurity often involves perimeter-based defense \cite{perimeter_based_defense_against_high_bandwidth_ddos_attacks}, ensuring that external threats are kept at bay.
However, with DS, where there might not always be a clear ``inside'' or ``outside'', these approaches might fall short.
The alternative, which we pursue, is that of modeling the adversary as an agent that is an integral part of the DS.
Adversary modeling \cite{dolev_yao_on_the_security_of_public_key_protocols} has been initially introduced to reason about cryptographic protocols \cite{sok_computer_aided_cryptography} but has since been extended to various fields in computer science and security research \cite{the_role_of_the_adversary_model_in_applied_security_research}.
The use of adversary models can facilitate the evaluation of security properties.
Indeed, like any other model, an adversary model, provided it has a well-defined semantics, can be used in formal verification (e.g., model checking) or in testing (e.g., via simulation).

In this paper, we propose an adversary model that builds upon the notions of assumptions, goals, and capabilities, as introduced in \cite{the_role_of_the_adversary_model_in_applied_security_research}, while being specifically tailored to assess the security of distributed systems at the protocol level. Importantly, our model targets attacks on distributed systems that combine multiple distributed protocols.

In this context, adversary goals capture the attacker's intent, while capabilities define the set of actions available to achieve those goals. Our focus is on adversaries aiming to compromise classical properties of distributed protocols, such as safety and liveness \cite{recognizing_safety_and_liveness,order_fairness_for_byzantine_consensus}.
Adversary capabilities are modeled as atomic actions that can be executed at any time during system execution. These capabilities are bounded by a set of assumptions reflecting the adversary’s environment and resources.
In contrast to the all-powerful network adversaries commonly used in cryptographic protocol verification \cite{sok_computer_aided_cryptography}, our model considers realistic, resource-limited adversaries constrained by the assumptions of the distributed protocols under study. These constraints stem from the underlying communication model \cite{impossibility_of_distributed_consensus_with_one_faulty_process,consensus_in_the_presence_of_partial_synchrony}, failure model \cite{schneider_failure_models}, and fault tolerance thresholds, as explored in the resource-limited adversary framework \cite{a_secure_control_framework_for_resource_limited_adversaries}.

In this paper, we apply our model to demonstrate the feasibility of attacks on a distributed network of blockchain nodes implementing HyperLedger Fabric (HF) \cite{an_in_depth_investigation_of_the_performance_characteristics_of_hyperledger_fabric}. Our choice of HF is motivated by two factors. On one hand, HF stands out as a popular choice for industrial blockchain applications. On the other hand, its structure that combines two services (an endorsing service for transaction validation and an ordering service for sequencing and appending transactions to the blockchain) enables us to demonstrate our approach by combining attacks on the different protocols implementing these services in a non-trivial manner.

Let us note that although HF is a permissioned blockchain, which means that participants must be authorized to be part of the system, HF nodes can be deployed in a wide-area network (e.g., the Internet) as opposed to a private intranet. As a result, HF is vulnerable to attacks \cite{a_framework_for_classifying_denial_of_service_attacks,yataglass_network_level_code_emulation_for_analyzing_memory_scanning_attacks,systematic_classification_of_side_channel_attacks} that can consist in either or both the adversary taking control of some of its constituting nodes, or the adversary otherwise manipulating exchanges between these nodes (e.g., increasing transmission delays via e.g., having control over routers, or via performing Denial of Service \cite{a_framework_for_classifying_denial_of_service_attacks}).

In terms of attacks, we focus on a specific class of liveness attacks known as fairness attacks. Fairness refers to ensuring that all users, or clients, have equally fair access to the blockchain, meaning that no client’s transactions are systematically favored over others. In consortium blockchains, censorship attacks targeting specific participants are a genuine concern. As such, evaluating the robustness of distributed protocols against fairness attacks is a critical aspect of ensuring system security. More specifically, we examine four fairness properties: an application-specific form of client fairness, whose violation is the adversary’s primary goal, and three types of order fairness \cite{order_fairness_for_byzantine_consensus,quick_order_fairness}, which relate the order in which transactions are received by individual nodes to the order in which they are ultimately committed to the blockchain.

In this paper, we show that even when the adversary operates within the tolerance thresholds of the underlying protocols — such as the proportion of compromised participants and assumptions about the communication and failure models — it is still possible to violate the client-fairness property in HF. Moreover, the different strategies used to achieve this violation have varying impacts on the associated order-fairness properties.

In \cite{adversary_augmented_simulation_to_evaluate_order_fairness_on_HF}, we introduced our adversary model and implemented it within the MAX \cite{max_tool} multi-agent simulation tool. We then defined several attacks on HF and used our simulator to demonstrate that even a weak adversary can negatively impact the fairness of a system based on HF.

Our framework also enabled the quantification of order-fairness violations, allowing us to assess the effects of our attacks on both the ordering and endorsing services of HF. To the best of our knowledge, this was the first time a blockchain simulator was augmented with a programmatic adversary \cite{a_systematic_review_and_empirical_analysis_of_blockchain_simulators}, that these specific attacks on HF were described, and that an empirical evaluation of order-fairness violations was explored.

This paper extends our previous work \cite{adversary_augmented_simulation_to_evaluate_order_fairness_on_HF} in five key ways:
\textbf{(1)} we broaden the discussion of related work, particularly regarding order fairness;
\textbf{(2)} we enhance our initial simulation framework by incorporating more realistic network assumptions—especially in terms of probabilistic delay distributions—and by collecting and analyzing additional metrics;
\textbf{(3)} we introduce a mitigation mechanism aimed at improving HF's resilience to transaction reordering attacks;
\textbf{(4)} we observe and evaluate the effects of a specific implementation of this mitigation mechanism within our simulations; and
\textbf{(5)} we vary additional simulation parameters—such as the number of peers, orderers, and clients—to demonstrate the generalizability of our findings.

This paper is organized as follows. 
Sec.\ref{sec:prelim} introduce preliminary notions and discusses related works.
In Sec.\ref{sec:adversarial_for_ds}, we recall in details the adversary model from \cite{adversary_augmented_simulation_to_evaluate_order_fairness_on_HF}.
Sec.\ref{sec:usecase} presents the HF system on which we apply our approach as well as the specific fairness properties that we consider.
The attacks that can be performed by the adversary are defined in Sec.\ref{sec:attacks}, and simulated in Sec.\ref{sec:sim_attacks}.
In Sec.\ref{sec:mitig_def}, we define our mitigation mechanism, and then observe its impact in Sec.\ref{sec:sim_mitig}.
Finally, after studying the impact of the network composition in Sec.\ref{sec:sim_params}, we conclude in Sec.\ref{sec:conclusion}.

%% file: 2_0_prelims.tex
\section{Preliminaries and Related Works\label{sec:prelim}}

\input{2_1}

\input{2_2}

\input{2_3}

\input{2_4}

\input{2_5}

\input{2_6}

\input{2_7}

%% file: 2_1.tex
\subsection{Communication and failure models}

Distributed protocols specify patterns of {\em communications} between distant systems with the aim of {\em performing} a service.
These services are characterized by {\em properties} often related to {\em safety} and {\em liveness} \cite{recognizing_safety_and_liveness}.
Here, {\em communications} involve message passing between sub-systems of a Distributed System (DS) over a network. Three distinct {\em communication models} \cite{consensus_in_the_presence_of_partial_synchrony} define assumptions that hold over message passing. 
In the {\em synchronous model} \cite{impossibility_of_distributed_consensus_with_one_faulty_process}, there is a finite time bound $\Delta$ (that can be known by the involved parties) s.t., if a message is sent at time $t$, it must be received before $t + \Delta$.
By contrast, the {\em asynchronous model} \cite{impossibility_of_distributed_consensus_with_one_faulty_process} allows an arbitrary delay between emission and reception.
With the {\em eventually synchronous model} \cite{consensus_in_the_presence_of_partial_synchrony}, communications are initially asynchronous, but there is an unknown Global Stabilization Time (GST) after which they become synchronous.
Another equivalent definition of the {\em eventually synchronous model} (without GST) is to consider that the bound $\Delta$ is not known by any of the involved parties \cite{consensus_in_the_presence_of_partial_synchrony,themis_fast_strong_order_fairness_in_byzantine_consensus}.

%% file: 2_2.tex
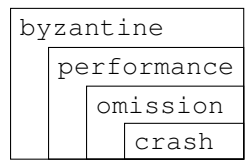
\begin{wrapfigure}{l}{3cm}
\vspace*{-.1cm}
    \centering

\begin{tikzpicture}
\node (n0) at (0,0) {\texttt{byzantine}};
\node[below right=.25cm and .5cm of n0.west] (n1) {\texttt{performance}};
\node[below right=.25cm and .5cm of n1.west] (n2) {\texttt{omission}};
\node[below right=.25cm and .5cm of n2.west] (n3) {\texttt{crash}};
\draw (n1.east |- n3.south) rectangle (n3.west |- n3.north);
\draw (n1.east |- n3.south) rectangle (n2.west |- n2.north);
\draw (n1.east |- n3.south) rectangle (n1.west |- n1.north);
\draw (n1.east |- n3.south) rectangle (n0.west |- n0.north);
\end{tikzpicture}

    \caption{Failures}
    \label{fig:failure_models}
\vspace*{-.1cm}
\end{wrapfigure}

Distributed protocols are deployed in an environment consisting of a DS with various sub-systems, each corresponding to a running process. 
The individual failure of such processes may negatively impact the service performed by the protocol (i.e., the associated properties may not be upheld).
{\em Failure models} \cite{schneider_failure_models} (see Fig.\ref{fig:failure_models}) define assumptions on the types of failures that may occur.

A {\em crash} failure involves a process terminating prematurely.
An {\em omission} failure occurs when it never delivers an event (e.g., never sends a message it is expected to send). As illustrated on Fig.\ref{fig:failure_models}, a crash is a specific omission where, after a certain time, all subsequent events are never delivered.
With the {\em performance} failure model, only expected events (considering the protocols that nodes implement) occur, but the time of their occurrence may be overdue.
Omission failures are infinitely late performance failures. Finally, {\em Byzantine} failures authorize any arbitrary behavior.

Some distributed protocols are designed to uphold specific properties, even in cases where a number of processes (sub-systems) fail.
These {\em Fault Tolerant} (FT) protocols \cite{schneider_failure_models} are characterized by the nature of the failures they can withstand (i.e., a failure model) and a threshold (usually a proportion of involved processes) of failures below which they maintain their properties. 
For instance, Tendermint \cite{dissecting_tendermint} is a Byzantine FT (BFT) consensus algorithm (the property it upholds being that of maintaining consensus agreeement) that can withstand lessup to one-third of faulty processes.

%% file: 2_3.tex
\subsection{Ledgers \& Transaction Reordering}

A Distributed Ledger is a network of replicated state machines that maintain a consistent state. By its decentralized nature, it has no single point of control/failure, which improves security. The replication of the data it carries also improves transparency and auditability.
Blockchains are a means to implement a Distributed Ledger.
In a Blockchain, individual machines are referred to as ``nodes'' and changes in the state machine occur when a transaction is ``delivered''.
We call ``clients'' the entities that submit these transactions to the nodes.
By agreeing on the initial state and the order in which transactions are delivered, the nodes maintain a consensus on the current state.
To improve throughput (i.e., the rate at which transactions can be delivered), transactions are batched into consecutive blocks.
The order of delivery of transactions then corresponds to the order in which
they appear on and within blocks.

The properties of distributed ledgers which are mostly studied can be categorized as either ``\textit{consistency properties}'', ``\textit{liveness properties}'' or ``\textit{fairness properties}''.
While the first may correspond to consensus agreement and validity which are safety properties \cite{recognizing_safety_and_liveness}, the second may refer to various distinct notions such as starvation-freedom, local progress or consensus wait-freedom \cite{safety_liveness_exclusion_in_distributed_computing}.
Fairness, in the context of Blockchains \cite{on_fairness_in_commitee_based_blockchains,on_fairness_in_voting_consensus_protocols} may refer to various notions, including the fairness in committee selection \cite{on_fairness_in_commitee_based_blockchains}, rewarding \cite{do_the_rich_get_richer_fairness_analysis_for_blockchain_incentives} or the ability to take decisions \cite{on_fairness_in_voting_consensus_protocols,fairledger_a_fair_blockchain_protocol_for_financial_institutions} w.r.t.~individual nodes' voting power.
Consistency \textbf{(1)}, liveness \textbf{(2)} and the aforementioned fairness properties (in e.g., committee selection or rewards) \textbf{(3)} may resp.~enforce that \textbf{(1)} all nodes deliver the exact same transactions in the same order, \textbf{(2)} it is always possible for an honest client to submit a transactions that will eventually be delivered or \textbf{(3)} the rules for nodes to participate in the consensus process are fair and they are rewarded according to their participation.

However, as explained in \cite{order_fairness_for_byzantine_consensus}, these properties enforce no constraint on the agreed upon order of transactions.
In that case, even though the involved algorithms may be BFT, an adversary may still be able to manipulate the order of transactions.
A malicious adversary which causes the order of transactions to change (i.e., the consensus yields a different total order as the one which would have occurred without the intervention of the adversary) is said to perform a ``transaction reordering attack'' \cite{sok_preventing_transaction_reordering_manipulations_in_decentralized_finance}.
A typical example is that of front-running : if the adversary is aware that a transaction $x$ has been submitted but not yet delivered, it may submit a new transaction $x'$ and make so that $x'$ is delivered before $x$ (e.g., by leveraging reward mechanisms \cite{flash_boys_frontrunning_in_decentralized_exchanges_miner_extractable_value_and_consensus_instability} or by coordinating Byzantine nodes that it controls \cite{adversary_augmented_simulation_to_evaluate_order_fairness_on_HF}).
In decentralized finance \cite{flash_boys_frontrunning_in_decentralized_exchanges_miner_extractable_value_and_consensus_instability}, front-running can be leveraged to extract profits via manipulating the value of financial assets.
Maximum Extractable Value (MEV) bots can continuously monitor a distributed ledger and perform such attacks opportunistically \cite{flash_boys_frontrunning_in_decentralized_exchanges_miner_extractable_value_and_consensus_instability,sok_preventing_transaction_reordering_manipulations_in_decentralized_finance}.
It is estimated that MEV bots have extracted around $\sim675$ million \$ on the Ethereum blockchain alone in the span between January 2020 and September 2022, which underlines the importance of being concerned with transaction ordering.

%% file: 2_4.tex
\subsection{Order Fairness\label{ssec:related_works_order_fairness}}

The rise in potential applications of transaction reordering attacks lead to an increased interest in properties related to the fairness with which the order of transactions is decided.
In the context of decentralized finance, in \cite{sok_preventing_transaction_reordering_manipulations_in_decentralized_finance}, ``fairness'' is achieved whenever participants cannot include, exclude or front-run a transaction after having seen its content.
Formal definitions of similar concepts exist in the literature.
Using an older notion of ``order'' \cite{implementing_fault_tolerant_services_using_the_state_machine_approach_a_tutorial} in State Machine Replication, \cite{sok_a_consensus_taxonomy_in_the_blockchain_era} defines ``serializability'' in ledger consensus.
This ``serializability'' coïncides \footnote{in \cite{sok_a_consensus_taxonomy_in_the_blockchain_era}, the ``majority'' is defined as the set of all honest nodes and ``reception'' of a transaction $x$ by a node signifies $x$ enters the node's local view of the mempool} with the ``\textit{receive-order fairness}'' of \cite{order_fairness_for_byzantine_consensus} which is defined as follows : for any two transactions $x$ and $x'$, if a majority of nodes receive $x$ before $x'$ then $x$ is delivered before $x'$.
In the following, we will use the terminology {\em order-fairness} (OF) to refer to such properties.
OF properties relate the order in which transactions are delivered to some partial orders on communication events occurring in the distributed system.
\textit{receive-order fairness} is a particular case of OF property that is defined with regards to reception events.

\paragraph{OF w.r.t.~receptions events.}
As per
\cite{order_fairness_for_byzantine_consensus}, \textit{receive-order fairness} is impossible to achieve. Indeed, one might notice that for three transactions $x_1$, $x_2$ and $x_3$, it occur that a majority of nodes receive $x_1$ before $x_2$, $x_2$ before $x_3$ and $x_3$ before $x_1$, hence forming a Condorcet cycle \cite{condorcet_attack_against_fair_transaction_ordering}.

As a result, \cite{order_fairness_for_byzantine_consensus} proposes a weaker property of ``\textit{block-order fairness}'' and introduces the ``Aequitas'' consensus protocol that upholds it.
Block-order fairness is defined as follows : for any two transactions $x$ and $x'$, if a majority of nodes receive $x$ before $x'$ then no honest node can deliver $x$ in a block after the block in which $x'$ is delivered.
However, in \cite{order_fairness_for_byzantine_consensus}, the meaning of these blocks is more specific than the generic notion of block in a Blockchain. Indeed, they rather correspond to batching together exactly the transactions that are in the same Condorcet cycle (i.e., transactions that are not in a Condorcet cycle together must not be put in the same block).
\cite{themis_fast_strong_order_fairness_in_byzantine_consensus} clarifies this via defining ``\textit{$\gamma$-(all)-batch-order fairness}'' as follows : for $\gamma > 1/2$, if a proportion of nodes greater than $\gamma$ receives $x$ before $x'$ and if $x$ and $x'$ are not in a Condorcet cycle, then $x$ must be delivered before $x'$. \cite{themis_fast_strong_order_fairness_in_byzantine_consensus} also introduces the Themis algorithm that upholds \textit{$\gamma$-(all)-batch-order fairness}.

\cite{quick_order_fairness} introduces ``\textit{differential-order fairness}'' and a ``quick order-fair atomic broadcast'' protocol that upholds it.
This property is defined as follows: for any two transactions $x$ and $x'$, if the number of honest nodes that receive $x$ first exceeds by more than $2*f$ (where $n = 3*f +1$ is the total number of nodes) the number of nodes that receive $x'$ first, then $x$ must be delivered before $x'$.

These properties encode an intuitive notion of fairness in consensus ordering, as the delivery order, if ``fair'', should mimic the ``reception'' order. 
Protocols that uphold such OF properties are called Algorithmic Committee Ordering algorithms in \cite{sok_preventing_transaction_reordering_manipulations_in_decentralized_finance}.

However, most existing consensus algorithms and Blockchain systems were not designed with these properties in mind. As a result, assessing the propensity of an existing Blockchain system to violate such OF properties can be difficult.
In particular, the notion of ``reception'' may be ambiguous. 
For instance, it may be so that clients do not broadcast their transactions to all the nodes (then the instants of ``reception'' do not necessarily exist). 
In Blockchains using the execute-order-validate architecture, different subsets of nodes might perform different tasks and thus do not receive transactions in the same manner (in HyperLedger Fabric \cite{vulnerabilities_on_hyperledger_fabric}, peers receive transactions before orderers do).

\paragraph{Limitations of OF defined w.r.t.~receptions.}
If a protocol upholds an order-fairness property that is defined w.r.t.~reception events, it only implies that the delivery order cannot be tampered with given certain reception orders (one per participating node). 
However, an adversary may actually tamper with the reception orders themselves, in order to influence the delivery order.
For instance, if the adversary benefits from a quicker network connection than honest nodes and clients (lower latency when the adversary receives and sends transactions), it may listen to an incoming transaction $x$ and front-run it via submitting $x'$ and ensuring that most nodes receive $x'$ before $x$ \cite{sok_preventing_transaction_reordering_manipulations_in_decentralized_finance}.
In that case, upholding receive-order-fairness guarantees the front-running will succeed.

But even without the reduced latency requirement, an adversary may still succeed. 
Let us consider a consensus algorithm that upholds a form of \textit{$\gamma$-(all)-batch-order fairness}.
Then, as described in \cite{condorcet_attack_against_fair_transaction_ordering} the adversary may artificially create Condorcet cycles by submitting transactions in a specific manner : for any transaction $x''$, it immediately (instant \textbf{(1)}) sends $x''$ to half of the nodes and then waits before (instant \textbf{(2)}) sending $x''$ again to the other half of the nodes. Indeed, it is then likely that $x''$ will form a Condorcet cycle, ``trapping'' a number of transactions that were submitted between instants \textbf{(1)} and \textbf{(2)}.
Considering our previous front-running example, if both $x$ and $x'$ are trapped in the Condorcet cycle formed by $x''$, it is more likely for the front-running to succeed (given that no constraint is enforced on the ordering within a block and that $x$ and $x'$ are, as a result, in the same block).

As a result, as shown in \cite{condorcet_attack_against_fair_transaction_ordering}, existing Algorithmic Committee Ordering algorithms are not robust against all forms of transaction reordering attacks.

\paragraph{Blind-order fairness.} \cite{maximal_extractable_value_protection_on_a_DAG,fairblock_preventing_blockchain_front_running_with_minimal_overheads} (called ``on-chain commit and reveal'' in \cite{sok_preventing_transaction_reordering_manipulations_in_decentralized_finance}) solves the problem of transaction order manipulation by hiding the content of the transactions until their delivery order has been decided. As the adversary ignores the content of the transactions, it cannot easily identify and choose the transactions it wants to reorder (except through some leakage of metadata or if it just wants to front-run blindly \cite{fairblock_preventing_blockchain_front_running_with_minimal_overheads}).
Blind-order fairness can be achieved in various ways. Time-lock puzzles \cite{time_lock_puzzles_and_timed_release_crypto} or delay encryption \cite{delay_encryption} enable time-based implementations (i.e., decryption is only possible after an amount of time has passed, e.g., that of solving a puzzle). Verifiable secret sharing and threshold encryption \cite{how_to_share_a_secret} enable communication-based implementations (i.e., knowledge of the key for decrypting transactions is split in shares that are only gathered after delivery).
In any case, blind-order fairness requires significant overhead (computational and communication costs) \cite{fairblock_preventing_blockchain_front_running_with_minimal_overheads} and may negatively impact latency and throughput \cite{sok_preventing_transaction_reordering_manipulations_in_decentralized_finance}.
Moreover, even though it may limit the ability of an adversary to perform transaction reordering attacks, blind-order fairness does not reduce the number of order-fairness violations that may naturally occur (due to e.g.,~network delays and non-deterministic communications) in the system.
In other words, implementing blind-order fairness does not make the system ``fair'' but only prevents an adversary to make it ``more unfair'' than it already is (see also the notion of ``\textit{bounded unfairness}'' in \cite{ordering_transactions_with_bounded_unfairness}).

\paragraph{Send-order fairness.} Instead of considering the orders with which nodes receive transactions, ``\textit{send-order fairness}'' \cite{order_fairness_for_byzantine_consensus} relates the orders in which transaction are emitted (by clients) to the order in which they are delivered. 
Unlike the variants of \textit{receive-order fairness}, upholding \textit{send-order fairness} may prevent an adversary with a better network connection to perform front-running as it cannot possibly make so, after receiving a transactions $x$ and sending $x'$ to front-run it, that the time of emission of $x'$ is lower than that of $x$.
However, actually implementing a system that upholds \textit{send-order fairness} remains an unsolved problem \cite{order_fairness_for_byzantine_consensus}.
Indeed, it requires keeping track of the global order in which emission events occur across all the clients.
Because distant machines have uncorrelated local clocks, and because, in any case, Byzantine clients may falsify timestamps, one cannot simply rely on trusting clients to report the time at which they send transactions.
A potential solution would be to use a Byzantine Fault Tolerant vector clock algorithm.
However, such algorithms do not exist (it is impossible in an asynchronous network as per \cite{byzantine_fault_tolerant_causal_ordering}, at most, we have Crash Fault Tolerance for vector clocks \cite{cft_vector_clock}).
A partial solution would be to combine Trusted Execution Environments and a Clock Synchronization algorithm \cite{an_overview_of_clock_synchronization} such as NTP \cite{rfc5905}.
In addition of the communications overhead they incur, such algorithms only guarantee that at any given time, the difference in the readings of any pair of local clocks do not exceed a certain maximal skew $\delta$ \cite{distributed_runtime_verification_under_partial_synchrony}.
This is problematic in cases where the difference between the timestamps of events is smaller than $\delta$.
Still, the notion of \textit{send-order fairness} is particularly pertinent as it is the most accurate description of order fairness.
Moreover, in a controlled environment (e.g., in a simulator), it can actually be monitored.

\paragraph{Fairness to clients.} 
\textit{Send-order fairness} is, intrinsically, a form of ``fairness to clients''. More generally, {\em client-fairness} refers to nodes treating clients fairly, which includes the order with which transactions coming from distinct clients should be handled \cite{byzid_byzantine_fault_tolerane_from_intrusion_detection,rbft_redundant_byzantine_fault_tolerance}. 
Jain's fairness index \cite{a_quantitative_measure_of_fairness_and_discrimination_for_resource_allocation_in_shared_computer_systems} quantifies fairness between $m$ client sharing resources. Jain's index is $1$ if all clients receive the same allocation and $k/m$ (for any $k \in [1,m]$) if $m-k$ clients receive no allocation and the $k$ others equally share what remains. Jain's index represents a global notion of fairness which may also be used to quantify ``fair access to transactions from all clients''. In that context, value $1$ would mean that transactions from all clients are treated equally and value $0$ that the transactions coming from one specific client are always prioritized over the others.
The ``client-fairness score'' from \cite{adversary_augmented_simulation_to_evaluate_order_fairness_on_HF} describes a similar notion.
However, whereas Jain's fairness index is global, this score is defined on a ``per client'' basis, as it evaluates the fairness towards a specific client.

\paragraph{Bounded unfairness.} The ``strongest'' (i.e., closest to fairness to the clients) notion of order-fairness is send-order fairness. However, we have seen that it is not achievable due to lacking mechanisms to maintain a BFT global clock across distant clients. The second strongest is receive-order fairness which is also impossible to uphold due to the Condorcet paradox \cite{order_fairness_for_byzantine_consensus}. The variants of batch-order fairness \cite{themis_fast_strong_order_fairness_in_byzantine_consensus,quick_order_fairness} simply ignore the problem by ignoring pairs of transactions that violate receive-order fairness if they are in the same Condorcet cycle. However, in practice, distributed ledgers must output a total order on transactions (i.e., transactions from the same Condorcet cycle cannot remain unordered). We have also seen that an adversary may perform Condorcet attacks \cite{condorcet_attack_against_fair_transaction_ordering} to force sets of transactions into the same cycle. In turn, this render protocols upholding batch-order fairness powerless as they only guarantee order-fairness between transactions that are not in the same cycle.
As for blind-order fairness \cite{fairblock_preventing_blockchain_front_running_with_minimal_overheads}, it may only prevent an adversary from making the system more unfair than it already is, provided there is no metadata leakage and at the cost of computation and/or communication overheads.
As a result, there are no ideal solution to the problem of providing an order-fair total order of transactions.
Still, instead of trying to provide a fair order, one can approximate a solution by minimizing unfairness in the order.

This is what is proposed in \cite{ordering_transactions_with_bounded_unfairness} via the notion of ``{\em bounded unfairness}''. Given a threshold $\gamma > 1/2$ and a bound $\Omega \in \mathbb{N}$, a ledger satisfies ``{\em $\gamma$-$\Omega$-bounded-unfairness}'' if, for any pair of transaction $x$ and $x'$, if a proportion of nodes greater than $\gamma$ receives $x$ before $x'$ then $x$ cannot be delivered at an index that is greater than $\Omega$ plus the index at which $x'$ is delivered. We can see that when $\Omega = 0$, bounded-unfairness coincides with the classical formulation of {\em receive-order fairness} and is therefore impossible to uphold.
\cite{ordering_transactions_with_bounded_unfairness} proves that there exists a lower bound for the value of $\Omega$ from which it becomes possible to guarantee {\em $\gamma$-$\Omega$-bounded-unfairness}.
\cite{ordering_transactions_with_bounded_unfairness} also shows that the Algorithmic Committee Ordering algorithms from \cite{order_fairness_for_byzantine_consensus,themis_fast_strong_order_fairness_in_byzantine_consensus,quick_order_fairness} (resp.~Aequitas, Themis and the quick-order-fair broadcast) as well as other timestamp-based methods (Pompe from \cite{byzantine_ordered_consensus_without_byzantine_oligarchy} and Wendy from \cite{wendy_the_good_little_fairness_widget}) do not uphold {\em $\gamma$-$\Omega$-bounded-unfairness} with this minimal bound.

Moreover, \cite{ordering_transactions_with_bounded_unfairness} proves that calculating this bound (which depends on the actual transactions that are considered and their dependencies i.e., wether or not a $\gamma$-fraction of nodes has received one before the other) and a total order of transactions such that {\em $\gamma$-$\Omega$-bounded-unfairness} is upheld corresponds to solving a NP-hard problem.
Thus, any protocol that would guarantee the minimal unfairness in its ordering would not satisfy liveness.
\cite{ordering_transactions_with_bounded_unfairness} details one such protocol called Taxis$_{WL}$ ($WL$ standing for Weak Liveness).
To conclude, any attempt at providing order-fairness in transaction ordering either \textbf{(1)} only solves part of the problem (via batch-order fairness, which only provides a partial instead of a total order), \textbf{(2)} only focuses on making the order impervious to manipulations (via blind-order fairness) or \textbf{(3)} yields a protocol that does not satisfy liveness.

Reconciling order-fairness and liveness necessary requires upholding weaker forms of order-fairness. This means that the total order that is output is not necessarily the ``best'' but is chosen according to a certain heuristics.
In the literature, the most common mechanisms used to implement such heuristics involve reasoning on timestamps. 
The Pompe algorithm from \cite{byzantine_ordered_consensus_without_byzantine_oligarchy} uses the median timestamps of the times at which nodes in a quorum receive a transaction to determine its delivery order. Because there are at most $f$ Byzantine nodes (which may maliciously provide exceedingly low or high values), using the median value guarantees that the retained value is greater or equal than a timestamp send by an honest node and also lower or equal than a timestamp send by another honest node.
In the live version of Taxis$_{WL}$ (simply called Taxis) \cite{ordering_transactions_with_bounded_unfairness}, a timeout mechanism is used to break long cycles of dependencies between transactions and transactions within the same Condorcet cycle are sorted according to the median timestamp.
In \cite{condorcet_attack_against_fair_transaction_ordering}, the Themis algorithm from \cite{themis_fast_strong_order_fairness_in_byzantine_consensus} is extended with a method to sort transactions within Condorcet cycles. Instead of relying on a median timestamp, it uses Tideman's Ranked Pairs algorithm.

%% file: 2_5.tex
\subsection{Adversary models}

To assess the robustness of a DS, it is a common practice (from Cybersecurity) to consider an attacker actively trying to harm it.
Adversary models describe such attackers \cite{the_role_of_the_adversary_model_in_applied_security_research}.
The level of abstraction of these models may vary from natural language to concrete algorithms and implementations.
Historically, adversary models such as the symbolic Dolev-Yao \cite{dolev_yao_on_the_security_of_public_key_protocols} model were central to the design of provably-secure cryptographic schemes.
In formal verification of (cryptographic) security protocols, whether a symbolic or computational model is used, the adversary is all powerful on the network (i.e., it can intercept all messages and compute and send new messages to any node) but has no power outside of it (i.e., side channels attacks or code injection are not modeled) \cite{sok_computer_aided_cryptography}.
However, in other fields of computer sciences, in which the use of adversary models remain limited \cite{the_role_of_the_adversary_model_in_applied_security_research}, these assumptions on the power of the adversary may not be desirable.

In \cite{the_role_of_the_adversary_model_in_applied_security_research}, a more generic description of adversary models according to three aspects is discussed. These correspond to the adversary's \textbf{(1)} assumptions, \textbf{(2)} goals and \textbf{(3)} capabilities.
Assumptions involve the conditions under which the adversary may act. 
Goals correspond to the adversary's intentions (which are related to information retrieval in most of the literature on cryptography).
Capabilities synthesize all the actions the adversary may perform.
A passive attacker may only eavesdrop on message passing without any tampering while an active attacker may, among other things, intercept and modify messages (Man-In-The-Middle attack).

Certain assumptions may bind the capabilities of adversaries.
Adaptability \cite{efficient_multiparty_computations_secure_against_an_adaptive_adversary} refers to the ability of the adversary to update its plan i.e., the choice of its victims and of which adversarial actions to perform. 
While static adversaries have a fixed plan (established before the execution of the system), adaptive adversaries may make choices at runtime.
Threshold cryptography \cite{how_to_share_a_function_securely} was introduced as a means to share a secret securely among a fixed set of participants, a threshold number of which being required to access it. Hence, adversaries attacking such protocols within its assumptions must not be able to infect more participants than the threshold, thus bounding their power.
By extension, adversarial actions can be limited by a corresponding resource as in \cite{a_secure_control_framework_for_resource_limited_adversaries} (bounded resource threshold adversaries), or via a more abstract notion of budget.

Various adversary models have been designed specifically for Blockchain systems. For instance, that from \cite{xiao2020modeling} focuses on network connectivity (i.e.,~the adversary only performs network related actions) and how this can be exploited to impact the consensus mechanism.
The limitation of adversary models to only represent network related actions stem from their origin in cryptographic protocol verification \cite{sok_computer_aided_cryptography}, in which the adversary has a total control over the network.

The adversary model from \cite{adversary_augmented_simulation_to_evaluate_order_fairness_on_HF} (which this present paper extends) allows modeling an adversary which manipulates the network. 
However, it may additionally target individual subsystems (thus modeling e.g., side channel or code injection). Moreover, it incorporates notions of communication models \cite{consensus_in_the_presence_of_partial_synchrony}, failure models \cite{schneider_failure_models} and Fault Tolerance thresholds to further limit the capabilities of the adversary \cite{a_secure_control_framework_for_resource_limited_adversaries}, which enables parameterizing attacks depending on assumptions under which involved distributed protocols are to be used. 
This approach also allows observing the side-effects of attacks (whether or not the adversary succeeds) on various metrics.
If the side effects of an attack are known, their observation might allow detecting the attack.

%% file: 2_6.tex
\subsection{Simulation}

Validating systems can either involve formal verification or testing which are two orthogonal approaches \cite{schneider_failure_models}.
The former involves techniques such as model checking or automated theorem proving which do not scale well with the complexity of verified systems and properties.
Indeed, in complex and dynamic DS involving several protocols we must then consider all possible correct as well as incorrect behaviors (due to failures up to the relevant failure model) combined with all possible interleaving and side effects between these behaviors (due to communications delays up to the relevant communication model and/or performance failures).
Although techniques can be used to negate (e.g., universal composability) or mitigate (e.g., partial order reduction or symbolic execution) this complexity, this is not always possible.
In this context, empirical security evaluations, in the form of testing, is more adapted.

Tests can be performed against a concrete implementation of the DS.
However, it may involve unexpected side-effects due to executing the whole implementation and hardware dependent protocol stack.
Similarly to software integration tests being performed via code isolation using mockups, one can focus on and isolate specific aspects of the DS via the use of a simulator in which parts of the protocol stack are abstracted away. 
Additionally, this allows a finer control over communications because they occur within the simulator and not on a network on which control is lacking.
Simulation also has the added advantage of being able to closely and easily monitor the simulated system e.g., by collecting metrics (which is more difficult on a live network or testnet deployed on a WAN).

As highlighted by the recent review \cite{a_systematic_review_and_empirical_analysis_of_blockchain_simulators}, blockchain simulators are important tools for understanding these complex systems. 
Yet, to our knowledge, none have been fitted with a programmatic adversary to simulate adversarial attacks.
Most simulators address specific blockchain systems (Bitcoin, Ethereum) and/or are oriented towards performance evaluations (latency \& throughput) rather than security aspects.
Shadow-Bitcoin \cite{miller2015shadow} allows replicating Bitcoin networks on a large scale, but lacks flexibility for non-Bitcoin blockchains. 
While lacking an adversary, some simulators can be adapted for security analyses such as eVIBES \cite{deshpande2018evibes} for Ethereum networks.

Multi-Agent-Systems (MAS) is an agent-oriented modeling paradigm which is particularly adapted to DS with large numbers of agents.
The behavior of each agent can be proactive (following a specific plan regardless of their environment) and reactive (reacting to stimuli e.g., incoming messages).
Agent Group Role (AGR) \cite{from_agents_to_organizations_an_organizational_view_of_multi_agent_systems} is a MAS framework which focuses on the interactions agents have by playing roles within groups.
MAX \cite{max_tool} is a simulator based on AGR that leverages MAS for blockchain networks.

%% file: 2_7.tex
\subsection{HyperLedger Fabric}

\cite{adversary_augmented_simulation_to_evaluate_order_fairness_on_HF} demonstrates an approach to computational studies of order-fairness via adversary-augmented simulation on a specific Blockchain system : HyperLedger Fabric (HF) \cite{an_in_depth_investigation_of_the_performance_characteristics_of_hyperledger_fabric}.
Although generally secure, HF has some known vulnerabilities.
If the addresses of peers are known by malicious entities, DoS \cite{vulnerabilities_on_hyperledger_fabric} might occur. To mitigate this, \cite{vulnerabilities_on_hyperledger_fabric} recommends anonymizing peers (e.g., using random verifiable functions and pseudonyms).
HF chaincode is vulnerable to (smart contract) programming errors \cite{dabholkar2019ripping} which can be mitigated by formal verification of smart contracts \cite{vulnerabilities_on_hyperledger_fabric} and updating deployed contracts \cite{putz2020detecting}. 
HF, like any other permissioned blockchain, is vulnerable to the compromise of the Membership Service Provider (MSP) \cite{dabholkar2019ripping}. 
Potential solutions include using secure hardware for registration and transactions \cite{vulnerabilities_on_hyperledger_fabric}, and monitoring requests to detect potential attacks \cite{putz2020detecting}.
HF's flexible consensus protocols have distinct strengths and weaknesses. \cite{putz2020detecting} points out the vulnerability to Network Partitioning from internal attackers affecting network routing, identifiable via methods like BGP hijacking and DNS attacks. Its Gossip protocol, essential for block delivery, is susceptible to Eclipse attacks \cite{dabholkar2019ripping}.

The attacks on HF defined in \cite{adversary_augmented_simulation_to_evaluate_order_fairness_on_HF} have not been described previously \cite{evaluating_blockchain_systems_a_comprehensive_study_of_security_and_dependability_attributes} and correspond to transaction reordering attacks \cite{sok_preventing_transaction_reordering_manipulations_in_decentralized_finance}.

%% file: 3_0_adv_model.tex
\section{Our adversary model\label{sec:adversarial_for_ds}}

\input{3_1}

\input{3_2}

\input{3_3}

\input{3_4}

\input{3_5}

%% file: 3_1.tex
In this section we define our adversary model.
Fig.\ref{fig:adversary_model_description} illustrates it following the approach from \cite{the_role_of_the_adversary_model_in_applied_security_research}.

%% file: 3_2.tex
In our context, the adversary's environment is the DS it aims to harm.
We formalize it as a set $S$ of sub-systems.
At any given time, each sub-system $s \in S$ has a certain state (defined by e.g., the current values of its internal state variables) in a state space $\Gamma_s$.
The state space of the overall system, which is the product of its sub-systems', is $\Gamma = \prod_{s \in S} \Gamma_s$.
At any given time during its execution, the state of the DS is a certain $\eta \in \Gamma$.

\begin{figure}[h]
    \centering

\begin{tabular}{|c|c|c|}
\hline 
{\small\textbf{Assumptions}}
&
{\small\textbf{Goals}}
&
{\small\textbf{Capabilities}}
\\
\hline 
\makecell[l]{
\scriptsize Environment (system \& assumptions):\\
\scriptsize - Communication Model\\
\scriptsize - Failure Model\\
\scriptsize Resources (binding capabilities):\\
\scriptsize - Knowledge\\
\scriptsize - Fault-Tolerance thresholds\\
}
&
\makecell[c]{
\scriptsize property\\\scriptsize violation
}
&
\makecell[l]{
\scriptsize - \underline{process discovery}\\
\scriptsize - \underline{adaptation}\\
\scriptsize - adversarial actions\\
}
\\
\hline 
\end{tabular}

    \caption{Our adversary model {\scriptsize(\underline{adaptive adversaries} underlined})}
    \label{fig:adversary_model_description}
\end{figure}

The goals of the adversary must be clearly defined so that the success or failure of attacks can be ascertained.
In the following, we consider that goal to be to invalidate a property $\phi$ of the system, and we denote it as a First Order Logic \cite{an_introduction_to_first_order_logic} formula.
Given a state $\eta \in \Gamma$ of the system, the property can be either satisfied (i.e.~$\eta \models \phi$) or not satisfied (i.e.~$\eta \not\models \phi$).
Thus, the goal of the adversary is to lead the system to a state $\eta$ s.t.~$\eta \not\models \phi$.

For instance, let us consider a DS $S$ with two sub-systems $s_1$ and $s_2$ which must agree on a value $x$ stored as $x_1$ in $s_1$ and $x_2$ in $s_2$. Before agreement is reached, the value of $x$ is undefined which we may denote as $x = \emptyset$.
After consensus, the values of $x_1$ and $x_2$ must be the same.
This safety property of correct consensus can be described using $\phi = (x_1 = \emptyset) \vee (x_2 = \emptyset) \vee (x_1=x_2)$.
Given a state $\eta \in \Gamma$, in order to check whether or not $\phi$ holds it suffices to verify that $\eta(\phi)$ (i.e., $(\eta(x_1) = \emptyset) \vee (\eta(x_2) = \emptyset) \vee (\eta(x_1)=\eta(x_2))$) holds.

The expressiveness of this approach is only limited by the expressiveness of the language that is used to define $\phi$ and the state variables of $\Gamma$. Both global and local variables can be used. If the adversary has several goals we can use disjunctions (resp.~conjunctions) to signify that it suffices for one of these to be (resp.~requires that all of these are) fulfilled.

%% file: 3_3.tex
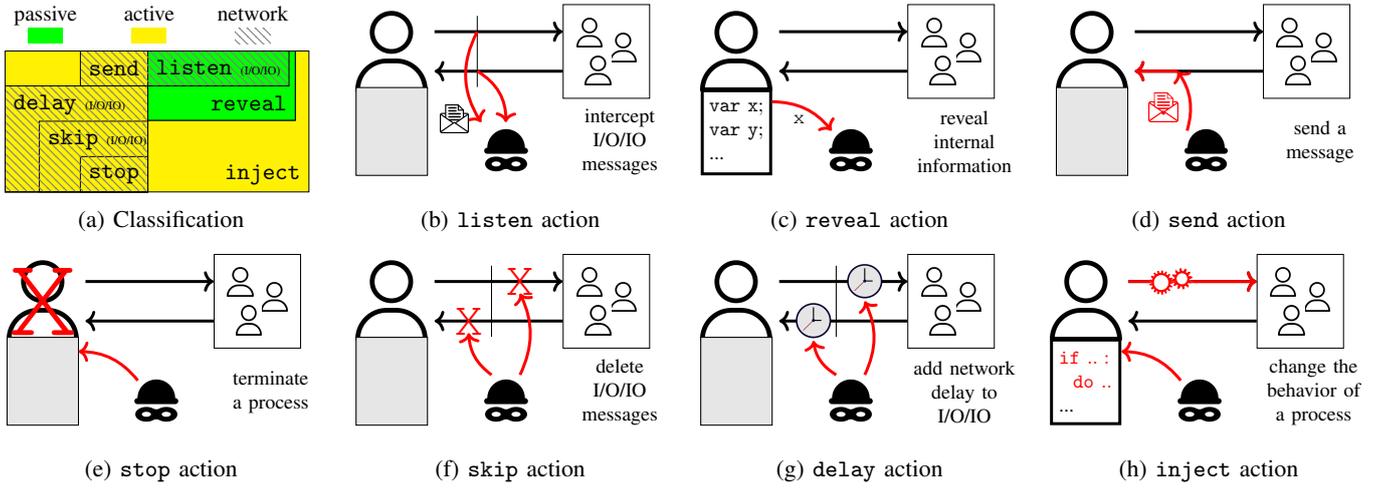
\begin{figure*}[h]
\centering 

\input{figures/actions/table}

    \caption{Adversarial actions}
    \label{fig:adversarial_actions}
\vspace*{-.25cm}
\end{figure*}

\subsection{Adversarial actions\label{ssec:advact}}

We propose in Fig.\ref{fig:adversarial_actions_classification} a novel classification of adversarial actions.
We name 7 types of actions, each one being illustrated by a diagram on Fig.\ref{fig:adversarial_actions}.
The process target of the action is represented on the left, the other processes of the DS on the right, and the adversary below them. The black horizontal arrows represent message passing. The effects of the action is represented in red.

Actions of type \advactReveal~and \advactListen~are passive actions. While \advactReveal~allows the adversary to read an internal state variable of a target process (e.g., the $x$ variable on Fig.\ref{fig:action_reveal}), \advactListen~only allows reading incoming and/or outgoing messages (red arrows on Fig.\ref{fig:action_listen}).
Because message buffers are a specific kind of state variables, a \advactListen~action can be performed via a \advactReveal~action. Hence, on Fig.\ref{fig:adversarial_actions_classification}, \advactListen~is a subtype of \advactReveal.
\advactListen~actions can be further specialized depending on the nature of the messages that are observed e.g., whether they correspond to inputs, outputs or both (as indicated by I/O/IO on Fig.\ref{fig:adversarial_actions_classification}).

In the real world, \advactListen~actions correspond to network eavesdropping (also called sniffing or snooping), a common vulnerability in open networks, particularly wireless ones as discussed in \cite{intercepting_mobile_communications}. 
\advactReveal~actions can mean access with read permissions. It may also involve passive side-channels attacks \cite{systematic_classification_of_side_channel_attacks} where a process, despite being software secure, leaks information (e.g., memory footprint, power consumption etc.), or more active tampering with certain types of memory scanning attacks \cite{yataglass_network_level_code_emulation_for_analyzing_memory_scanning_attacks}
in which an attacker reads and interprets memory addresses associated with a process.

While actions of type \advactListen~and \advactReveal~are passive (i.e., have no direct impact on system execution), those of types \advactSend, \advactDelay, \advactSkip, \advactStop~and \advactInject~are active. 
\advactSend~allows the adversary to send messages to a target process (see Fig.\ref{fig:action_send}), which, combined with specific knowledge (see resources on Fig.\ref{fig:adversary_model_description}), can be used to impersonate third parties (with e.g., knowledge of private keys).
With \advactStop, the adversary forces a process to crash (terminate prematurely). With \advactSkip, it prevents message exchange between the target and the rest of the system. 
If \advactSkip~concerns every messages, it is equivalent (from the point of view of the system) to \advactStop~(hence on Fig.\ref{fig:adversarial_actions_classification}, \advactSkip~contains \advactStop).
\advactDelay~makes so that message exchanges with the target are slowed down. 
As a result, it delays the reception of the messages that it receives and emits. If the added delay is infinite, then \advactDelay~is equivalent to \advactSkip.
In the real world, \advactDelay~may be implemented via Denial of Service \cite{a_framework_for_classifying_denial_of_service_attacks}.
\advactInject~modifies the behavior of the target process either by forcing it to express a given behavior at a given time or by changing the manner with which it reacts to events (e.g., to incoming messages). This may realistically correspond to code injection attacks \cite{defining_code_injection_attacks} or the adversary having user access to the target's information system.

Network actions (hatched on Fig.\ref{fig:adversarial_actions_classification}) include actions of types \advactListen, \advactSend, \advactStop, \advactSkip~and \advactDelay~because they may only require tampering with the network environment and not the with the target sub-system hardware or software directly.

%% file: figures/actions/table.tex
\begin{subfigure}{0.24\textwidth}
    \centering
\scalebox{.9}{
    \input{figures/actions/classif}
}
    \caption{Classification}
    \label{fig:adversarial_actions_classification}
\end{subfigure}
\hfill
\begin{subfigure}{0.24\textwidth}
    \centering
    \scalebox{.75}{\input{figures/actions/listen}}
    \caption{\advactListen~action}
    \label{fig:action_listen}
\end{subfigure}
\hfill
\begin{subfigure}{0.24\textwidth}
    \centering
    \scalebox{.75}{\input{figures/actions/reveal}}
    \caption{\advactReveal~action}
    \label{fig:action_reveal}
\end{subfigure}
\hfill
\begin{subfigure}{0.24\textwidth}
    \centering
    \scalebox{.75}{\input{figures/actions/send}}
    \caption{\advactSend~action}
    \label{fig:action_send}
\end{subfigure}

\vspace*{.25cm}

\begin{subfigure}{0.24\textwidth}
    \centering
    \scalebox{.75}{\input{figures/actions/stop}}
    \caption{\advactStop~action}
    \label{fig:action_stop}
\end{subfigure}
\hfill
\begin{subfigure}{0.24\textwidth}
    \centering
    \scalebox{.75}{\input{figures/actions/skip}}
    \caption{\advactSkip~action}
    \label{fig:action_skip}
\end{subfigure}
\hfill
\begin{subfigure}{0.24\textwidth}
    \centering
    \scalebox{.75}{\input{figures/actions/delay}}
    \caption{\advactDelay~action}
    \label{fig:action_delay}
\end{subfigure}
\hfill
\begin{subfigure}{0.24\textwidth}
    \centering
    \scalebox{.75}{\input{figures/actions/inject}}
    \caption{\advactInject~action}
    \label{fig:action_inject}
\end{subfigure}

%% file: figures/actions/classif.tex
\begin{tikzpicture}
\node[minimum height=.5cm,text centered] (nA1) at (0,0) {\advactDelay~{\tiny(I/O/IO)}};
\node[minimum height=.5cm,text centered,below right=.25cm and .5cm of nA1.west] (nA2) {\advactSkip~{\tiny(I/O/IO)}};
\node[minimum height=.5cm,text centered,below right=.25cm and .6cm of nA2.west] (nA3) {\advactStop};
\node[minimum height=.5cm,text centered,above left=0cm and 0cm of $(nA3.east |- nA1.north)$] (nC1) {\advactSend};
\node[minimum height=.5cm,text centered,right=0cm of nC1.east] (nB2) {\advactListen~{\tiny(I/O/IO)}};
\node[minimum height=.5cm,text centered,below right=.25cm and .8cm of nB2.west] (nB1) {\advactReveal};
\node[minimum height=.5cm,text centered,right = 1cm of nA3] (n0) {\advactInject};
\begin{pgfonlayer}{bg}    
\draw[fill=yellow] (n0.east |- nA3.south) rectangle (nA1.west |- nC1.north);
\draw (nA1.north west) rectangle (nA3.south east);
\draw (nA2.north west) rectangle (nA3.south east);
\draw (nA3.north west) rectangle (nA3.south east);
\draw (nB2.north west) rectangle (nB2.south east);
\draw (nB2.north west) rectangle (nB1.south east);
\draw[fill=green] (nB2.north west) rectangle (nB1.south east);
%
\draw[pattern=north west lines, pattern color=gray] (nA1.north west) rectangle (nA3.south east);
%
\draw[pattern=north west lines, pattern color=gray] (nC1.north west) rectangle (nB2.south east);
\end{pgfonlayer}
\node[fill=green,above left=.1cm and .25cm of nC1,minimum height=.2cm,minimum width=.5cm] (leg1) {};
\node[fill=yellow,right=1cm of leg1.east, anchor=west,minimum height=.2cm,minimum width=.5cm] (leg2) {};
\node[pattern=north west lines, pattern color=gray,right=1cm of leg2.east, anchor=west,minimum height=.2cm,minimum width=.5cm] (leg3) {};
\node[above=-.1cm of leg1,anchor=south,minimum height=.5cm,text centered] (leg1_lab) {\small passive};
\node[above=-.1cm of leg2,anchor=south,minimum height=.5cm,text centered] (leg2_lab) {\small \vphantom{p}active};
\node[above=-.1cm of leg3,anchor=south,minimum height=.5cm,text centered] (leg3_lab) {\small \vphantom{p}network};
\end{tikzpicture}

%% file: figures/actions/listen.tex
\begin{tikzpicture}
\input{figures/actions/base}
\draw[-] ($(ag1.320) + (.75,0)$) -- ($(ag1.40) + (.75,0)$);
\draw[red,line width=1.5pt] ($(ag1.25) + (.75,0)$) edge[bend right=30,->] ($(adv.120) + (0,-.25)$);
\draw[red,line width=1.5pt] ($(ag1.335) + (.75,0)$) edge[bend left=30,->] ($(adv.90) + (0,-.25)$);
\node (cl1) at ($(ag1.1) + (.35,-1.2)$) {\includegraphics[width=.5cm]{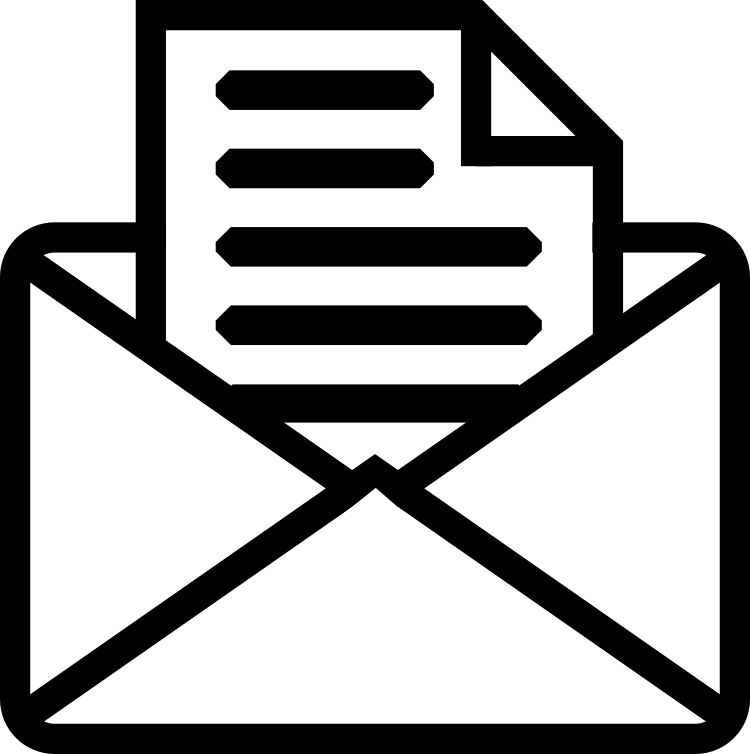}};
\node[align=left,align=center] (leg) at (4,-1.6) {intercept\\I/O/IO\\messages};
\draw[fill=gray!20] (attr.north west) rectangle (attr.south east);
\end{tikzpicture}

%% file: figures/actions/send.tex
\begin{tikzpicture}
\input{figures/actions/base}
\draw[red,line width=1.5pt] ($(ag1.335) + (.75,0)$) edge[bend left=25,<-] ($(adv.120) + (.1,-.4)$);
\draw[red,line width=1.5pt] ($(ag1.335) + (.75,0)$) edge[->] (ag1.335);
\node (cl1) at ($(ag1.1) + (.5,-1)$) {\includegraphics[width=.5cm]{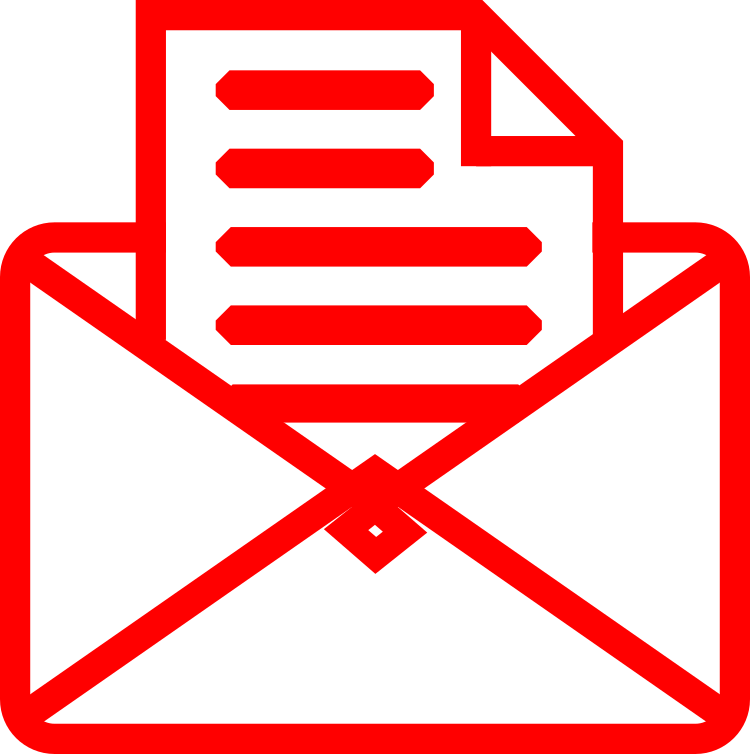}};
\node[align=center] (leg) at (4,-1.6) {send a\vphantom{g}\\message};
\draw[fill=gray!20] (attr.north west) rectangle (attr.south east);
\end{tikzpicture}

%% file: figures/actions/skip.tex
\begin{tikzpicture}
\input{figures/actions/base}
\draw[-] ($(ag1.320) + (1,0)$) -- ($(ag1.40) + (1,0)$);
\node (cl1) at ($(ag1.25) + (1.5,0)$) {\Huge\textcolor{red}{\texttt{X}}};
\node (cl2) at ($(ag1.335) + (.6,0)$) {\Huge\textcolor{red}{\texttt{X}}};
\node[align=center] (leg) at (4,-1.6) {delete\\I/O/IO\\messages};
\draw[fill=gray!20] (attr.north west) rectangle (attr.south east);
\draw[red,line width=1.5pt] ($(adv.north) + (.25,-.3)$) edge[bend right=25,->] ($(cl1.south) + (0,.15)$);
\draw[red,line width=1.5pt] ($(adv.north) + (-.25,-.3)$) edge[bend left=25,->] ($(cl2.south) + (0,.15)$);
\end{tikzpicture}

%% file: figures/actions/delay.tex
\begin{tikzpicture}
\input{figures/actions/base}
\draw[-] ($(ag1.320) + (1,0)$) -- ($(ag1.40) + (1,0)$);
\node (cl1) at ($(ag1.25) + (1.5,0)$) {\includegraphics[width=.75cm]{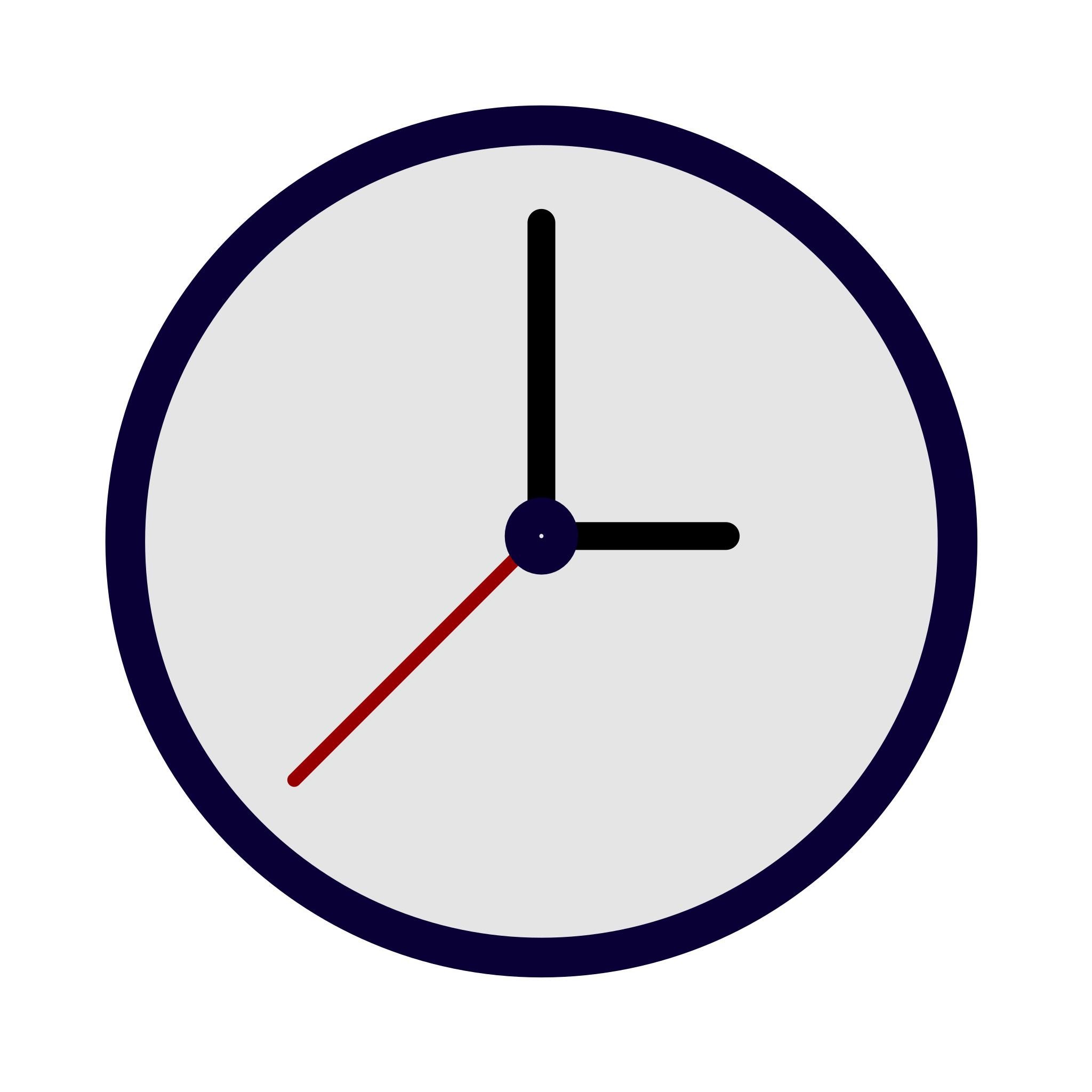}};
\node (cl2) at ($(ag1.335) + (.6,0)$) {\includegraphics[width=.75cm]{figures/icons/icon_clock_tp.png}};
\node[align=center] (leg) at (4,-1.6) {add network\\ delay to\\I/O/IO};
\draw[fill=gray!20] (attr.north west) rectangle (attr.south east);
\draw[red,line width=1.5pt] ($(adv.north) + (.25,-.3)$) edge[bend right=25,->] ($(cl1.south) + (0,.15)$);
\draw[red,line width=1.5pt] ($(adv.north) + (-.25,-.3)$) edge[bend left=25,->] ($(cl2.south) + (0,.15)$);
\end{tikzpicture}

%% file: figures/actions/inject.tex
\begin{tikzpicture}
\input{figures/actions/base}
\draw[red,line width=1.5pt] (attr.40) edge[bend left=25,<-] ($(adv.120) + (.1,-.4)$);
\node[align=center] (leg) at (4,-1.6) {change the\\behavior of\\a process};
\node[align=left] (internal) at (attr) {
\textcolor{red}{$\mathtt{if~..:}$}
\\
\textcolor{red}{$~~\mathtt{do~..}$}
\\...
};
\draw[->,red,line width=1.5pt] (ag1.25) -- (ds.west |- ag1.25);
\node (cl1) at ($(ag1.25) + (.75,0)$) {\includegraphics[width=.75cm]{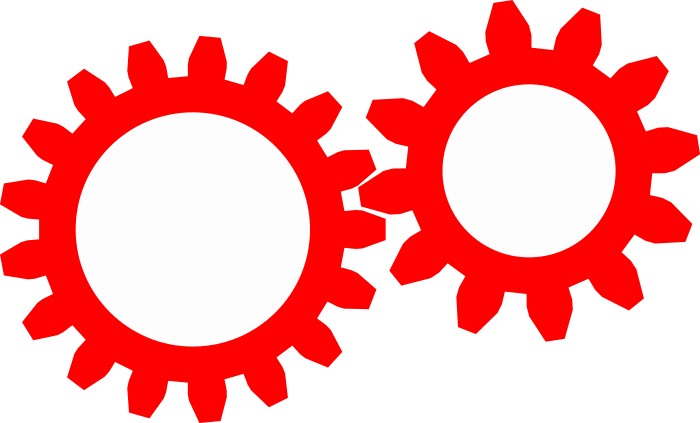}};
\end{tikzpicture}

%% file: 3_4.tex
\subsection{Capabilities binding assumptions}

The adversary's assumptions (Fig.\ref{fig:adversary_model_description}) include a communication model and a failure model for individual processes.
These models bind the capabilities of the adversary as they do not allow certain classes of adversarial actions.
Fig.\ref{fig:actions_wrt_failure_and_comm} summarizes these limitations.

It is always possible to perform \advactReveal~(and thus \advactListen) actions.
The asynchronous communication model always enables the unrestricted use of \advactDelay~actions.
While \advactSkip~is allowed under the omission failure model, only \advactStop~is available under crash failures.
Under both failure models and with the synchronous communication model, the use of \advactDelay~actions is limited to the addition of a maximum delay $\delta$ so that the total retransmission time (i.e., between the output $o$ and the input $i$) of the affected message does not exceed a certain $\Delta$ time. Given $t$ the retransmission time without intervention from the adversary we hence have $i - o = t + \delta < \Delta$.
Under the eventually synchronous communication model, this condition is only required after the GST (hence $o \geq GST$ on Fig.\ref{fig:actions_wrt_failure_and_comm}).

\begin{figure}[h]
    \centering
    \input{figures/binding}
    \caption{Enabled actions w.r.t.~assumptions}
    \label{fig:actions_wrt_failure_and_comm}
\end{figure}

The adversary's assumptions also include its knowledge, which represents the information it possesses about the system. 
This includes it being aware of the existence of the various sub-systems that are part of the DS.
In the case of an adaptive adversary, which may update its plan of actions according to new information, its capabilities can include process discovery.
Knowledge can directly bind adversarial capabilities when certain action requires specific knowledge (e.g., authentication).

We use the concept of resource limited adversary \cite{a_secure_control_framework_for_resource_limited_adversaries} to bind the capabilities of the adversary w.r.t.~FT thresholds.
We abstract away adversarial actions as a set $A$.
Each action $a \in A$ has a target sub-system $s(a) \in S$, and a baseline cost $\kappa(a) \in \mathbb{K}$, where $\mathbb{K}$ is the ordered vector space in which the budget of the adversary is represented.
The adversary is bound by a certain initial budget $B \in \mathbb{K}$ which limits its capabilities. 
For instance, let us suppose the initial budget of the adversary is the vector $(f_x,f_y)$, representing the maximal number of nodes it can infect on protocol $x$ and resp.~$y$. 
Then, if an action $a_x$ involves sabotaging a node participating in protocol $x$, we have $\kappa(a_x) = (1,0)$ and the remaining budget is $(f_x-1,f_y)$ after performing $a_x$.
Because it might cost less to target a sub-system that has already been victim of a previous action, we consider a protection level function $\psi \in \mathbb{K}^S$ (which may vary during the simulation) to modulate this cost. 
Then, given a current budget $b \leq B$, the adversary can perform an action $a \in A$ if the associated cost is within its budget i.e., 
iff $\kappa(a) \odot \psi(s(a)) \leq b$, with $\odot$ the Hadamard product (element-wise product). 
After performing this action, the remaining budget is then $b - \kappa(a) \odot \psi(s(a))$.
For instance, in our previous example we have an initial protection level $\psi(s(a_x)) = (1,1)$ and therefore $\kappa(a_x) \odot \psi(s(a_x) = (1,0) \odot (1,1) = (1,0)$. 
After having performed $a_x$, the protection level for $s(a_x)$ becomes $\psi'(s(a_x)) = (0,1)$ and therefore performing another action $a'_x$ on that same node w.r.t.~protocol $x$ (i.e., s.t., $s(a'_x) = s(a_x)$) has no cost (i.e., $\kappa(a'_x) \odot \psi(s(a'_x) = (1,0) \odot (0,1) = (0,0)$).

%% file: figures/binding.tex
\begin{tabular}{|l||*{3}{c|}}\hline
\diagbox{\scriptsize Fail.}{\scriptsize Comm.}
&
Synch.
&
Async.
&
Event. Synch.
\\
\hline\hline
Crash  
&
\makecell[l]{\scriptsize \advactReveal\\[-.1cm]\scriptsize \advactStop\\[-.1cm]\scriptsize \advactDelay\\[-.1cm]\tiny~~$t + \delta < \Delta$}
&
\makecell[l]{\scriptsize \advactReveal\\[-.1cm]\scriptsize \advactDelay}
& 
\makecell[l]{\scriptsize \advactReveal\\[-.1cm]\scriptsize \advactStop\\[-.1cm]\scriptsize \advactDelay\\[-.1cm]\tiny~~$o \geq GST \Rightarrow t + \delta < \Delta$}
\\
\hline
Omission
&
\makecell[l]{\scriptsize \advactReveal\\[-.1cm]\scriptsize \advactSkip\\[-.1cm]\scriptsize \advactDelay\\[-.1cm]\tiny~~$t + \delta < \Delta$}
&
\makecell[l]{\scriptsize \advactReveal\\[-.1cm]\scriptsize \advactDelay}
& 
\makecell[l]{\scriptsize \advactReveal\\[-.1cm]\scriptsize \advactSkip\\[-.1cm]\scriptsize \advactDelay\\[-.1cm]\tiny~~$o \geq GST \Rightarrow t + \delta < \Delta$}
\\
\hline
Performance
&
\makecell[l]{\scriptsize \advactReveal\\[-.1cm]\scriptsize \advactDelay}
&
\makecell[l]{\scriptsize \advactReveal\\[-.1cm]\scriptsize \advactDelay}
& 
\makecell[l]{\scriptsize \advactReveal\\[-.1cm]\scriptsize \advactDelay}
\\
\hline
Byzantine 
&
\makecell[l]{\scriptsize \advactInject}
& 
\makecell[l]{\scriptsize \advactInject}
& 
\makecell[l]{\scriptsize \advactInject}
\\
\hline
\end{tabular}

%% file: 3_5.tex
\subsection{System simulation and success of attack}

Combining a model of the DS and of the adversary, we can simulate attacks and test whether or not the adversary's goal is met.
The simulation's state at any time is given by a tuple $(\eta,b,\psi)$ where $\eta \in \Gamma$ gives the current state of the system, $b \in \mathbb{K}$ correspond to the remaining budget of the adversary and $\psi \in \mathbb{K}^S$ gives the current protection levels of sub-systems (for each type of resource and each target sub-system).
Protection levels of individual sub-systems may vary during the simulation for several reasons.
It might cost less to target a process that has already been victim of an action.
Inversely, the system might heal and reset the sub-systems' protection level or apply countermeasures to increase it further.

We distinguish between two kinds of events: adversarial actions in $A$ and system events in $E$ (which correspond to the system acting spontaneously). 
Let us consider a relation $\rightarrow_E \subseteq (\Gamma \times \mathbb{K}^S) \times E \times (\Gamma \times \mathbb{K}^S)$ s.t., for any $(\eta,\psi) \xrightarrow{e} (\eta',\psi')$, $\eta'$ and $\psi'$ describe the state and protection levels of the system after the occurrence of $e \in E$.
Similarly, let us consider $\rightarrow_A \subseteq (\Gamma \times \mathbb{K}^S) \times A \times (\Gamma \times \mathbb{K}^S)$.
Then, the space of simulations is the graph with vertices in $\mathbb{G} = \Gamma \times \mathbb{K} \times \mathbb{K}^S$ and edges defined by the transition relation $\leadsto \subseteq \mathbb{G}^2$ s.t.:

\vspace*{-.2cm}

\begin{prooftree}
\AxiomC{$\kappa(a) \odot \psi(s(a)) \leq b$}
\AxiomC{$(\eta,\psi) \xrightarrow{a}_A (\eta',\psi')$}
\LeftLabel{$attack$~}
\BinaryInfC{$(\eta,~b,~\psi) \leadsto (\eta',~b - \kappa(a) \odot \psi(s(a)),~\psi')$}
\end{prooftree}

\vspace*{-.3cm}

\begin{prooftree}
\AxiomC{$(\eta,\psi) \xrightarrow{e}_E (\eta',\psi')$}
\LeftLabel{$exec$~}
\UnaryInfC{$(\eta,~b,~\psi) \leadsto (\eta',~b,~\psi')$}
\end{prooftree}

Any simulation
\textbf{(1)} starts from a node $(\eta_0,B,\psi_0)$ where $\eta_0$ is the initial state of the system, $B$ is the initial budget of the adversary and $\psi_0$ gives the initial protection level of sub-systems 
and \textbf{(2)} corresponds to a finite path $(\eta_0,B,\psi_0) \overset{*}{\leadsto} (\eta_j,b_j,\psi_j)$ in graph $\mathbb{G}$, its length $j$ being related to the duration of the simulation.
To assess the success of the adversary, we then check if and how property $\phi$ is invalidated in that path.

%% file: 4_0_usecase.tex
\section{Use-case and fairness properties\label{sec:usecase}}

\input{4_1}

\input{4_2}

\input{4_3}

\input{4_4}

%% file: 4_1.tex
\subsection{Hyperledger Fabric system}

A system using HyperLedger Fabric consists of a set of subsystems $S = S_c \cup S_p \cup S_o$ deployed over a network, where:
\begin{itemize}
    \item $S_c$ is a set of clients\footnote{for the sake of simplicity, we do not distinguish between clients and the applications through which they interact with HF}, with $|S_c| = m$
    \item $S_p$ is a set of peers, with $|S_p| = n$
    \item $S_o$ is a set of orderers, with $|S_o| = n'$
\end{itemize}

\begin{figure}[h]
    \centering
    \includegraphics[width=0.475\textwidth]{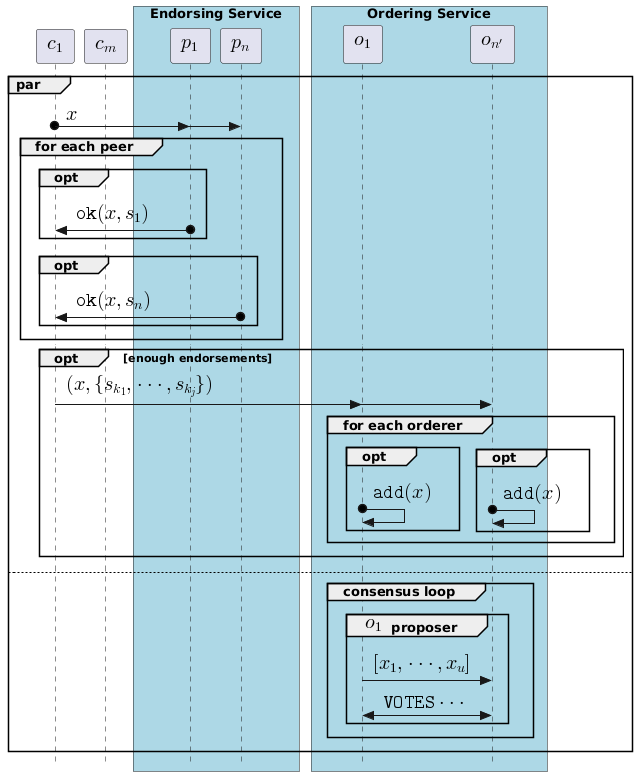}
    \caption{Schematic behavior of HF}
    \label{fig:hf_tm_sd}
\end{figure}

Fig.\ref{fig:hf_tm_sd} roughly describes the behavior of such a system.
Clients may send transactions (denoted as $x$) to the peers, which are tasked with endorsing them.
Upon receiving a transaction each peer $p_i$ may send a corresponding endorsement with a cryptographic signature $s_i$.
Once it has collected ``enough'' endorsements (which depends on the HF system's endorsing policy parameterization), a client may then forward its endorsed transaction to the orderers (with the corresponding set of cryptographic signatures attesting its endorsement)

The orderers run a consensus algorithm to agree on the order of endorsed transactions. 
It is possible to use Crash Fault Tolerant algorithms in cases where the network is deemed safe or Byzantine Fault Tolerant ones if this is not the case.
In any case, this allows orderers to order transactions in batches that are called blocks.

For the Tendermint \cite{dissecting_tendermint} BFT consensus algorithms, which we will consider in the remainder of the paper, orderers may successively take the role of a proposer, which proposes a value for the next block of transactions ($\lbrack x_1,\cdots,x_u \rbrack$ on Fig.\ref{fig:hf_tm_sd}). 
For honest proposers, the content of that block corresponds to the set of endorsed transactions it has received up to that moment and which have not already been put in a previous block. As for the order of transactions within the block, it corresponds to the local order with which that orderer received them.

Hyperledger Fabric (HF) thus implements a Distributed Ledger, and, more specifically, a non-revocable permissioned Blockchain.
We say that a transaction is delivered once it has been put in a block. 
The delivery order of transactions corresponds to the order with which they are delivered on and within blocks.
We may refer to the set of peers (resp.~orderers) as the ``endorsing service'' (resp.~``ordering service'').

%% file: 4_2.tex
\subsection{OF properties as evaluation metrics}

In the following, we define several metrics that can be used to monitor the effect of transaction reordering attacks on our HF system.
These metrics correspond to counting the number of pairs of transactions $(x,x')$ for which specific OF properties are violated.

In the case of HF, there is no single notion of ``node'' because there is a distinction between peers and orderers. 
Thus, the notion of ``reception'' can be interpreted in two manners. 
Indeed, we can either consider the receptions of not-yet endorsed transactions by the peers, or the receptions of endorsed transactions by the orderers.

Let us consider properties $OF^\beta_\alpha$ of the form: for any transactions $x$ and $x'$, \textbf{if} $\alpha(x,x')$ \textbf{then} $\beta(x,x')$.

The premise of the property, which involves partial orders between communication events, is represented by the $\alpha$ predicate.
We consider three variants : $\alpha \in \{S_{ND},E_{DS},O_{RD}\}$.
If we have $S_{ND}(x,x')$, then the emission of $x$ precedes that of $x'$.
If we have $E_{DS}(x,x')$ then, a majority of peers receive an endorsement request for $x$ before they do for $x'$.
Finally, if $O_{RD}(x,x')$ then, a majority of orderers receive $x$ with sufficient endorsements before they do for $x'$.

The property's right-hand side, represented by the $\beta$, involves partial orders between the delivery events.
We consider two variants : $\beta \in \{D_{LV}, B_{LC}\}$.
$D_{LV}(x,x')$ (resp.~$B_{LC}(x,x')$) corresponds to $x$ being delivered before $x'$ (resp.~$x$ being delivered in a block that is either before that or the same as the one in which $x'$ is delivered).

Let us remark that $OF^{D_{LV}}_{S_{ND}}$ corresponds to {\em send-order fairness}, while $OF^{D_{LV}}_{E_{DS}}$ and resp.~$OF^{D_{LV}}_{O_{RD}}$ correspond to {\em receive-order fairness} from the perspective of the peers and resp.~the orderers.
Because blocks in HF do not coincide with Condorcet cycles, $OF^{B_{LC}}_{E_{DS}}$ and resp.~$OF^{B_{LC}}_{O_{RD}}$ do not correspond to the \textit{$\gamma$-batch-order fairness} from \cite{themis_fast_strong_order_fairness_in_byzantine_consensus} (which is a generalization of the \textit{block-order fairness} from \cite{order_fairness_for_byzantine_consensus}).
Still, counting the violations of $OF^{B_{LC}}_{S_{ND}}$, $OF^{B_{LC}}_{E_{DS}}$ and $OF^{B_{LC}}_{O_{RD}}$ is pertinent as, if one observes an increase of these numbers as the adversarial power increases, this means that the adversary pushes transactions $x$ it front-runs into blocks after the one in which the corresponding $x'$ is delivered.

%% file: 4_3.tex
\subsection{Application layer \& client fairness}

Let us consider that several clients repeatedly compete to solve puzzles.
For a given puzzle $k$, a client $c$ wins iff the first delivered transaction that contains the solution of $k$ was sent by $c$.
All clients having the same aptitude, the game is {\em client-fair} iff every client has the same likelihood of winning, which is $1/m$ where $m$ is the number of clients.
Let us denote by $g$ the total number of resolved puzzle competitions.

\begin{figure}[h]
    \centering
\scalebox{.9}{
\begin{tikzpicture}
\node (plot) at (0,0) {\includegraphics[width=8cm]{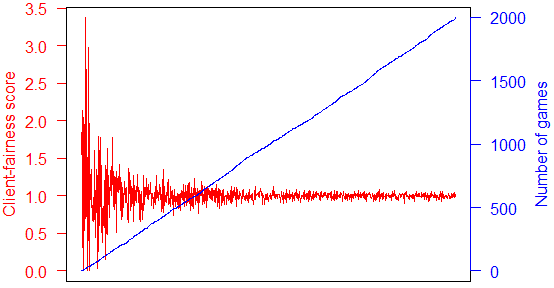}};
\draw[->] (-2.25,-1.75) -- (2.6,-1.75);
\node (xax) at (1.2,-1.5) {length of simulation};
\end{tikzpicture}
}
    \caption{Convergence of client-fairness score towards $1$}
    \label{fig:client_fairness_score_convergence}
\end{figure}

As blocks are regularly delivered, $g$ increases with the length of the simulation, as illustrated on Fig.\ref{fig:client_fairness_score_convergence} (in blue on the right axis).
For a client $c$, we denote by \%$g(c)$ the percentage of games it has won during a simulation.
The game is {\em client-fair} if, for all clients $c$, the longer the simulation is, the more \%$g(c)$ is close to $1/m$. 
Via defining a client-fairness score $\mathtt{score}(c) = \%g(c) * m$, we obtain an independent metric, that converges towards $1$, as illustrated on Fig.\ref{fig:client_fairness_score_convergence} (in red, on the left axis) if the game is client-fair.

We consider the goal of the adversary to be to diminish the likelihood for a target client $c \in S_c$ to win puzzles.
We formalize this as the {\em client-fairness} property:
$\phi(c) = (g > 5000) \wedge (\mathtt{score}(c) < 0.75)$.
This signifies that, the adversary wins if, after more than 5000 competitions have been resolved, the client's $\mathtt{score}$ is less than $0.75$.

Using this specific application use-case provides us with an additional client-fairness $\mathtt{score}$ metric as well as a clear goal for the adversary, which success can thus be measured.
Additionally, it is particularly convenient for quantifying the number of $OF$ violations.
Indeed, counting the violations amounts to comparing, for each pair of delivered transactions $x$ and $x'$, their delivery order and either the order of their emission or the order of their reception in each of the peers or orderers.
If we consider $X$ to be the total number of delivered transactions, there are $X*(X-1)/2$ such pairs to consider. 
Moreover, for our results to be statistically significant, we need a high number $k$ of puzzle games, which yields a large number of transactions $X = m*k$ having $m$ (the number of clients) transactions per game.
Thus, comparing reception orders on every pair of transactions on every peer amounts to $n*m*k*(m*k -1)/2$ comparisons, which quickly become intractable.
Thanks to our puzzle use-case, the order between two transactions only matters if they concern the same puzzle game as they may commute otherwise.
Hence, we only need to consider $n*k*m*(m-1)/2$ comparisons.

%% file: 4_4.tex
\subsection{Parameterization of the system}

\begin{figure}[h]
    \centering
    \resizebox{.425\textwidth}{!}{\input{figures/network_with_delays}}
    \caption{Fabric network with delays}
    \label{fig:network_with_delays}
\end{figure}
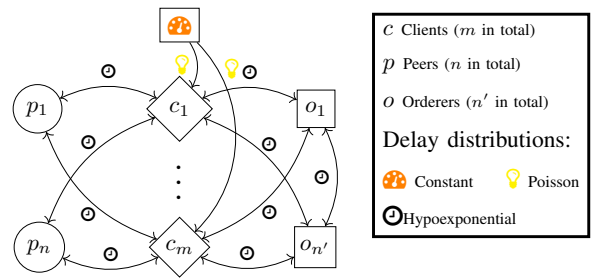

For the experiments, we consider the network described on Fig.\ref{fig:network_with_delays}.
The peers ($p_1$ to $p_n$ on the left) exclusively communicate with the clients ($c_1$ to $c_m$ in the middle).
The orderers ($o_1$ to $o_{n'}$ on the right) communicate with the clients and among themselves.
Peer to peer communications is simulated with no loss and delays for the transmission of individual messages are sampled from a probability distribution denoted as \faClockO.
As recommended in \cite{models_of_network_delay}, we consider hypoexponential distributions for these delays and we bound the maximal value a delay may take with a bound $\Delta$, which corresponds to that of a partially synchronous communication model \cite{consensus_in_the_presence_of_partial_synchrony} (in its formulation without the GST, with $\Delta$ not known by any of the involved parties).

We suppose that puzzles are revealed at regular intervals, represented by \textcolor{orange}{\faDashboard} on Fig.\ref{fig:network_with_delays}, and that all clients are aware of the puzzles at the same time.
Any client can solve a puzzle in a certain amount of time that is sampled from a Poisson distribution, represented by \textcolor{yellow}{\faLightbulbO} on Fig.\ref{fig:network_with_delays}.
We use the same \textcolor{yellow}{\faLightbulbO} distribution for all clients so that they all have the same ability to solve puzzles.

Upon solving a puzzle, a client submits the corresponding transaction to HF (via the peers and then the orderers).
In our use-case, the HF system also receives third party transactions (that do not correspond to puzzle solutions) at regular intervals. These heartbeat transactions simulate a concurrent usage of the HF system by other applications (besides the puzzle game).

\begin{figure}[h]
    \centering
    \includegraphics[width=.475\textwidth]{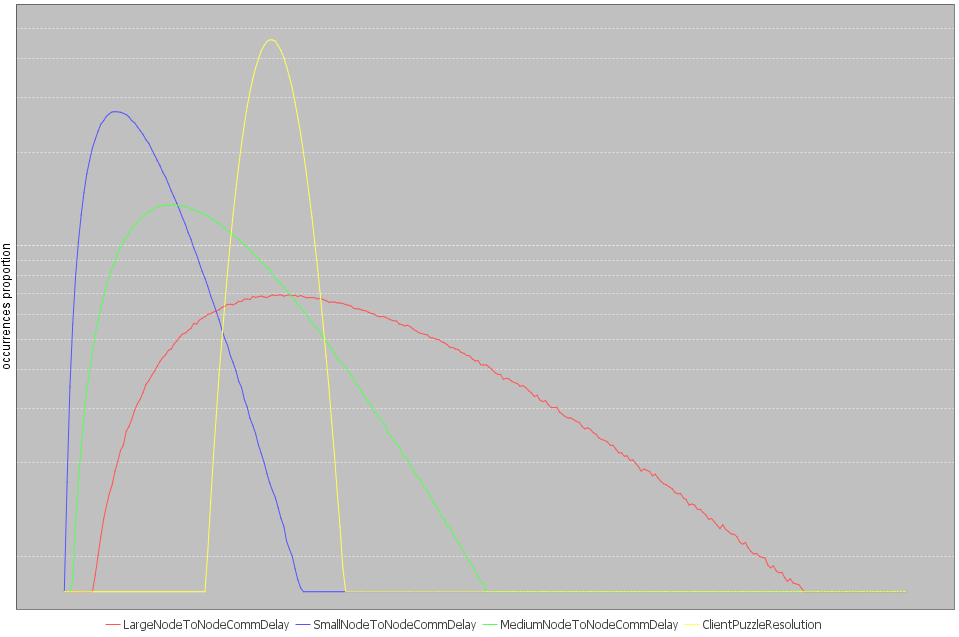}
    \caption{The delay distributions}
    \label{fig:distros}
\end{figure}

We consider three variants of the \faClockO~ delay distribution: one with small delays denoted as \textcolor{blue}{\faClockO} (in blue), another with medium delays denoted as \textcolor{green}{\faClockO} (in green) and a third with large delays denoted as \textcolor{red}{\faClockO} (in red).
Fig.\ref{fig:distros} represents these distributions in logarithmic scale.

The endorsing policy of HF is parameterized as follows: for a transaction to be endorsed and forwarded to the ordering service, it suffices to obtain at least $n/2$ endorsements from distinct peers, $n$ being the total number of peers.

Finally, we parameterize the cost of actions and the budget $B$ of the adversary so that it cannot apply adversarial actions to more than $(n' - 1)/3$ orderers (where $n'$ is the total number of orderers) and $n - n/2$ peers.
As a result, for the ordering service, which runs Tendermint, we remain in the hypotheses of Byzantine Fault Tolerance (below the one third threshold).
Likewise, for the endorsing service, the adversary cannot guarantee censorship of the transactions.

%% file: figures/network_with_delays.tex
\begin{tikzpicture}
\node[draw,diamond] at (0,0) (client1) {$c_1$};
\node[below=.1cm of client1] (clients_dots) {\rotatebox{90}{\Large$\cdots$}};
\node[draw,diamond,below=.1cm of clients_dots,inner sep=2] (client2) {$c_{m}$};
\node[draw,circle, left=1.25cm of client1] (peer1) {$p_1$};
\node[draw,circle, left=1.25cm of client2] (peer2) {$p_n$};
\node[draw,regular polygon,regular polygon sides=4,inner sep=1,right=1.25cm of client1] (orderer1) {$o_{1}$};
\node[draw,regular polygon,regular polygon sides=4,inner sep=-.3,right=1.25cm of client2] (orderer2) {$o_{n'}$};
\node[draw,above = .5cm of client1] (provider) {\textcolor{orange}{\faDashboard}};
\draw (provider) edge[->, bend left = 30] node[midway,left] {\textcolor{yellow}{\faLightbulbO}} (client1);
\draw (provider) edge[->, bend left = 45] node[pos=.15,right] {\textcolor{yellow}{\faLightbulbO}} (client2);
\draw (client1) edge[<->, bend right = 25] node[midway,above] {{\scriptsize\faClockO}} (peer1);
\draw (client1) edge[<->, bend right = 25] node[midway,above] {{\scriptsize\faClockO}} (peer2);
\draw (client2) edge[<->, bend left = 25] node[midway,below] {{\scriptsize\faClockO}} (peer1);
\draw (client2) edge[<->, bend left = 25] node[midway,above] {{\scriptsize\faClockO}} (peer2);
\draw (client1) edge[<->, bend left = 25] node[midway,above] {{\scriptsize\faClockO}} (orderer1);
\draw (client1) edge[<->, bend left = 25] node[midway,above] {{\scriptsize\faClockO}} (orderer2);
\draw (client2) edge[<->, bend right = 25] node[midway,below] {{\scriptsize\faClockO}} (orderer1);
\draw (client2) edge[<->, bend right = 25] node[midway,above] {{\scriptsize\faClockO}} (orderer2);
\draw (orderer1) edge[<->, bend left = 25] node[midway,left] {{\scriptsize\faClockO}} (orderer2);
%
\node[draw,line width=1.25,inner sep=2] at (4.5,-0.25) {
\begin{tikzpicture}
\node (leg1) at (0,0) {$c$ {\scriptsize Clients ($m$ in total)}};
\node[below=0.15cm of leg1.south west, anchor=north west] (leg2) {$p$ {\scriptsize Peers ($n$ in total)}};
\node[below=0.15cm of leg2.south west, anchor=north west] (leg3) {$o$ {\scriptsize Orderers ($n'$ in total)}};
\node[below=0.15cm of leg3.south west, anchor=north west] (leg4) {Delay distributions:};
\node[below=0.1cm of leg4.south west, anchor=north west] (leg5) {\textcolor{orange}{\faDashboard} {\scriptsize Constant} ~~ \textcolor{yellow}{\faLightbulbO} {\scriptsize Poisson}};
\node[below=0.1cm of leg5.south west, anchor=north west] (leg6) {\faClockO {\scriptsize Hypoexponential}};
\end{tikzpicture}
};
\end{tikzpicture}

%% file: 5_0_attacks.tex
\section{Basic attack scenarios\label{sec:attacks}}

We consider several basic attack scenarios that can be combined by the adversary to harm the target client $c$.

\input{5_1}

\input{5_2}

%% file: 5_1.tex
\subsection{Peer sabotage\label{ssec:endorser_sabotage}}

Peer sabotage consists in applying an \advactInject~action to a peer so that it never endorses transactions from $c$.
If the adversary were to sabotage more than $n/2$ peers, it would guarantee that no transactions from $c$ are ever delivered because there are $n$ peers and at least $n/2$ distinct endorsements are required.
However, sabotaging fewer peers still has an effect, particularly in a slow network.

Let us indeed denote by $t$ the time required for $c$ to receive an endorsement (for a given transaction) from any given peer $p \in S_p$.
We represent the probability of receiving the endorsement from $p$ before a certain timestamp $z$ using $\mathcal{P}(t < z)$.
If we suppose all events to be independent (i.e., we have i.i.d.~variables) and have the same likelihood (i.e., peer to peer channels of communications have equally probable delays), for any honest peer $p$ we may denote by $X$ this probability $\mathcal{P}(t < z)$. 
On the contrary, the endorsement from $p$ being received after $t$ has a probability $\mathcal{P}(t \geq z) = 1 - X$.

Among $n$ trials, the probability of having exactly $k \leq n$ peers endorsing the transaction before timestamp $z$ is:
\[
\mathcal{P}(k~ {\scriptstyle\mathtt{endorsement}} < z) = \binom{n}{k} * X^k * (1-X)^{n - k}
\]

\begin{figure}[h!]
    \centering

\scalebox{.75}{
\input{figures/theory/peer_sabotage_theo_v3}
}

    \caption{Theoretical effect of peer sabotage (for $n=20$)}
    \label{fig:peer_sabotage_delay_theory}
\end{figure}
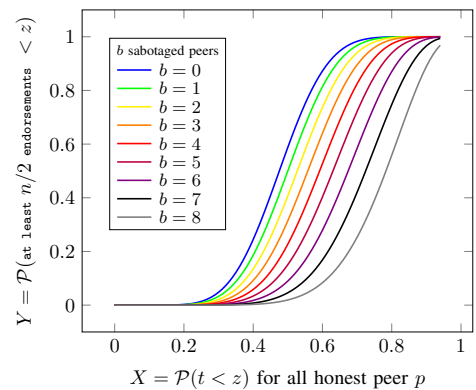

Sabotaged peers never endorse transactions (we always have $X=0$ for any timestamp $z$) and can therefore be ignored when counting the numbers of endorsements.
Therefore, given $b \leq n/2$ the number of sabotaged peers, the probability $Y$ of having at least $n/2$ endorsements from distinct peers before $z$ is:
\[
Y = \sum_{k = n/2}^{n-b} \binom{n - b}{k} * X^k * (1-X)^{n - b - k}
\]

On Fig.\ref{fig:peer_sabotage_delay_theory}, we plot this probability $Y$ w.r.t.~$X$ which corresponds to the probability $\mathcal{P}(t < z)$ for honest peers. On this plot, we consider a system with $n=20$ nodes, $n/2=10$ endorsements being required.
We can see that the more peers are sabotaged, the smaller is the probability of collecting enough endorsements before timestamp $z$.
We conclude that, infecting a minority of peers statistically delays the endorsement of transactions from $c$.
This delay might in turn be sufficient to force these transactions into later blocks in comparison to transactions from other clients emitted at the same time.

%% file: figures/theory/peer_sabotage_theo_v3.tex
\begin{tikzpicture}
\begin{axis}[ 
    xlabel={$X = \mathcal{P}(t < z)$ for all honest peer $p$},
    ylabel={$Y = \mathcal{P}({\scriptstyle\mathtt{at~least}} ~n/2~ {\scriptstyle\mathtt{endorsements}}~ < z)$},
    ylabel near ticks,
]
  
\addplot[blue,thick]
coordinates {
(0.0,0.0)(0.01,1.6863605465421854e-15)(0.02,1.5749451894225846e-12)(0.03,8.276588747996223e-11)(0.04,1.3383368025328806e-09)(0.05,1.1340718060930293e-08)(0.060000000000000005,6.383682365460939e-08)(0.07,2.708916039594493e-07)(0.08,9.345538562177697e-07)(0.09,2.751988316311776e-06)(0.09999999999999999,7.150904021083776e-06)(0.10999999999999999,1.678975545022509e-05)(0.11999999999999998,3.6249190234450166e-05)(0.12999999999999998,7.29279628522013e-05)(0.13999999999999999,0.000138139095180199)(0.15,0.00024838198962403224)(0.16,0.00042674530885820715)(0.17,0.0007043756190429915)(0.18000000000000002,0.0011219298345378364)(0.19000000000000003,0.0017309169328500213)(0.20000000000000004,0.002594827400674021)(0.21000000000000005,0.003789948170497047)(0.22000000000000006,0.005405766701379852)(0.23000000000000007,0.007544880184928614)(0.24000000000000007,0.010322344038612251)(0.25000000000000006,0.013864416943761202)(0.26000000000000006,0.01830668646909758)(0.2700000000000001,0.02379158836566988)(0.2800000000000001,0.030465362387513538)(0.2900000000000001,0.03847451642437541)(0.3000000000000001,0.047961897331343616)(0.3100000000000001,0.05906248974670391)(0.3200000000000001,0.07189908224797921)(0.3300000000000001,0.08657795250409996)(0.34000000000000014,0.10318472902451586)(0.35000000000000014,0.12178058637740144)(0.36000000000000015,0.1423989233550971)(0.37000000000000016,0.16504265981433378)(0.38000000000000017,0.1896822683999197)(0.3900000000000002,0.21625463290696284)(0.4000000000000002,0.24466279668360752)(0.4100000000000002,0.2747766334103172)(0.4200000000000002,0.3064344400961659)(0.4300000000000002,0.33944541953052804)(0.4400000000000002,0.3735929880266531)(0.45000000000000023,0.4086388153283738)(0.46000000000000024,0.44432747813996976)(0.47000000000000025,0.48039158783737557)(0.48000000000000026,0.5165572372851384)(0.49000000000000027,0.5525496018535783)(0.5000000000000002,0.5880985260009773)(0.5100000000000002,0.622943929206897)(0.5200000000000002,0.6568408734228287)(0.5300000000000002,0.6895641481233215)(0.5400000000000003,0.7209122478639008)(0.5500000000000003,0.7507106401715887)(0.5600000000000003,0.7788142476558106)(0.5700000000000003,0.805109096374747)(0.5800000000000003,0.8295131116080632)(0.5900000000000003,0.8519760711405469)(0.6000000000000003,0.8724787538527837)(0.6100000000000003,0.8910313468193359)(0.6200000000000003,0.9076711963209119)(0.6300000000000003,0.9224600064227556)(0.6400000000000003,0.9354806024721436)(0.6500000000000004,0.9468333856370542)(0.6600000000000004,0.956632608269889)(0.6700000000000004,0.9650025984735162)(0.6800000000000004,0.9720740560199868)(0.6900000000000004,0.9779805311688219)(0.7000000000000004,0.9828551835687419)(0.7100000000000004,0.9868279010648064)(0.7200000000000004,0.9900228387408536)(0.7300000000000004,0.9925564178408158)(0.7400000000000004,0.9945358032918383)(0.7500000000000004,0.9960578583359165)(0.7600000000000005,0.9972085561389933)(0.7700000000000005,0.9980628119550828)(0.7800000000000005,0.9986846861040541)(0.7900000000000005,0.9991278981297295)(0.8000000000000005,0.9994365863023399)(0.8100000000000005,0.9996462441751983)(0.8200000000000005,0.9997847670552662)(0.8300000000000005,0.9998735456642556)(0.8400000000000005,0.9999285514439309)(0.8500000000000005,0.9999613672517919)(0.8600000000000005,0.9999801278612076)(0.8700000000000006,0.9999903459366491)(0.8800000000000006,0.9999956102200738)(0.8900000000000006,0.999998152820404)(0.9000000000000006,0.9999992911393668)(0.9100000000000006,0.9999997566472125)(0.9200000000000006,0.999999927187021)(0.9300000000000006,0.9999999816939082)(0.9400000000000006,0.9999999963345036)
};
\label{b0}

\addplot[green,thick] 
coordinates {
(0.0,0.0)(0.01,8.509104732905513e-16)(0.02,8.020345794548888e-13)(0.03,4.254066587798824e-11)(0.04,6.943443179805265e-10)(0.05,5.9393390596643796e-09)(0.060000000000000005,3.3751160046339135e-08)(0.07,1.44598847820022e-07)(0.08,5.036834175747612e-07)(0.09,1.4976705518773051e-06)(0.09999999999999999,3.929882327127996e-06)(0.10999999999999999,9.318467523122158e-06)(0.11999999999999998,2.031948508032941e-05)(0.12999999999999998,4.129101310153026e-05)(0.13999999999999999,7.900561698626853e-05)(0.15,0.00014350736891607002)(0.16,0.0002490969324636223)(0.17,0.0004154149773937475)(0.18000000000000002,0.0006685813896358016)(0.19000000000000003,0.0010423363788258264)(0.20000000000000004,0.001579120549167106)(0.21000000000000005,0.0023310250203837863)(0.22000000000000006,0.003360540298662829)(0.23000000000000007,0.004741034114922995)(0.24000000000000007,0.006556893949809551)(0.25000000000000006,0.008903279303922339)(0.26000000000000006,0.011885441588629747)(0.2700000000000001,0.015617585262427049)(0.2800000000000001,0.020221261823330068)(0.2900000000000001,0.025823307679784625)(0.3000000000000001,0.03255335688130105)(0.3100000000000001,0.040540979288941154)(0.3200000000000001,0.04991251311399648)(0.3300000000000001,0.06078767701529203)(0.34000000000000014,0.07327606037731364)(0.35000000000000014,0.08747360037017389)(0.36000000000000015,0.10345916044147702)(0.37000000000000016,0.12129132669578953)(0.38000000000000017,0.14100553603950433)(0.3900000000000002,0.16261164304381404)(0.4000000000000002,0.18609202141541234)(0.4100000000000002,0.21140028113488576)(0.4200000000000002,0.23846066424405193)(0.4300000000000002,0.2671681615778913)(0.4400000000000002,0.297389370185422)(0.45000000000000023,0.3289640875783932)(0.46000000000000024,0.36170761513803523)(0.47000000000000025,0.395413719857028)(0.48000000000000026,0.42985818193115577)(0.49000000000000027,0.46480283632334163)(0.5000000000000002,0.5000000000000009)(0.5100000000000002,0.5351971636766606)(0.5200000000000002,0.5701418180688463)(0.5300000000000002,0.6045862801429742)(0.5400000000000003,0.6382923848619667)(0.5500000000000003,0.6710359124216085)(0.5600000000000003,0.7026106298145799)(0.5700000000000003,0.7328318384221105)(0.5800000000000003,0.7615393357559498)(0.5900000000000003,0.7885997188651158)(0.6000000000000003,0.8139079785845889)(0.6100000000000003,0.8373883569561873)(0.6200000000000003,0.8589944639604967)(0.6300000000000003,0.8787086733042114)(0.6400000000000003,0.8965408395585239)(0.6500000000000004,0.9125263996298271)(0.6600000000000004,0.9267239396226872)(0.6700000000000004,0.9392123229847085)(0.6800000000000004,0.9500874868860042)(0.6900000000000004,0.9594590207110593)(0.7000000000000004,0.9674466431186993)(0.7100000000000004,0.9741766923202158)(0.7200000000000004,0.9797787381766703)(0.7300000000000004,0.9843824147375732)(0.7400000000000004,0.9881145584113704)(0.7500000000000004,0.9910967206960778)(0.7600000000000005,0.9934431060501905)(0.7700000000000005,0.9952589658850771)(0.7800000000000005,0.9966394597013374)(0.7900000000000005,0.9976689749796165)(0.8000000000000005,0.9984208794508329)(0.8100000000000005,0.9989576636211741)(0.8200000000000005,0.9993314186103643)(0.8300000000000005,0.9995845850226063)(0.8400000000000005,0.9997509030675364)(0.8500000000000005,0.9998564926310838)(0.8600000000000005,0.9999209943830139)(0.8700000000000006,0.9999587089868986)(0.8800000000000006,0.9999796805149197)(0.8900000000000006,0.9999906815324768)(0.9000000000000006,0.9999960701176728)(0.9100000000000006,0.9999985023294481)(0.9200000000000006,0.9999994963165824)(0.9300000000000006,0.9999998554011522)(0.9400000000000006,0.9999999662488399)
}; 
\label{b1}

\addplot[yellow,thick] 
coordinates {
(0.0,0.0)(0.01,4.067583875640052e-16)(0.02,3.8693758162046577e-13)(0.03,2.071471818293994e-11)(0.04,3.412782628531393e-10)(0.05,2.9468853193508532e-09)(0.060000000000000005,1.6905883669928008e-08)(0.07,7.312586754869924e-08)(0.08,2.571900316233835e-07)(0.09,7.722120415045632e-07)(0.09999999999999999,2.0462439096099987e-06)(0.10999999999999999,4.900201451505992e-06)(0.11999999999999998,1.079214946183614e-05)(0.12999999999999998,2.2151902544560503e-05)(0.13999999999999999,4.281621784677621e-05)(0.15,7.85695232764649e-05)(0.16,0.00013778842595072455)(0.17,0.00023218058192815195)(0.18000000000000002,0.0003776003595318002)(0.19000000000000003,0.0005949156159770969)(0.20000000000000004,0.0009108923573862414)(0.21000000000000005,0.001359057565278349)(0.22000000000000006,0.001980495492510992)(0.23000000000000007,0.0028245296241467695)(0.24000000000000007,0.003949241533741203)(0.25000000000000006,0.005421779205789802)(0.26000000000000006,0.007318411090430719)(0.2700000000000001,0.009724288144011158)(0.2800000000000001,0.012732884217932797)(0.2900000000000001,0.01644509511893156)(0.3000000000000001,0.020967988121870552)(0.3100000000000001,0.0264132062471694)(0.3200000000000001,0.0328950447440408)(0.3300000000000001,0.04052823044120881)(0.34000000000000014,0.049425447421011987)(0.35000000000000014,0.059694664331932955)(0.36000000000000015,0.07143632909803951)(0.37000000000000016,0.08474050537703921)(0.38000000000000017,0.0996840314890668)(0.3900000000000002,0.11632778638881068)(0.4000000000000002,0.13471414837313583)(0.4100000000000002,0.15486473048775695)(0.4200000000000002,0.1767784720007544)(0.4300000000000002,0.20043015792818034)(0.4400000000000002,0.22576942860531743)(0.45000000000000023,0.25272032896597174)(0.46000000000000024,0.28118143287494135)(0.47000000000000025,0.3110265619818069)(0.48000000000000026,0.34210610161335964)(0.49000000000000027,0.374248898727638)(0.5000000000000002,0.40726470947265697)(0.5100000000000002,0.44094714699541826)(0.5200000000000002,0.47507706439123404)(0.5300000000000002,0.5094262936028314)(0.5400000000000003,0.543761649161813)(0.5500000000000003,0.5778490963397603)(0.5600000000000003,0.6114579768944471)(0.5700000000000003,0.6443651824213312)(0.5800000000000003,0.6763591655152059)(0.5900000000000003,0.7072436825680282)(0.6000000000000003,0.7368411690211746)(0.6100000000000003,0.7649956580855414)(0.6200000000000003,0.7915751670624149)(0.6300000000000003,0.8164734910587721)(0.6400000000000003,0.8396113616146352)(0.6500000000000004,0.8609369469873799)(0.6600000000000004,0.8804256909428079)(0.6700000000000004,0.8980795072130854)(0.6800000000000004,0.9139253665998486)(0.6900000000000004,0.9280133323277613)(0.7000000000000004,0.9404141160133617)(0.7100000000000004,0.9512162408781275)(0.7200000000000004,0.9605229100485061)(0.7300000000000004,0.968448685491486)(0.7400000000000004,0.9751160869934196)(0.7500000000000004,0.9806522204016802)(0.7600000000000005,0.9851855400659741)(0.7700000000000005,0.9888428421550871)(0.7800000000000005,0.9917465735704354)(0.7900000000000005,0.9940125259818386)(0.8000000000000005,0.9957479666837096)(0.8100000000000005,0.9970502382637665)(0.8200000000000005,0.9980058383621127)(0.8300000000000005,0.9986899700329802)(0.8400000000000005,0.9991665334083437)(0.8500000000000005,0.9994885115057928)(0.8600000000000005,0.9996986880740141)(0.8700000000000006,0.999830624170094)(0.8800000000000006,0.9999098133870508)(0.8900000000000006,0.999954933743352)(0.9000000000000006,0.9999791173719152)(0.9100000000000006,0.9999911671378432)(0.9200000000000006,0.9999966616426439)(0.9300000000000006,0.9999989058315571)(0.9400000000000006,0.99999970233951)
}; 
\label{b2}

\addplot[orange,thick] 
coordinates {(0.0,0.0)(0.01,1.8243915234857775e-16)(0.02,1.751533990518825e-13)(0.03,9.464229680337734e-12)(0.04,1.5738969247429196e-10)(0.05,1.3719096665542605e-09)(0.060000000000000005,7.945630278219959e-09)(0.07,3.469953406949345e-08)(0.08,1.232262349106782e-07)(0.09,3.7360846437668313e-07)(0.09999999999999999,9.997781220999994e-07)(0.10999999999999999,2.418029501159832e-06)(0.11999999999999998,5.378890587692235e-06)(0.12999999999999998,1.1152413718715813e-05)(0.13999999999999999,2.177586950986206e-05)(0.15,4.037079054728539e-05)(0.16,7.153336255019013e-05)(0.17,0.00012179841598502216)(0.18000000000000002,0.00020017290215131134)(0.19000000000000003,0.0003187299598976342)(0.20000000000000004,0.0004932497375232006)(0.21000000000000005,0.000743888289895161)(0.22000000000000006,0.0010958513860034038)(0.23000000000000007,0.0015800461885777913)(0.24000000000000007,0.0022336807336962336)(0.25000000000000006,0.0031007791403681114)(0.26000000000000006,0.004232579672728672)(0.2700000000000001,0.005687783268383841)(0.2800000000000001,0.007532621991962468)(0.2900000000000001,0.009840720076077291)(0.3000000000000001,0.012692724722277338)(0.3100000000000001,0.016175689550233352)(0.3200000000000001,0.020382200354367514)(0.3300000000000001,0.025409240460549674)(0.34000000000000014,0.03135680124199558)(0.35000000000000014,0.03832625199482455)(0.36000000000000015,0.04641849211097893)(0.37000000000000016,0.05573191702882463)(0.38000000000000017,0.0663602374967785)(0.3900000000000002,0.0783901989666768)(0.4000000000000002,0.09189925417123873)(0.4100000000000002,0.10695324688849515)(0.4200000000000002,0.12360416834273921)(0.4300000000000002,0.14188804946352165)(0.4400000000000002,0.16182305219450982)(0.45000000000000023,0.1834078211364976)(0.46000000000000024,0.2066201530017062)(0.47000000000000025,0.23141603568442842)(0.48000000000000026,0.25772910130778637)(0.49000000000000027,0.2854705285357719)(0.5000000000000002,0.3145294189453131)(0.5100000000000002,0.3447736605859873)(0.5200000000000002,0.37605127931038784)(0.5300000000000002,0.408192265368637)(0.5400000000000003,0.4410108494877331)(0.5500000000000003,0.47430818958215126)(0.5600000000000003,0.5078754167579328)(0.5700000000000003,0.5414969777692623)(0.5800000000000003,0.5749542009428914)(0.5900000000000003,0.6080290041569458)(0.6000000000000003,0.6405076570669066)(0.6100000000000003,0.6721845056872774)(0.6200000000000003,0.7028655658807281)(0.6300000000000003,0.7323718934297998)(0.6400000000000003,0.7605426422481233)(0.6500000000000004,0.7872377289267412)(0.6600000000000004,0.8123400311194557)(0.6700000000000004,0.8357570590742623)(0.6800000000000004,0.8574220536527305)(0.6900000000000004,0.8772944800966352)(0.7000000000000004,0.8953599041711324)(0.7100000000000004,0.9116292556331476)(0.7200000000000004,0.9261375026767841)(0.7300000000000004,0.9389417794802134)(0.7400000000000004,0.9501190265742829)(0.7500000000000004,0.9597632198128853)(0.7600000000000005,0.9679822775988568)(0.7700000000000005,0.9748947470898918)(0.7800000000000005,0.9806263778183855)(0.7900000000000005,0.9853066950347493)(0.8000000000000005,0.989065684765901)(0.8100000000000005,0.9920306978495356)(0.8200000000000005,0.9943236710058584)(0.8300000000000005,0.9960587494752562)(0.8400000000000005,0.9973403782233666)(0.8500000000000005,0.9982619077548227)(0.8600000000000005,0.9989047369704436)(0.8700000000000006,0.9993379902593081)(0.8800000000000006,0.9996187003542635)(0.8900000000000006,0.9997924437927843)(0.9000000000000006,0.999894353643127)(0.9100000000000006,0.9999504160733717)(0.9200000000000006,0.9999789449305287)(0.9300000000000006,0.9999921231915927)(0.9400000000000006,0.9999975030950943)
}; 
\label{b3}

\addplot[red,thick] 
coordinates {
(0.0,0.0)(0.01,7.581087654801807e-17)(0.02,7.345631258437552e-14)(0.03,4.006139442568316e-12)(0.04,6.724823640622949e-11)(0.05,5.917359685745241e-10)(0.060000000000000005,3.459896415036958e-09)(0.07,1.5255469057877616e-08)(0.08,5.470255372770871e-08)(0.09,1.6747862487939553e-07)(0.09999999999999999,4.526064684999996e-07)(0.10999999999999999,1.1055790317104412e-06)(0.11999999999999998,2.4840997459040504e-06)(0.12999999999999998,5.20272402251855e-06)(0.13999999999999999,1.0262683005394678e-05)(0.15,1.9222703223172175e-05)(0.16,3.4415679972859955e-05)(0.17,5.921462239572104e-05)(0.18000000000000002,9.834939719837658e-05)(0.19000000000000003,0.000158273514826051)(0.20000000000000004,0.0002475776081920004)(0.21000000000000005,0.000377443449756255)(0.22000000000000006,0.0005621294514588556)(0.23000000000000007,0.0008194757161139388)(0.24000000000000007,0.001171413984442384)(0.25000000000000006,0.0016444653738290108)(0.26000000000000006,0.0022702067520914424)(0.2700000000000001,0.003085685040743346)(0.2800000000000001,0.004133757791981861)(0.2900000000000001,0.005463338108982499)(0.3000000000000001,0.007129522436836522)(0.3100000000000001,0.009193580976022919)(0.3200000000000001,0.011722792472240662)(0.3300000000000001,0.01479010790170218)(0.34000000000000014,0.018473631060878723)(0.35000000000000014,0.02285590822135239)(0.36000000000000015,0.028023023738140276)(0.37000000000000016,0.03406350369682495)(0.38000000000000017,0.04106703522559189)(0.3900000000000002,0.049123014841597645)(0.4000000000000002,0.058318944993280206)(0.4100000000000002,0.06873870363883472)(0.4200000000000002,0.0804607170989135)(0.4300000000000002,0.09355607137092418)(0.4400000000000002,0.10808660142912528)(0.45000000000000023,0.12410300160325768)(0.46000000000000024,0.14164300278537928)(0.47000000000000025,0.16072966383325998)(0.48000000000000026,0.18136982501314997)(0.49000000000000027,0.20355277057326546)(0.5000000000000002,0.22724914550781303)(0.5100000000000002,0.2524101682359934)(0.5200000000000002,0.2789671762899504)(0.5300000000000002,0.3068315362230104)(0.5400000000000003,0.3358949418927663)(0.5500000000000003,0.36603011715589306)(0.5600000000000003,0.39709192998091175)(0.5700000000000003,0.42891891522253844)(0.5800000000000003,0.4613351930187298)(0.5900000000000003,0.4941527592087878)(0.6000000000000003,0.5271741135912971)(0.6100000000000003,0.5601951815264761)(0.6200000000000003,0.5930084746340569)(0.6300000000000003,0.625406427447005)(0.6400000000000003,0.6571848391546434)(0.6500000000000004,0.6881463432989913)(0.6600000000000004,0.7181038237514943)(0.6700000000000004,0.7468836927443898)(0.6800000000000004,0.7743289463775568)(0.6900000000000004,0.8003019150398972)(0.7000000000000004,0.8246866306931253)(0.7100000000000004,0.8473907400226325)(0.7200000000000004,0.8683469020520416)(0.7300000000000004,0.8875136208549015)(0.7400000000000004,0.9048754783043528)(0.7500000000000004,0.92044274811633)(0.7600000000000005,0.9342503904084305)(0.7700000000000005,0.9463564451662181)(0.7800000000000005,0.9568398628407389)(0.7900000000000005,0.9657978301673221)(0.8000000000000005,0.9733426684887042)(0.8100000000000005,0.9795983996099547)(0.8200000000000005,0.9846970896823178)(0.8300000000000005,0.9887750939521451)(0.8400000000000005,0.9919693335616688)(0.8500000000000005,0.9944137391243275)(0.8600000000000005,0.9962359937651646)(0.8700000000000006,0.9975547000845176)(0.8800000000000006,0.9984770806178427)(0.8900000000000006,0.9990972996192324)(0.9000000000000006,0.9994954655076524)(0.9100000000000006,0.9997373386120826)(0.9200000000000006,0.9998747289769094)(0.9300000000000006,0.9999465256119993)(0.9400000000000006,0.9999802541192839)
}; 
\label{b4}

\addplot[purple,thick] 
coordinates {
(0.0,0.0)(0.01,2.8689795070498007e-17)(0.02,2.8055827554238472e-14)(0.03,1.54437193841974e-12)(0.04,2.6168145750211434e-11)(0.05,2.3244544976806647e-10)(0.060000000000000005,1.3721213457363584e-09)(0.07,6.108395463703499e-09)(0.08,2.211656403472052e-08)(0.09,6.837774050569952e-08)(0.09999999999999999,1.8662024799999985e-07)(0.10999999999999999,4.6041377285189227e-07)(0.11999999999999998,1.0449281626286753e-06)(0.12999999999999998,2.2107823649480857e-06)(0.13999999999999999,4.405684056901095e-06)(0.15,8.337658276937434e-06)(0.16,1.5083553630500477e-05)(0.17,2.6226176979673774e-05)(0.18000000000000002,4.402283205580468e-05)(0.19000000000000003,7.160722504973282e-05)(0.20000000000000004,0.00011322566246400017)(0.21000000000000005,0.00017450722499578504)(0.22000000000000006,0.0002627661868905995)(0.23000000000000007,0.0003873334022140227)(0.24000000000000007,0.0005599117439179633)(0.25000000000000006,0.0007949490100145357)(0.26000000000000006,0.0011100200591471333)(0.2700000000000001,0.0015262083632190764)(0.2800000000000001,0.0020684757260214237)(0.2900000000000001,0.0027660076715121163)(0.3000000000000001,0.003652521008436011)(0.3100000000000001,0.004766519380056159)(0.3200000000000001,0.006151482253960518)(0.3300000000000001,0.007855972835290567)(0.34000000000000014,0.009933650827183834)(0.35000000000000014,0.012443176835361511)(0.36000000000000015,0.01544799653014509)(0.37000000000000016,0.019015994438491828)(0.38000000000000017,0.02321900942939167)(0.3900000000000002,0.028132206555167908)(0.4000000000000002,0.033833302884352136)(0.4100000000000002,0.040401648263027175)(0.4200000000000002,0.04791716551413121)(0.4300000000000002,0.05645915836125506)(0.4400000000000002,0.0661049992686686)(0.45000000000000023,0.07692871333818045)(0.46000000000000024,0.0889994783045588)(0.47000000000000025,0.10238006442781437)(0.48000000000000026,0.1171252415921818)(0.49000000000000027,0.13328018408582123)(0.5000000000000002,0.15087890625000042)(0.5100000000000002,0.16994276435207034)(0.5200000000000002,0.1904790615577808)(0.5300000000000002,0.212479793667241)(0.5400000000000003,0.235920573256249)(0.5500000000000003,0.26075976896369757)(0.5600000000000003,0.2869378948333057)(0.5700000000000003,0.3143772818174413)(0.5800000000000003,0.3429820597643946)(0.5900000000000003,0.3726384734409365)(0.6000000000000003,0.403215550414849)(0.6100000000000003,0.4345661319871155)(0.6200000000000003,0.4665282708961133)(0.6300000000000003,0.4989269913187002)(0.6400000000000003,0.5315763978952058)(0.6500000000000004,0.5642821112643039)(0.6600000000000004,0.596843998094191)(0.6700000000000004,0.6290591540494826)(0.6800000000000004,0.66072508877478)(0.6900000000000004,0.691643053064662)(0.7000000000000004,0.721621440204365)(0.7100000000000004,0.7504791862998723)(0.7200000000000004,0.7780490885758808)(0.7300000000000004,0.8041809564157384)(0.7400000000000004,0.8287445076578354)(0.7500000000000004,0.851631922647358)(0.7600000000000005,0.8727599710508823)(0.7700000000000005,0.8920716317244478)(0.7800000000000005,0.9095371341920095)(0.7900000000000005,0.9251543616935152)(0.8000000000000005,0.9389485703823365)(0.8100000000000005,0.9509713970846037)(0.8200000000000005,0.9612991489653787)(0.8300000000000005,0.9700303922382559)(0.8400000000000005,0.9772828833148393)(0.8500000000000005,0.9831899139520502)(0.8600000000000005,0.9878961712486679)(0.8700000000000006,0.9915532427655108)(0.8800000000000006,0.9943149253288086)(0.8900000000000006,0.9963325216562432)(0.9000000000000006,0.9977503299149522)(0.9100000000000006,0.9987015453974822)(0.9200000000000006,0.9993047979805549)(0.9300000000000006,0.9996615407395402)(0.9400000000000006,0.999854480337334)
};
\label{b5}

\addplot[violet,thick] 
coordinates {
(0.0,0.0)(0.01,9.650974271500003e-18)(0.02,9.525017337856001e-15)(0.03,5.292100810388832e-13)(0.04,9.051441310203905e-12)(0.05,8.116523132324222e-11)(0.060000000000000005,4.837064226297203e-10)(0.07,2.174170261908179e-09)(0.08,7.948742429073477e-09)(0.09,2.4816912209569396e-08)(0.09999999999999999,6.840414999999994e-08)(0.10999999999999999,1.704518587581625e-07)(0.11999999999999998,3.9075926113986844e-07)(0.12999999999999998,8.35177005145574e-07)(0.13999999999999999,1.6814984994622183e-06)(0.15,3.215284184591673e-06)(0.16,5.8777791817578705e-06)(0.17,1.0328130996036536e-05)(0.18000000000000002,1.752206857162327e-05)(0.19000000000000003,2.8809057258958427e-05)(0.20000000000000004,4.6049689600000075e-05)(0.21000000000000005,7.175470612972425e-05)(0.22000000000000006,0.00010924656403508354)(0.23000000000000007,0.00016284388849978048)(0.24000000000000007,0.00023806845943142633)(0.25000000000000006,0.00034187361598014913)(0.26000000000000006,0.00048289211701507445)(0.2700000000000001,0.0006717005947126272)(0.2800000000000001,0.0009210968004878477)(0.2900000000000001,0.0012463848898386612)(0.3000000000000001,0.0016656630493500054)(0.3100000000000001,0.002200106860655138)(0.3200000000000001,0.002874240949089847)(0.3300000000000001,0.003716190706089603)(0.34000000000000014,0.004757905231005112)(0.35000000000000014,0.006035342136290204)(0.36000000000000015,0.007588604525148099)(0.37000000000000016,0.009462020306216831)(0.38000000000000017,0.011704154077004437)(0.3900000000000002,0.014367742105050041)(0.4000000000000002,0.01750954147840007)(0.4100000000000002,0.02119008529637799)(0.4200000000000002,0.025473336834970977)(0.4300000000000002,0.030426236950960934)(0.4400000000000002,0.03611814058262809)(0.45000000000000023,0.04262014005448792)(0.46000000000000024,0.05000427498543251)(0.47000000000000025,0.058342630914270444)(0.48000000000000026,0.0677063312683601)(0.49000000000000027,0.07816442997802178)(0.5000000000000002,0.08978271484375026)(0.5100000000000002,0.10262243465090864)(0.5200000000000002,0.1167389659476394)(0.5300000000000002,0.1321804383006287)(0.5400000000000003,0.1489863396592773)(0.5500000000000003,0.16718612612619055)(0.5600000000000003,0.18679786288093644)(0.5700000000000003,0.20782692516153695)(0.5800000000000003,0.23026478999836097)(0.5900000000000003,0.2540879507405935)(0.6000000000000003,0.2792569872384008)(0.6100000000000003,0.3057158247672585)(0.6200000000000003,0.3333912143298567)(0.6300000000000003,0.362192465774587)(0.6400000000000003,0.3920114631624975)(0.6500000000000004,0.4227229889389469)(0.6600000000000004,0.4541853796738339)(0.6700000000000004,0.4862415313889891)(0.6800000000000004,0.5187202667713089)(0.6900000000000004,0.5514380698708099)(0.7000000000000004,0.5842011862193514)(0.7100000000000004,0.6168080777167548)(0.7200000000000004,0.6490522121813656)(0.7300000000000004,0.6807251572466078)(0.7400000000000004,0.7116199374324238)(0.7500000000000004,0.7415346018970025)(0.7600000000000005,0.770275938788302)(0.7700000000000005,0.7976632605213684)(0.7800000000000005,0.8235321730125017)(0.7900000000000005,0.8477382312672161)(0.8000000000000005,0.870160374169601)(0.8100000000000005,0.890704023347022)(0.8200000000000005,0.9093037251499584)(0.8300000000000005,0.9259252117349873)(0.8400000000000005,0.9405667576977652)(0.8500000000000005,0.9532597134926442)(0.8600000000000005,0.9640681069158202)(0.8700000000000006,0.9730872202454899)(0.8800000000000006,0.9804410743653622)(0.8900000000000006,0.9862787836090084)(0.9000000000000006,0.9907697875441501)(0.9100000000000006,0.9940980199992595)(0.9200000000000006,0.9964551429987814)(0.9300000000000006,0.9980330557540604)(0.9400000000000006,0.9990159884576684)
};
\label{b6}

\addplot[black,thick] 
coordinates {
(0.0,0.0)(0.01,2.7827128000000008e-18)(0.02,2.7718066175999997e-15)(0.03,1.554391468043999e-13)(0.04,2.6836197179392e-12)(0.05,2.429296875000001e-11)(0.060000000000000005,1.4616275883540487e-10)(0.07,6.6333109071182e-10)(0.08,2.448811681539687e-09)(0.09,7.720825751119589e-09)(0.09999999999999999,2.1492999999999986e-08)(0.10999999999999999,5.40947503738408e-08)(0.11999999999999998,1.2526863553564488e-07)(0.12999999999999998,2.704785815813245e-07)(0.13999999999999999,5.501922048447448e-07)(0.15,1.0630261625976555e-06)(0.16,1.9637594331428164e-06)(0.17,3.487319471235655e-06)(0.18000000000000002,5.979924196979788e-06)(0.19000000000000003,9.938612906765134e-06)(0.20000000000000004,1.606041600000003e-05)(0.21000000000000005,2.5302391994977062e-05)(0.22000000000000006,3.8953696427246544e-05)(0.23000000000000007,5.8720737426384286e-05)(0.24000000000000007,8.682631446594837e-05)(0.25000000000000006,0.00012612342834472683)(0.26000000000000006,0.00018022419127566383)(0.2700000000000001,0.0002536439565000786)(0.2800000000000001,0.00035196042869539907)(0.2900000000000001,0.00048198711233692304)(0.3000000000000001,0.000651960009000002)(0.3100000000000001,0.0008717359913999511)(0.3200000000000001,0.001153000767960293)(0.3300000000000001,0.0015094838141914352)(0.34000000000000014,0.001957177094544764)(0.35000000000000014,0.0025145538400971773)(0.36000000000000015,0.0032027830937881715)(0.37000000000000016,0.004045935197217398)(0.38000000000000017,0.0050711728832329855)(0.3900000000000002,0.006308922169383841)(0.4000000000000002,0.007793016832000033)(0.4100000000000002,0.009560809892837321)(0.4200000000000002,0.011653245283763942)(0.4300000000000002,0.014114882683859346)(0.4400000000000002,0.016993868461428783)(0.45000000000000023,0.020341845714427823)(0.46000000000000024,0.02421379659976696)(0.47000000000000025,0.028667810487354062)(0.48000000000000026,0.033764771980021015)(0.49000000000000027,0.039567963515977356)(0.5000000000000002,0.046142578125000146)(0.5100000000000002,0.05355513895035344)(0.5200000000000002,0.06187282338056988)(0.5300000000000002,0.07116269106156156)(0.5400000000000003,0.08149081668026197)(0.5500000000000003,0.09292133022340707)(0.5600000000000003,0.10551536941310423)(0.5700000000000003,0.11932995119483894)(0.5800000000000003,0.13441677148982895)(0.5900000000000003,0.15082094490405773)(0.6000000000000003,0.16857969868800055)(0.6100000000000003,0.1877210379358875)(0.6200000000000003,0.20826240176758526)(0.6300000000000003,0.2302093330099988)(0.6400000000000003,0.25355418664195334)(0.6500000000000004,0.2782749049334804)(0.6600000000000004,0.3043338897364838)(0.6700000000000004,0.3316770047001432)(0.6800000000000004,0.3602327422138633)(0.6900000000000004,0.3899115915368787)(0.7000000000000004,0.42060564576100123)(0.7100000000000004,0.4521884858656247)(0.7200000000000004,0.4845153800455038)(0.7300000000000004,0.5174238355943185)(0.7400000000000004,0.5507345387711438)(0.7500000000000004,0.5842527151107804)(0.7600000000000005,0.617769938397557)(0.7700000000000005,0.6510664108271454)(0.7800000000000005,0.6839137295392743)(0.7900000000000005,0.7160781455082039)(0.8000000000000005,0.7473243095040014)(0.8100000000000005,0.777419486246305)(0.8200000000000005,0.8061382017066636)(0.8300000000000005,0.8332672695012296)(0.8400000000000005,0.8586111201596536)(0.8500000000000005,0.8819973314464388)(0.8600000000000005,0.9032822285156983)(0.8700000000000006,0.9223563891465311)(0.8800000000000006,0.9391498512598663)(0.8900000000000006,0.9536367769621426)(0.9000000000000006,0.9658392790770007)(0.9100000000000006,0.9758300620698035)(0.9200000000000006,0.9837334689730066)(0.9300000000000006,0.9897244588893678)(0.9400000000000006,0.9940249653644195)
};
\label{b7}

\addplot[gray,thick] 
coordinates {
(0.0,0.0)(0.01,6.4805500000000015e-19)(0.02,6.514892800000001e-16)(0.03,3.687580525499998e-14)(0.04,6.426512588800001e-13)(0.05,5.872802734375003e-12)(0.060000000000000005,3.567383451648003e-11)(0.07,1.6346701422005514e-10)(0.08,6.093699599564801e-10)(0.09,1.9402386138584544e-09)(0.09999999999999999,5.4549999999999966e-09)(0.10999999999999999,1.386757375004764e-08)(0.11999999999999998,3.243974546423803e-08)(0.12999999999999998,7.076207457363232e-08)(0.13999999999999999,1.4543145542883315e-07)(0.15,2.8392823608398425e-07)(0.16,5.300525655182541e-07)(0.17,9.513374416523816e-07)(0.18000000000000002,1.6489131745994972e-06)(0.19000000000000003,2.7703529439186753e-06)(0.20000000000000004,4.5260800000000085e-06)(0.21000000000000005,7.209961952232289e-06)(0.22000000000000006,1.1224754570111636e-05)(0.23000000000000007,1.7113084649802376e-05)(0.24000000000000007,2.5594676828103057e-05)(0.25000000000000006,3.7610530853271566e-05)(0.26000000000000006,5.4374741903767504e-05)(0.2700000000000001,7.743462532513722e-05)(0.2800000000000001,0.00010873975698922437)(0.2900000000000001,0.000150720469865964)(0.3000000000000001,0.0002063762550000007)(0.3100000000000001,0.0002793743997588419)(0.3200000000000001,0.0003741590570419422)(0.3300000000000001,0.0004960707754322305)(0.34000000000000014,0.0006514763316261866)(0.35000000000000014,0.0008479084927868685)(0.36000000000000015,0.0010942150979420518)(0.37000000000000016,0.001400716584763463)(0.38000000000000017,0.0017793708019568299)(0.3900000000000002,0.0022439436394135004)(0.4000000000000002,0.002810183680000012)(0.4100000000000002,0.0034959987306257597)(0.4200000000000002,0.004321631728746942)(0.4300000000000002,0.005309833146962397)(0.4400000000000002,0.006486026636593997)(0.45000000000000023,0.007878464265443154)(0.46000000000000024,0.009518367320185454)(0.47000000000000025,0.011440048265631912)(0.48000000000000026,0.013681009087512675)(0.49000000000000027,0.016282010899359295)(0.5000000000000002,0.019287109375000073)(0.5100000000000002,0.02274365028438469)(0.5200000000000002,0.026702219170909917)(0.5300000000000002,0.031216539022892374)(0.5400000000000003,0.03634330967088715)(0.5500000000000003,0.04214198259757219)(0.5600000000000003,0.04867446489014459)(0.5700000000000003,0.05600474620972401)(0.5800000000000003,0.06419844291214971)(0.5900000000000003,0.0733222538447438)(0.6000000000000003,0.08344332288000032)(0.6100000000000003,0.09462850394664006)(0.6200000000000003,0.10694352519894433)(0.6300000000000003,0.12045205004568409)(0.6400000000000003,0.13521463406029152)(0.6500000000000004,0.1512875783352679)(0.6600000000000004,0.16872168164838408)(0.6700000000000004,0.18756089589935418)(0.6800000000000004,0.20784089167785774)(0.6900000000000004,0.22958754356275335)(0.7000000000000004,0.2528153478550009)(0.7100000000000004,0.27752578894134844)(0.7200000000000004,0.30370567440169954)(0.7300000000000004,0.331325463340997)(0.7400000000000004,0.3603376172785044)(0.7500000000000004,0.39067500829696794)(0.7600000000000005,0.42224942507608876)(0.7700000000000005,0.45495022394524165)(0.7800000000000005,0.4886431792270676)(0.7900000000000005,0.5231695949455482)(0.8000000000000005,0.5583457484800017)(0.8100000000000005,0.593962746002228)(0.8200000000000005,0.6297868795813732)(0.8300000000000005,0.665560586725197)(0.8400000000000005,0.7010041248940534)(0.8500000000000005,0.7358180862234527)(0.8600000000000005,0.7696868913725727)(0.8700000000000006,0.8022834161312433)(0.8800000000000006,0.8332749202201328)(0.8900000000000006,0.8623304646632165)(0.9000000000000006,0.8891300222550016)(0.9100000000000006,0.9133755050460224)(0.9200000000000006,0.9348039534892576)(0.9300000000000006,0.9532031539896189)(0.9400000000000006,0.9684299751426305)
};
\label{b8}

\end{axis}
\node [draw,fill=white] at (1.5,3.5) {\shortstack[l]{
{\scriptsize $b$ sabotaged peers}\\
\ref{b0} {\small $b = 0$}\\
\ref{b1} {\small $b = 1$}\\
\ref{b2} {\small $b = 2$}\\
\ref{b3} {\small $b = 3$}\\
\ref{b4} {\small $b = 4$}\\
\ref{b5} {\small $b = 5$}\\
\ref{b6} {\small $b = 6$}\\
\ref{b7} {\small $b = 7$}\\
\ref{b8} {\small $b = 8$}
}
};
\end{tikzpicture}

%% file: 5_2.tex
\subsection{Orderer sabotage\label{ssec:orderer_sabotage}}

Wining a puzzle requires a solution-carrying transaction to be ordered in a new block.
For this purpose, Tendermint \cite{dissecting_tendermint} consensus instances are regularly executed by the orderers.
Tendermint is based on rounds of communications, each one corresponding to an attempt to reach consensus.
These attempts rely on a proposer to PROPOSE a new block, which will then be voted upon.

The adversary can sabotage an orderer via an \advactInject~action to force it not to include transactions from $c$ whenever it proposes a new block.
Because there are unknown delays between emissions and corresponding receptions (which might be arbitrary in the asynchronous communication model, or bounded in the synchronous and partially synchronous communication models), and because some messages might even be lost (depending on the failure model) it is impossible for the other orderers to know whether these transactions were omitted on purpose or because they have not been received at the moment of the proposal (guaranteeing the discretion of the attack).

If there are sabotaged orderers, the likelihood of transactions from $c$ to be included in the next block diminishes, thus negatively impacting its client-fairness score.
However, because the proposer generally isn't the same from one round to the next, infecting less than $f'=(n'-1)/3$ orderers cannot reduce the score to $0$ i.e., total censorship is not possible.
Yet, because orderers (as Tendermint \cite{dissecting_tendermint} is used) require $2*f'+1$ PRECOMMIT messages to order a block, if the adversary sabotages more than $f'$ orderers and makes so that these orderers do not PREVOTE and PRECOMMIT blocks containing transactions from $c$, then, total censorship is possible.

%% file: 6_0_sim_attacks.tex
\section{Simulations of the attacks\label{sec:sim_attacks}}

\input{6_1}

\input{6_2}

\input{6_3}

\input{6_4}

%% file: 6_1.tex
Using MAX \cite{max_tool}, we have simulated attack scenarios which are combinations of basic attacks from Sec.\ref{sec:attacks}.
In these attacks, the adversary attempts to reduce the fairness $\mathtt{score}$ (as defined in Sec.\ref{sec:usecase}) of a specific client.
To do so, it infects a minority of peers and/or orderers.

For the simulations of this section, we consider $m=3$ clients, $n = 25$ peers and $n' = 25$ orderers (in Sec.\ref{sec:sim_params}, we consider additional simulations, varying $m$, $n$ and $n'$).
To be endorsed, a transaction requires $12$ endorsements from distinct peers.
We use an arbitrary unit of time denoted as ``tick'' in our discrete time simulations.
Heartbeat transactions are submitted every 20 ticks.
A new puzzle is revealed every 200 ticks and the \textcolor{yellow}{\faLightbulbO} Poisson distribution which determines the time required by a client to solve a puzzle has a mean of $75$ ticks.
The Tendermint consensus is parameterized so that empty blocks can be emitted and the timeout for each phase is set to 1500 ticks.
In every simulation $5000$ puzzles are revealed and the duration of the simulation is $1~200~000$ ticks (we wait $200~000$ ticks after the last puzzle is revealed so as to guarantee very puzzle to be solved).
The details of the experiments and the means to reproduce them are available at \cite{max_fabric_tendermint_of_exp_with_mitigation}.

We vary the following parameters : \textbf{(1)} the \faClockO~ delay distribution, which can either be small (\textcolor{blue}{\faClockO} in blue), medium (\textcolor{green}{\faClockO} in green) or large (\textcolor{red}{\faClockO} in red), \textbf{(2)} the proportion of infected peers (between $0\%$ and $50\%$) and \textbf{(3)} the proportion of infected orderers (between $0\%$ and $33\%$).

For each simulation, we measure $9$ metrics : the numbers of $OF_{S_{ND}}^{D_{LV}}$, $OF_{S_{ND}}^{B_{LC}}$, $OF_{E_{DS}}^{D_{LV}}$, $OF_{E_{DS}}^{B_{LC}}$, $OF_{O_{RD}}^{D_{LV}}$ and $OF_{O_{RD}}^{B_{LC}}$ violations, the client-fairness $\mathtt{score}(c)$ of the target client $c$, the total number of blocks that is delivered and a measure on the observed distribution of the sizes of the blocks (the third quartile).

%% file: 6_2.tex
\subsection{Analysis of the results}

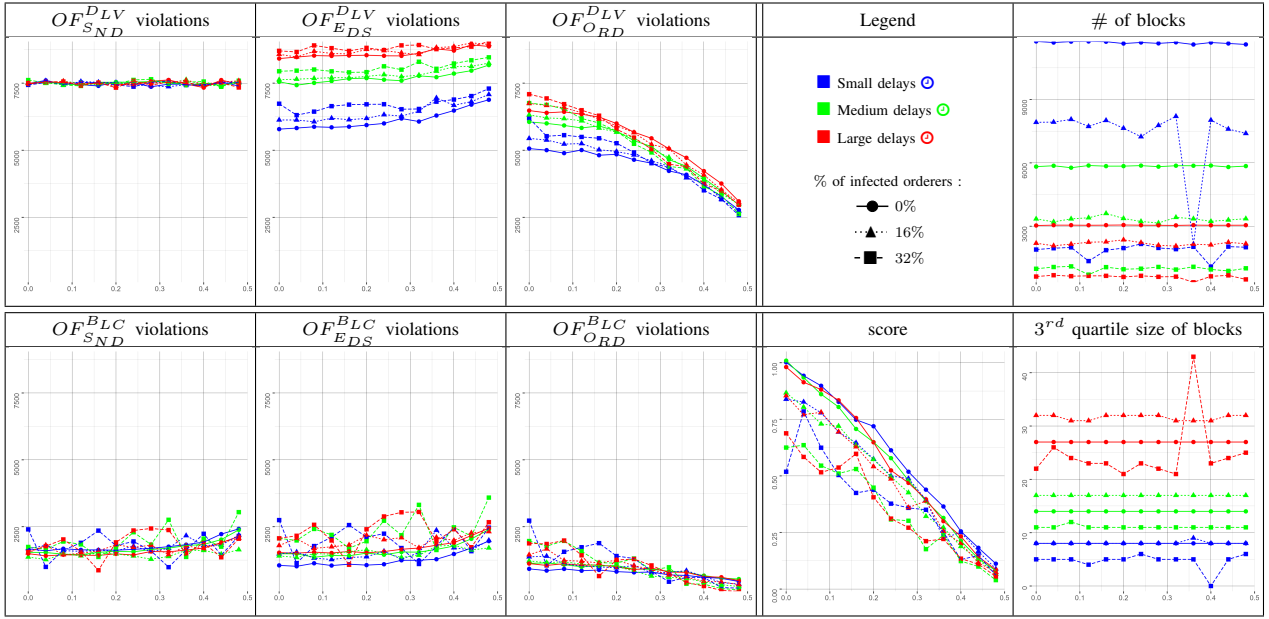
\begin{figure*}[h]
    \centering

\setlength\tabcolsep{1.5pt}
\begin{tabular}{|c|c|c||c|c|}
\hline
{\scriptsize$OF_{S_{ND}}^{D_{LV}}$ violations}
&
{\scriptsize$OF_{E_{DS}}^{D_{LV}}$ violations}
&
{\scriptsize$OF_{O_{RD}}^{D_{LV}}$ violations}
&
{\scriptsize Legend}
&
{\scriptsize $\#$ of blocks}
\\
\hline 
\includegraphics[scale=.3]{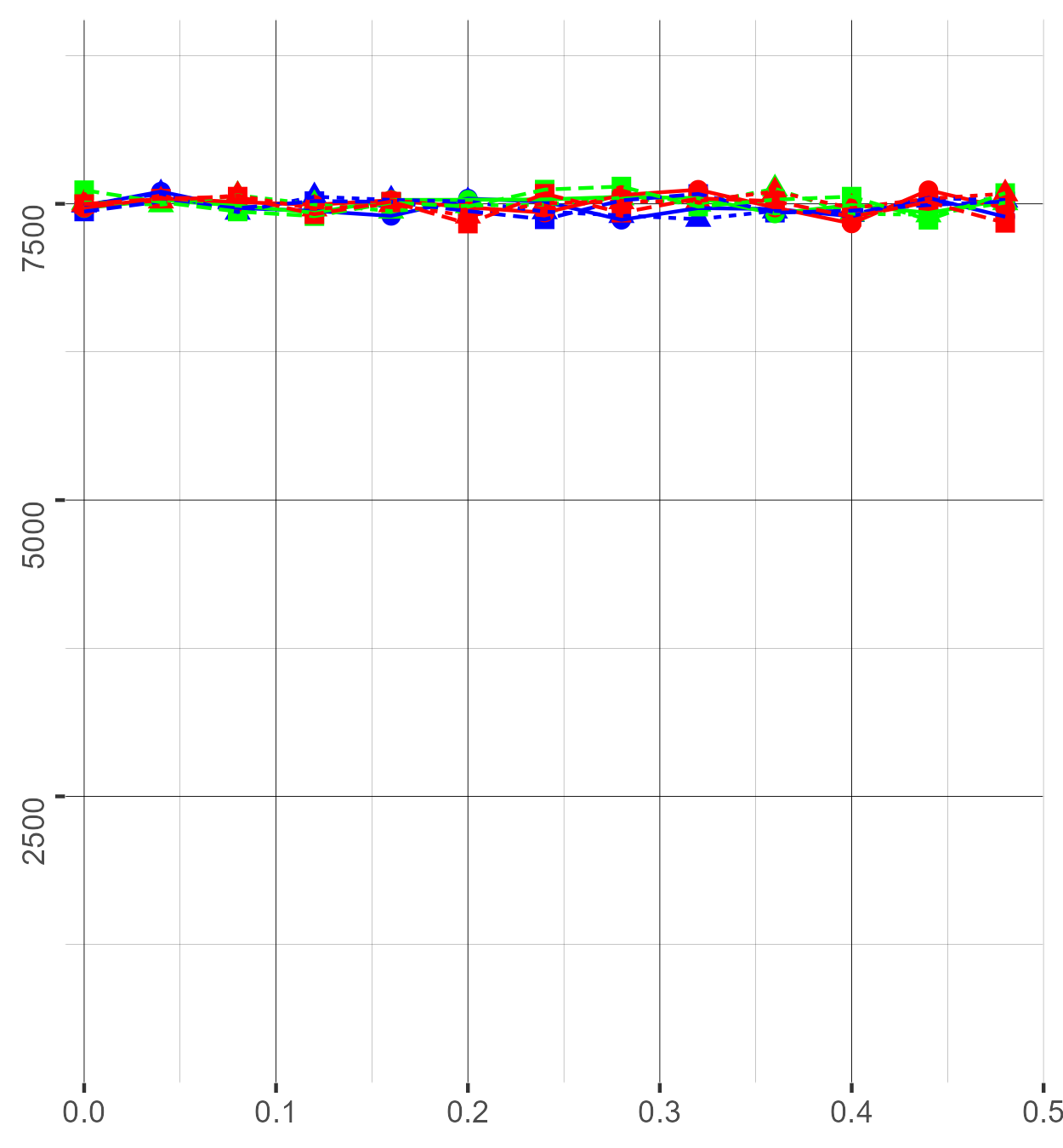}
&
\includegraphics[scale=.3]{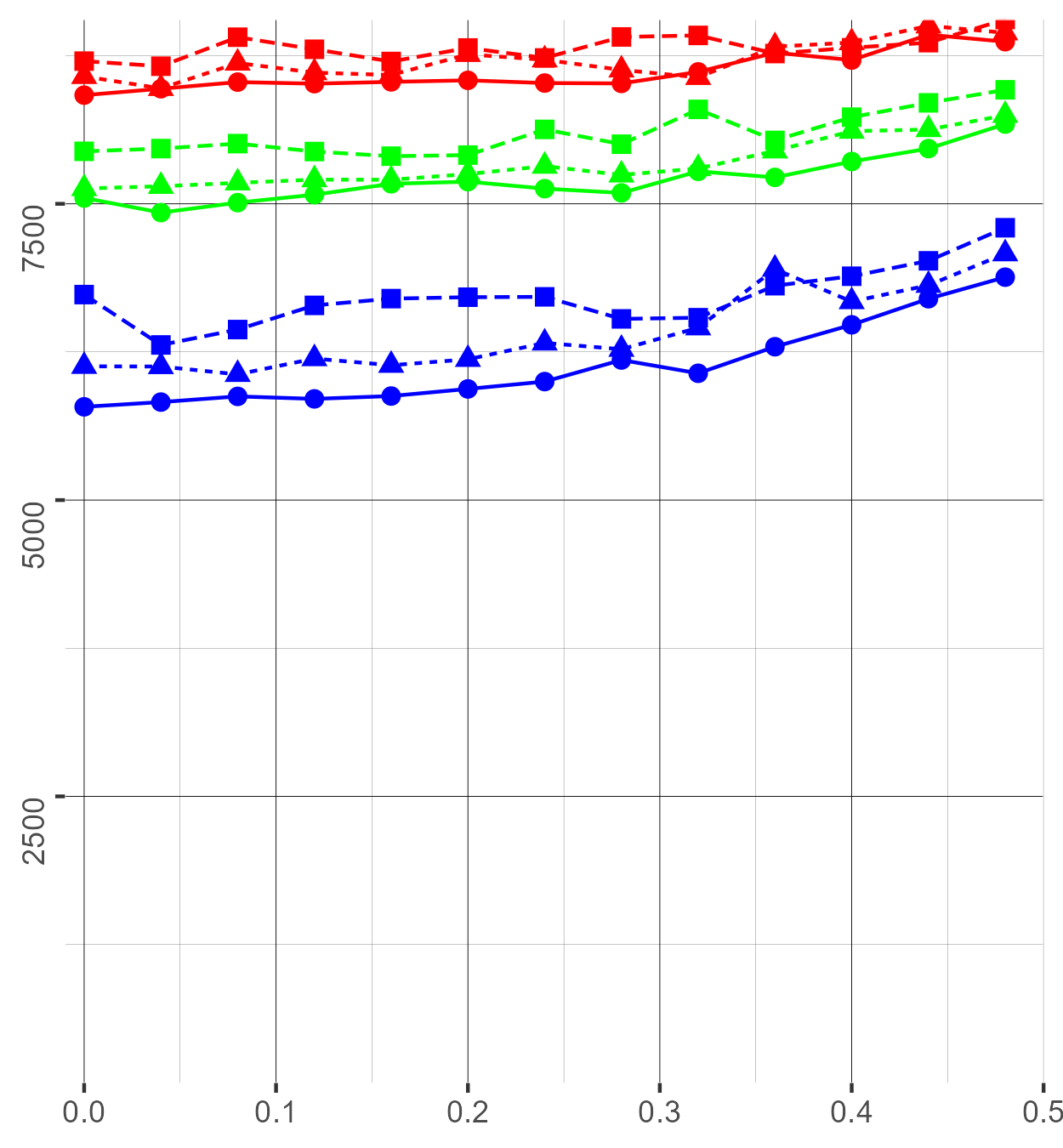}
&
\includegraphics[scale=.3]{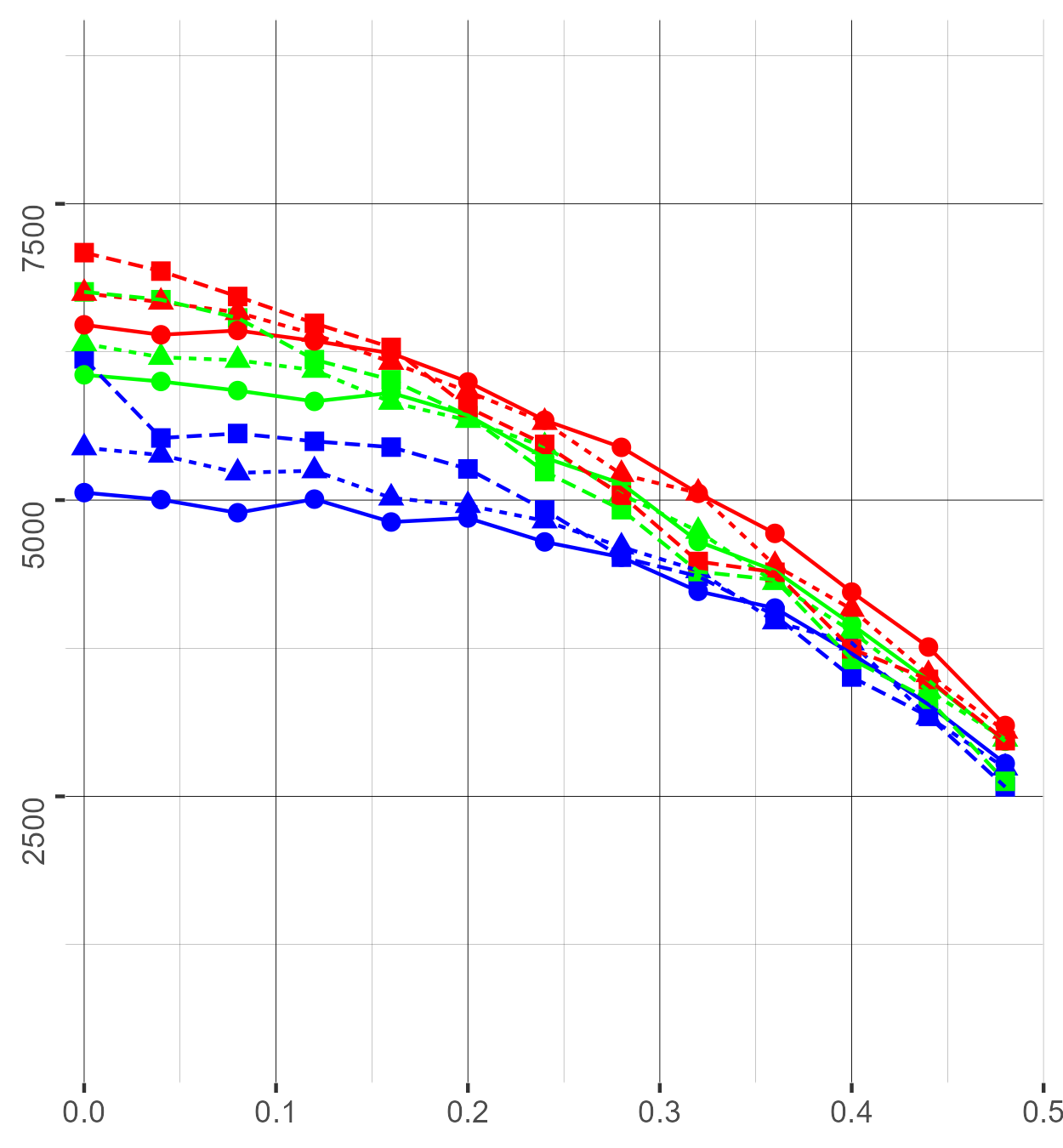}
&
\scalebox{.7}{
\begin{tikzpicture}
\node[align=center] (leg1) at (0,0) {\textcolor{blue}{$\blacksquare$} {\footnotesize Small delays 
 \textcolor{blue}{\faClockO}}}; 
\node[below=.25cm of leg1.south west,anchor=west,align=center] (leg2) {\textcolor{green}{$\blacksquare$} {\footnotesize Medium delays 
 \textcolor{green}{\faClockO}}};
\node[below=.25cm of leg2.south west,anchor=west,align=center] (leg3) {\textcolor{red}{$\blacksquare$} {\footnotesize Large delays 
 \textcolor{red}{\faClockO}}};
\node[below right=.3cm and -2.5cm of leg3] (leg4) {\footnotesize $\%$ of infected orderers :};
\node[below right=0cm and -1.5cm of leg4] (leg5) {\footnotesize $0\%$};
\node[draw=black,fill=black,left=.2cm of leg5,circle,inner sep=2pt] (leg5s) {};
\draw[thick] ($(leg5s) + (-.3,0) $) -- ($ (leg5s) + (.3,0) $);
\node[below=.25cm of leg5.south west,anchor=west,align=center] (leg6) {\footnotesize $16\%$};
\node[left=.2cm of leg6,inner sep=0pt] (leg6s) {$\blacktriangle$};
\draw[thick,dotted] ($(leg6s) + (-.3,0) $) -- ($ (leg6s) + (.3,0) $);
\node[below=.25cm of leg6.south west,anchor=west,align=center] (leg7) {\footnotesize $32\%$};
\node[left=.2cm of leg7,inner sep=0pt] (leg7s) {$\blacksquare$};
\draw[thick,dashed] ($(leg7s) + (-.3,0) $) -- ($ (leg7s) + (.3,0) $);
\node[below=.25cm of leg7] {};
\end{tikzpicture}
}
&
\includegraphics[scale=.3]{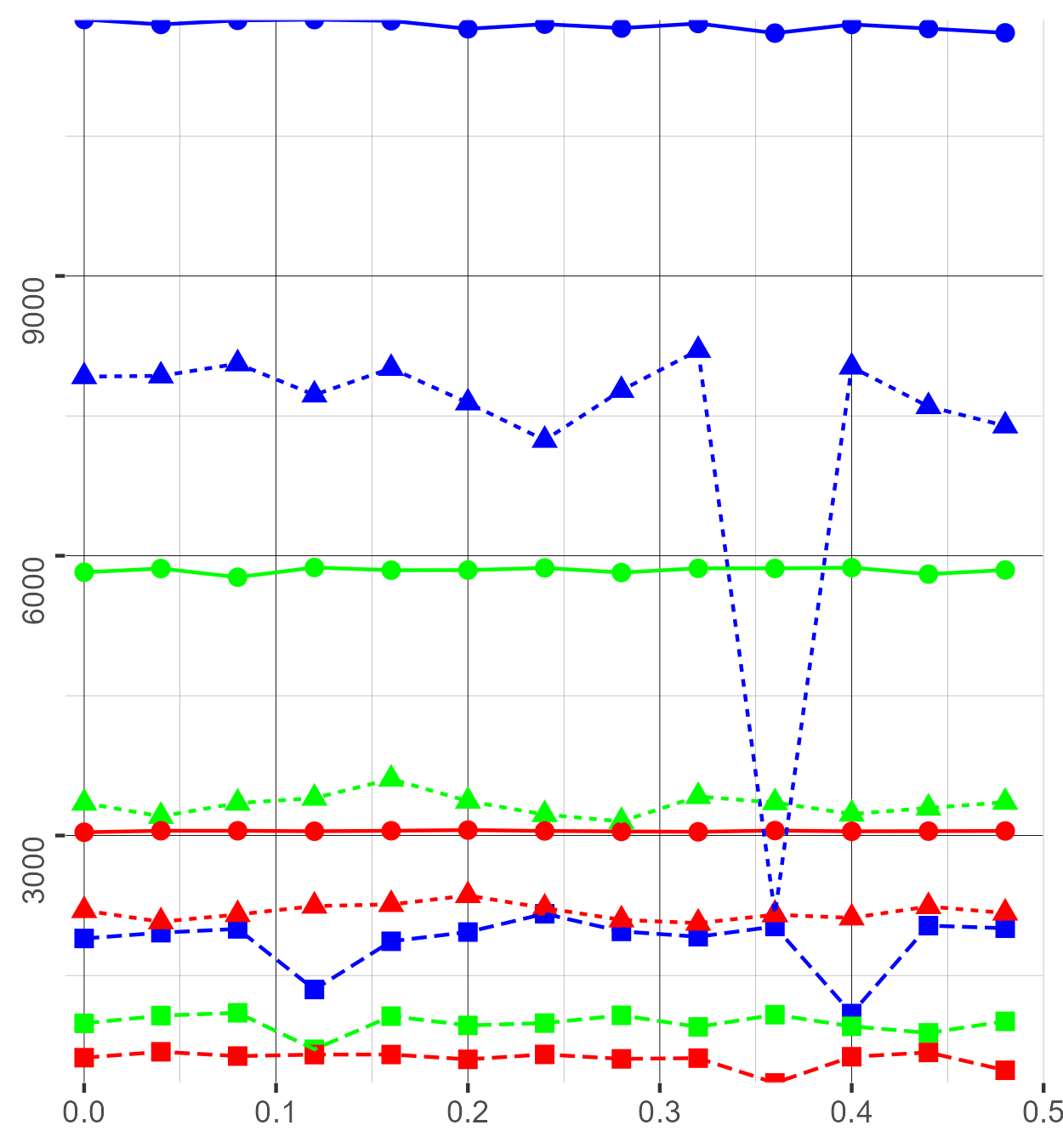}
\\
\hline 
\hline
{\scriptsize$OF_{S_{ND}}^{B_{LC}}$ violations}
&
{\scriptsize$OF_{E_{DS}}^{B_{LC}}$ violations}
&
{\scriptsize$OF_{O_{RD}}^{B_{LC}}$ violations}
&
{\scriptsize score}
&
{\scriptsize $3^{rd}$ quartile size of blocks}
\\
\hline 
\includegraphics[scale=.3]{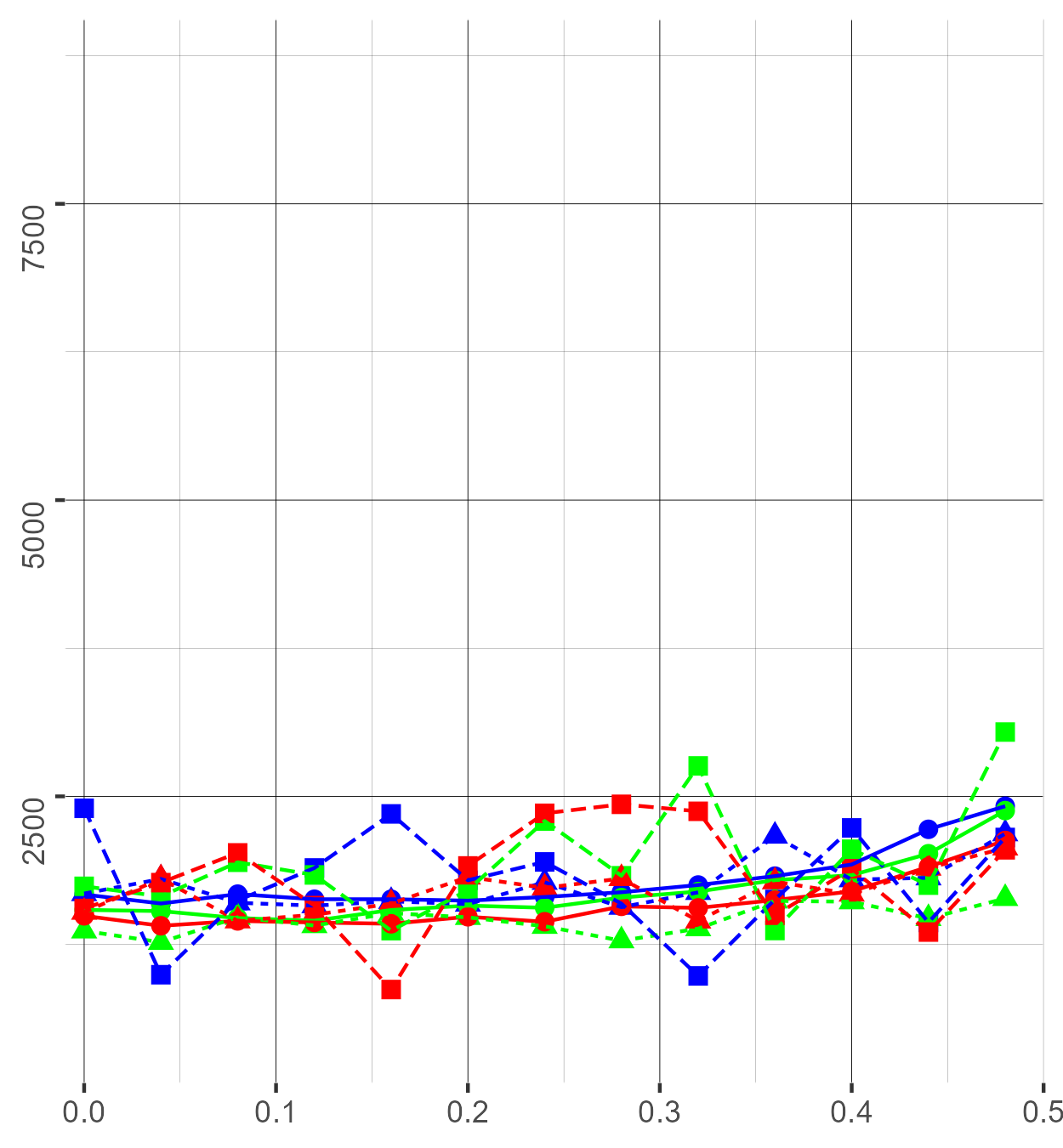}
&
\includegraphics[scale=.3]{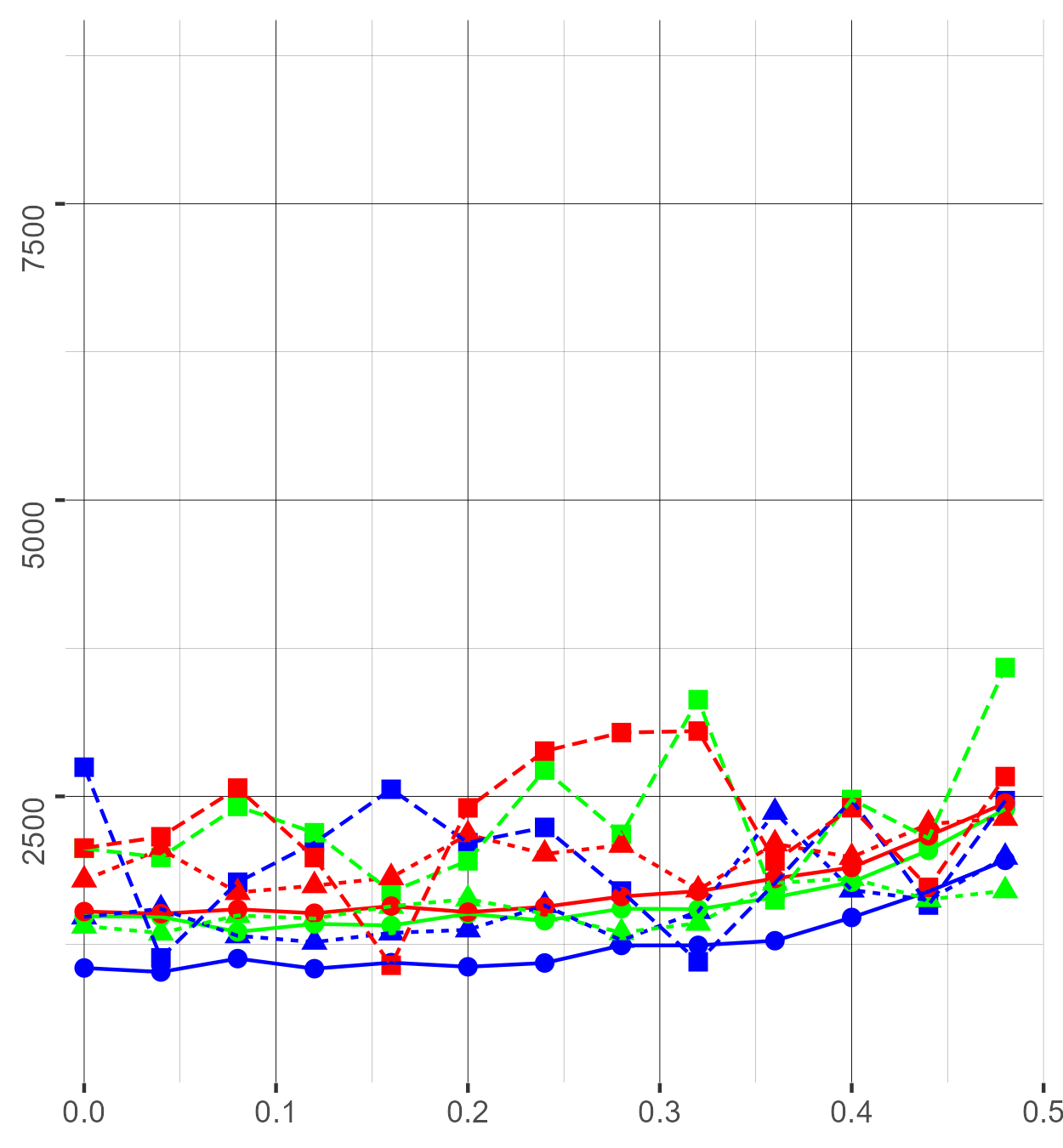}
&
\includegraphics[scale=.3]{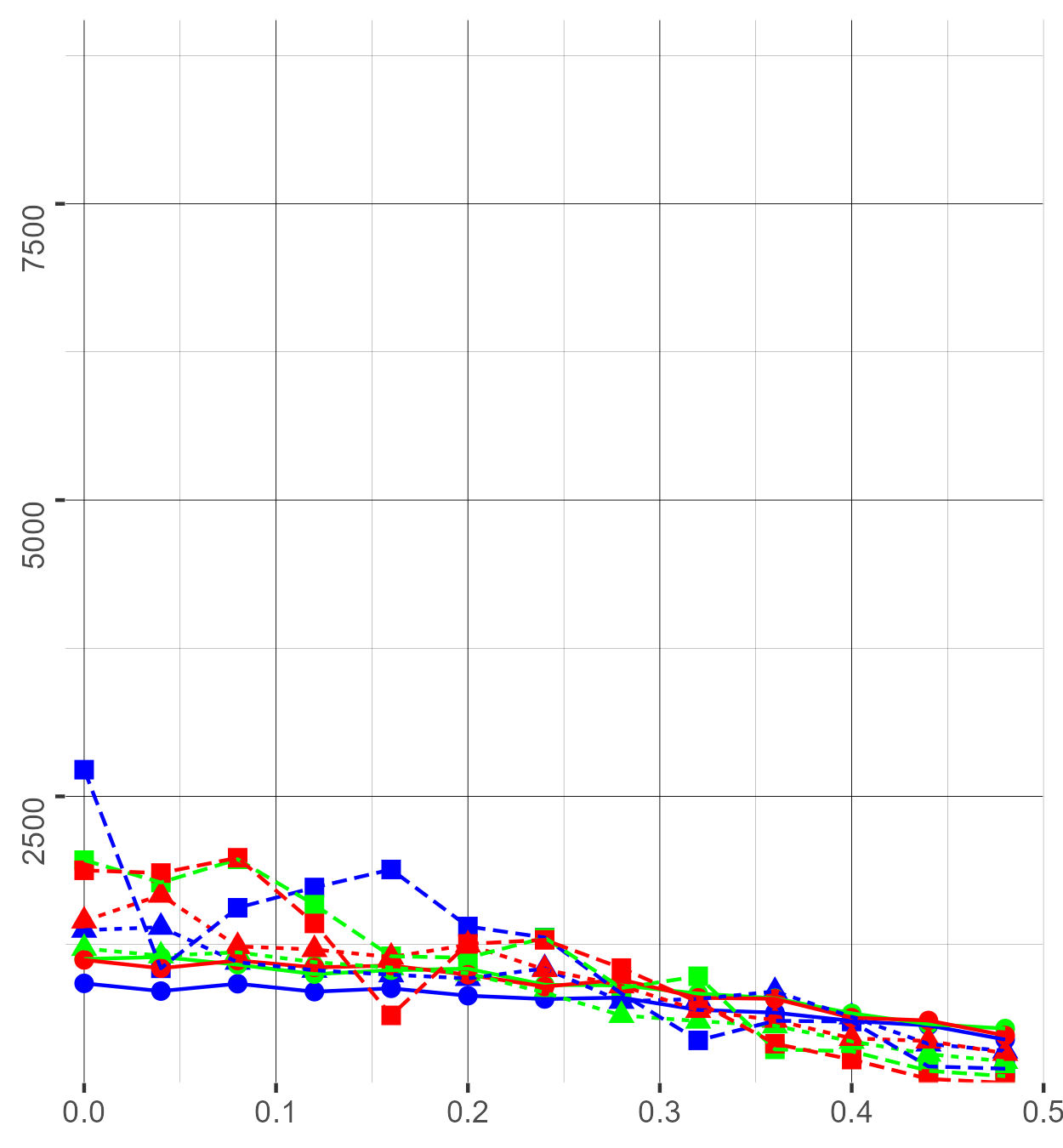}
&
\includegraphics[scale=.3]{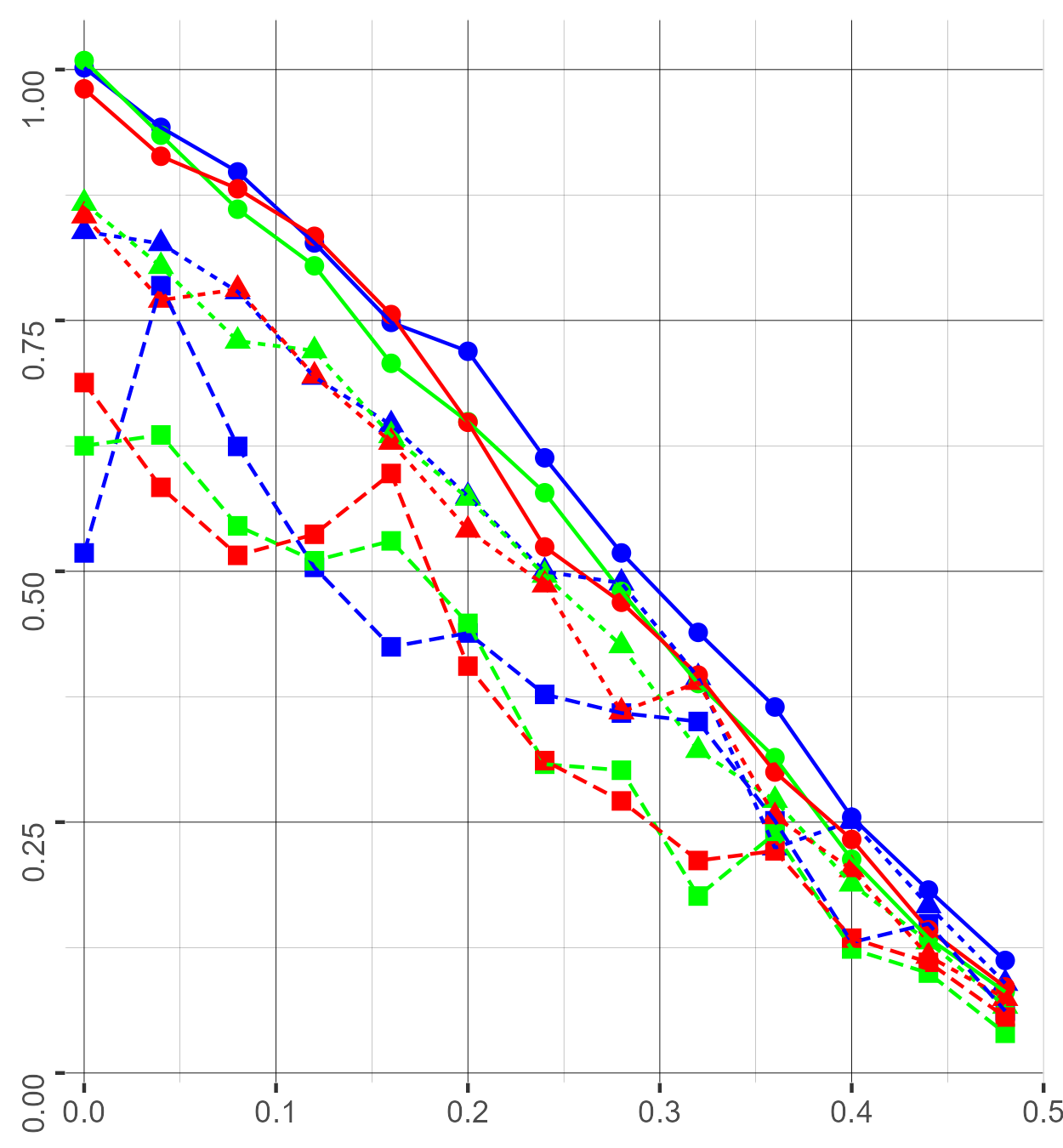}
&
\includegraphics[scale=.3]{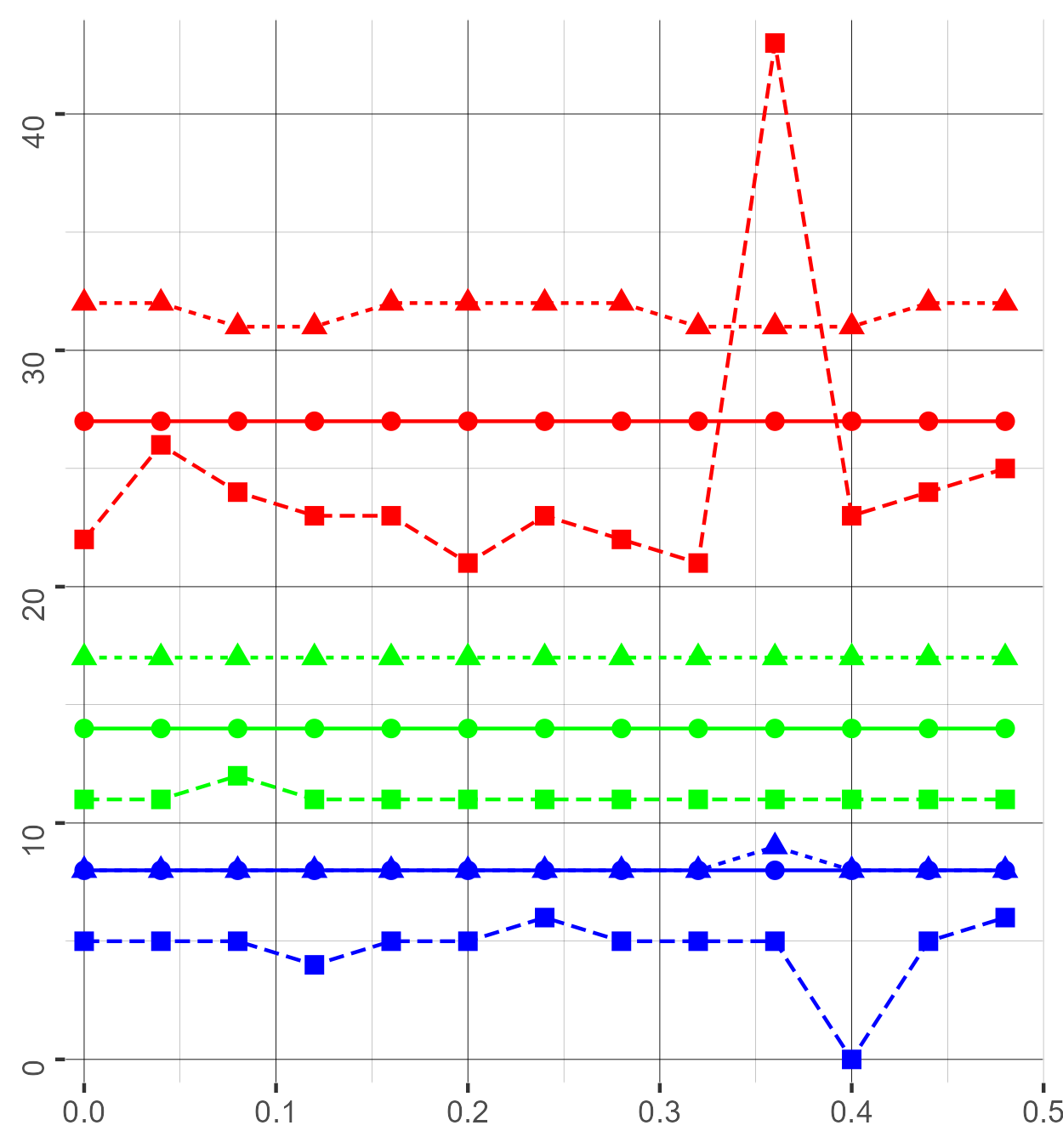}
\\
\hline 
\end{tabular}
\setlength\tabcolsep{6pt}
    
    \caption{Varying the number of infected Peers}
    \label{fig:exp1_peers}
\vspace*{-.5cm}
\end{figure*}

\paragraph{Diagrams presentation.}
Fig.\ref{fig:exp1_peers} focuses on the effect of the attack on the endorsing service (i.e., the effect of the adversary infecting increasingly more peers).
The 9 diagrams on Fig.\ref{fig:exp1_peers} plot our 9 metrics on the vertical axis, with, on the horizontal axis, the proportion of peers that are infected by the adversary.
In each diagram, we have 9 curves, which correspond to 9 combinations of parameters.
Each curve has 13 data points, which correspond to infecting between 0 and 12 peers.
Thus, Fig.\ref{fig:exp1_peers} reports metrics collected from $117$ distinct simulations.

As indicated in the legend, the three blue (resp.~green and red) curves correspond to simulations in which the small (resp.~medium and large) delay distribution \textcolor{blue}{\faClockO} (resp.~\textcolor{green}{\faClockO} and \textcolor{red}{\faClockO}) is considered.
The shape of the points and style of the lines correspond to proportions of infected orderers ($0\%$,$16\%$ or $32\%$), highlighting the effect of combining peer and orderer sabotage.

The six diagrams on the left correspond to the number of violations of our six $OF$ properties. These six diagram share the same scale on the vertical axis. The 3 diagrams on the right report on the $\mathtt{score}(c)$ metric of the targeted client $c$ as well as the total number of blocks that have been delivered and their size (we report the third quartile of the observed distribution of the size of the blocks as it is more stable and is not too much biased by the presence of empty blocks).

\paragraph{Discussion on peer sabotage.}
We observe that, when no peers and no orderers are infected (at horizontal position $0$, for the plain line curves), the $\mathtt{score}$ of the target client (i.e. the proportion of puzzles it has won, relative to the total number of puzzles ($\sim5000$) and the total number of clients) stays around $1$ which is expected as per Fig.\ref{fig:client_fairness_score_convergence}.

As discussed in Sec.\ref{ssec:endorser_sabotage}, peer sabotage statistically delays the endorsement of transactions from the target client.
Because collecting sufficiently many endorsements for these transactions is delayed, their delivery is likely to be delayed as well, resulting in other clients being more likely to win puzzle games.
Experimentally, this effect is observed on Fig.\ref{fig:exp1_peers}, especially on the diagrams corresponding to the $\mathtt{score}$ and $OF_{E_{DS}}^{D_{LV}}$ metrics.
Indeed, we observe that, as the proportion of infected peers increase, the score decreases and the number of receive-order fairness violations w.r.t.~the instants at which peers receive the transactions increases.
The attack is quite successful, as the score almost reaches $0$ (meaning the target client almost never wins) when the proportion of infected peers comes closer to the threshold (which is $50\%$ in our use-case, having a majority endorsing policy).

This statistical delay, because it is correlated to Y on Fig.\ref{fig:peer_sabotage_delay_theory}, depends on the distribution of the communication delays between the peers and the clients, which corresponds to either the blue \textcolor{blue}{\faClockO} (small delays), green \textcolor{green}{\faClockO} (medium delays) or red \textcolor{red}{\faClockO} (large delays)  from Fig.\ref{fig:distros} and is correlated to X on Fig.\ref{fig:peer_sabotage_delay_theory}.
Indeed, with higher communication delays, the likelihood that an honest peer receives an endorsement before a certain timestamp, which corresponds $X$ on Fig.\ref{fig:peer_sabotage_delay_theory}, decreases. At lower values of $X$, the values of $Y$ are also lowered.

On Fig.\ref{fig:exp1_peers} this translates into having, at a fixed proportion of infected peers (i.e., horizontal position), a lower score for larger delays. Indeed, if we look at, e.g., the top three curves on the $\mathtt{score}$ diagram, we observe that the $\mathtt{score}$ is higher in the case of the small delays \textcolor{blue}{\faClockO}.
For $OF_{E_{DS}}^{D_{LV}}$, we observe that the baseline number of $OF_{E_{DS}}^{D_{LV}}$ violations (whether or not peers are infected) is higher at higher values of delays.
This is due to the fact that larger randomly sampled delays are more likely to cause endorsed transactions to be mis-ordered due to overtaking in communications from the peers to the clients and then from the clients to the orderers.
Also, the number of violations tends to increase as more peers are infected and this increase is sharper for the smaller delays (at larger delays there are already many naturally occurring violations so the effect of the adversary is less noticeable).
Similar remarks may be made w.r.t.~$OF_{E_{DS}}^{B_{LC}}$.

As for $OF_{O_{RD}}^{D_{LV}}$, we also observe a higher number of violations at higher delays.
However, an increase in the proportion of infected peers results in a decrease in the number of $OF_{O_{RD}}^{D_{LV}}$ violations.
Indeed, unlike $OF_{E_{DS}}^{D_{LV}}$ which is defined w.r.t.~the instants at which peers receive transactions, $OF_{O_{RD}}^{D_{LV}}$ is defined w.r.t.~the instants at which orderers receive endorsed transactions.
Infecting peers results in tampering with the delivery order but also with the order with which orderers receive endorsed transactions (the attack targeting the endorsing service which is upstream w.r.t.~ the orderers).
Thus, the attack does not yield an increase in $OF_{O_{RD}}^{D_{LV}}$ violations.
On the contrary, it makes so that there is less competition between transactions from the target client and the other clients. Indeed, because delayed transactions are more likely to be both received and ordered after those of the other clients for the same puzzle, there are less risks of $OF_{O_{RD}}^{D_{LV}}$ violations, having less pairs of transactions susceptible to cause such violations.
Interestingly, in our simulations we have three clients, and, after neutralizing the target client on the endosring service, the number of $OF_{O_{RD}}^{D_{LV}}$ violations roughly decreases by two thirds.
Similar remarks may be made w.r.t.~$OF_{O_{RD}}^{B_{LC}}$.

Concerning $OF_{S_{ND}}^{D_{LV}}$, the effect of the attack is not noticeable.
It may be so that the non-determinism of the network already saturates the number of violations.
However, even if there still is some noise, we can observe an increase in $OF_{S_{ND}}^{B_{LC}}$ violations with the proportion of infected peers.
The statistical delay caused by peer sabotage is indeed likely to force transactions from the target client into later blocks, which is observable when considering $OF_{S_{ND}}^{B_{LC}}$.

Finally, at higher delays, there are less blocks but these blocks are bigger.
Infecting peers do not have a noticeable effect on the number and size of blocks, as the ordering service is not directly tampered with.

%% file: 6_3.tex
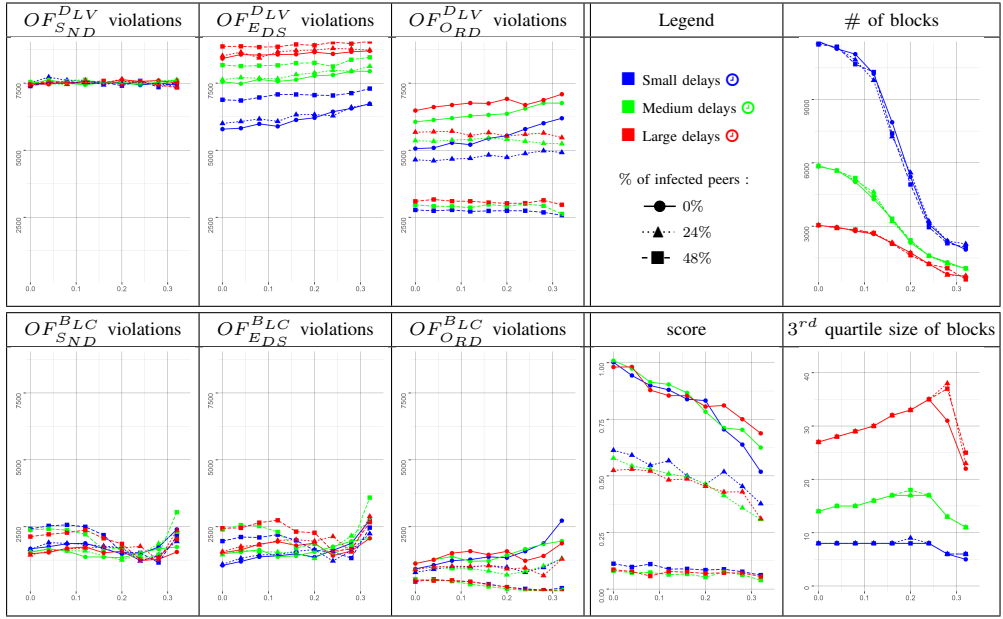
\begin{figure*}[h]
    \centering

\setlength\tabcolsep{1.5pt}
\begin{tabular}{|c|c|c||c|c|}
\hline
{\scriptsize$OF_{S_{ND}}^{D_{LV}}$ violations}
&
{\scriptsize$OF_{E_{DS}}^{D_{LV}}$ violations}
&
{\scriptsize$OF_{O_{RD}}^{D_{LV}}$ violations}
&
{\scriptsize Legend}
&
{\scriptsize $\#$ of blocks}
\\
\hline 
\includegraphics[scale=.3]{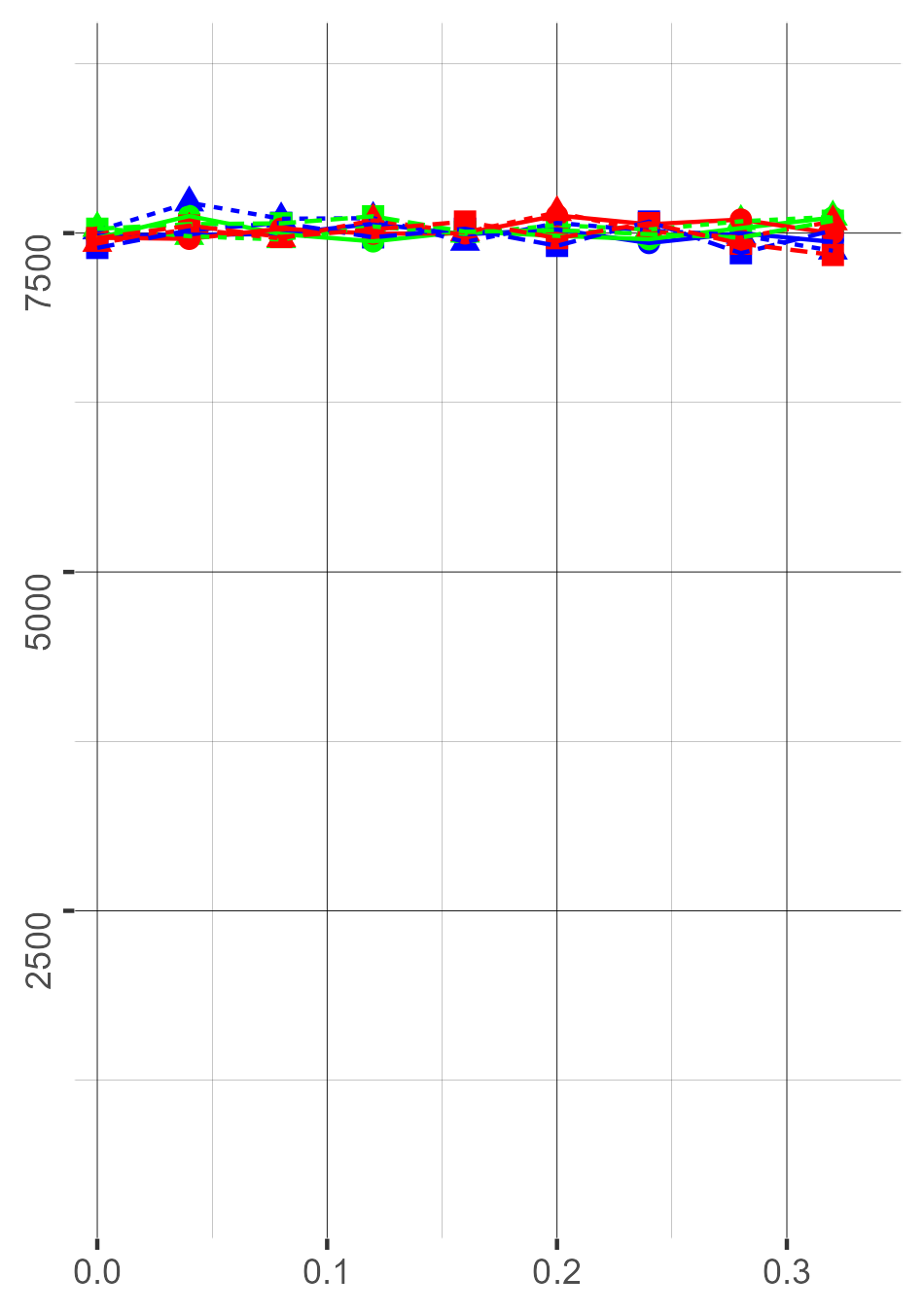}
&
\includegraphics[scale=.3]{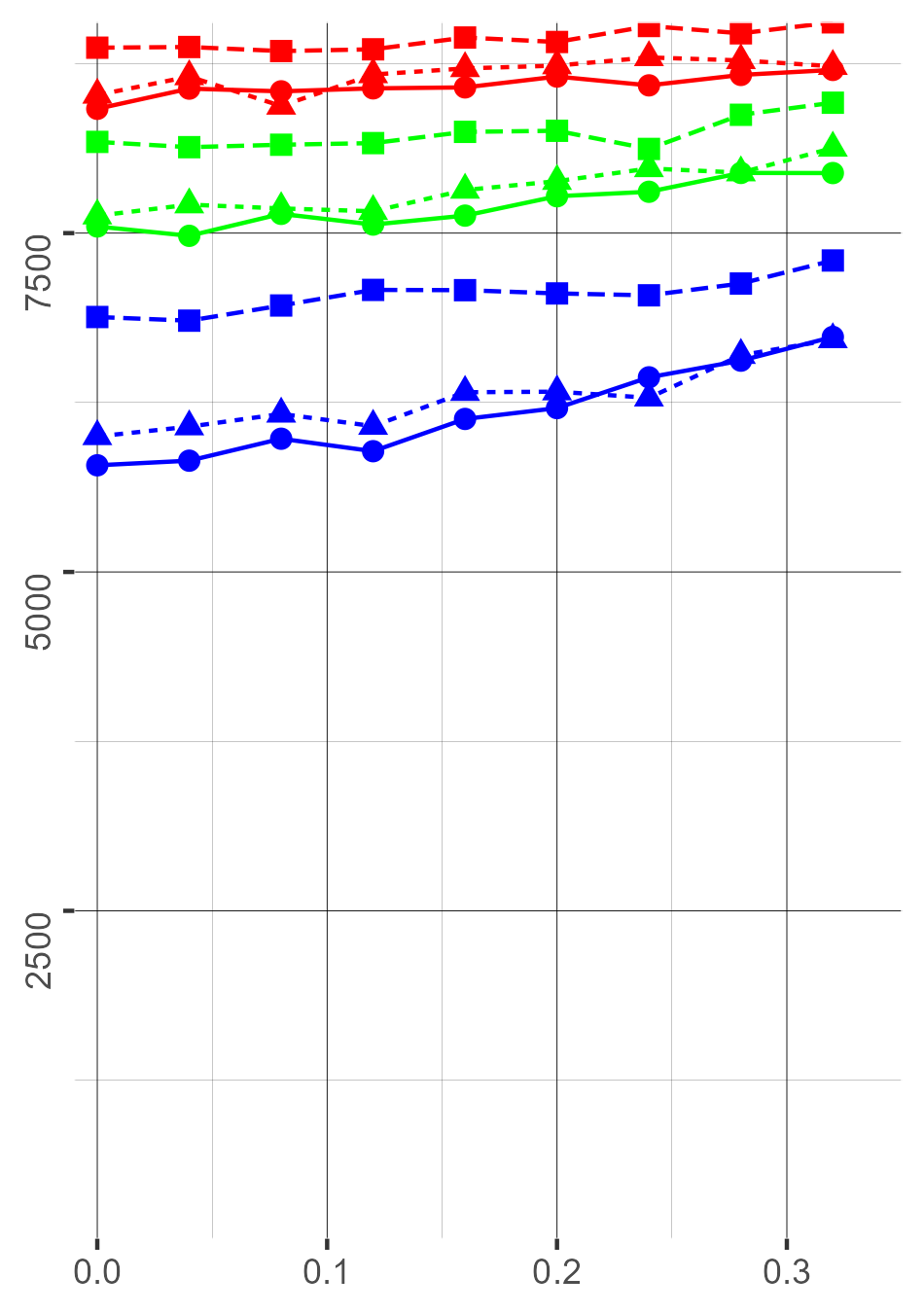}
&
\includegraphics[scale=.3]{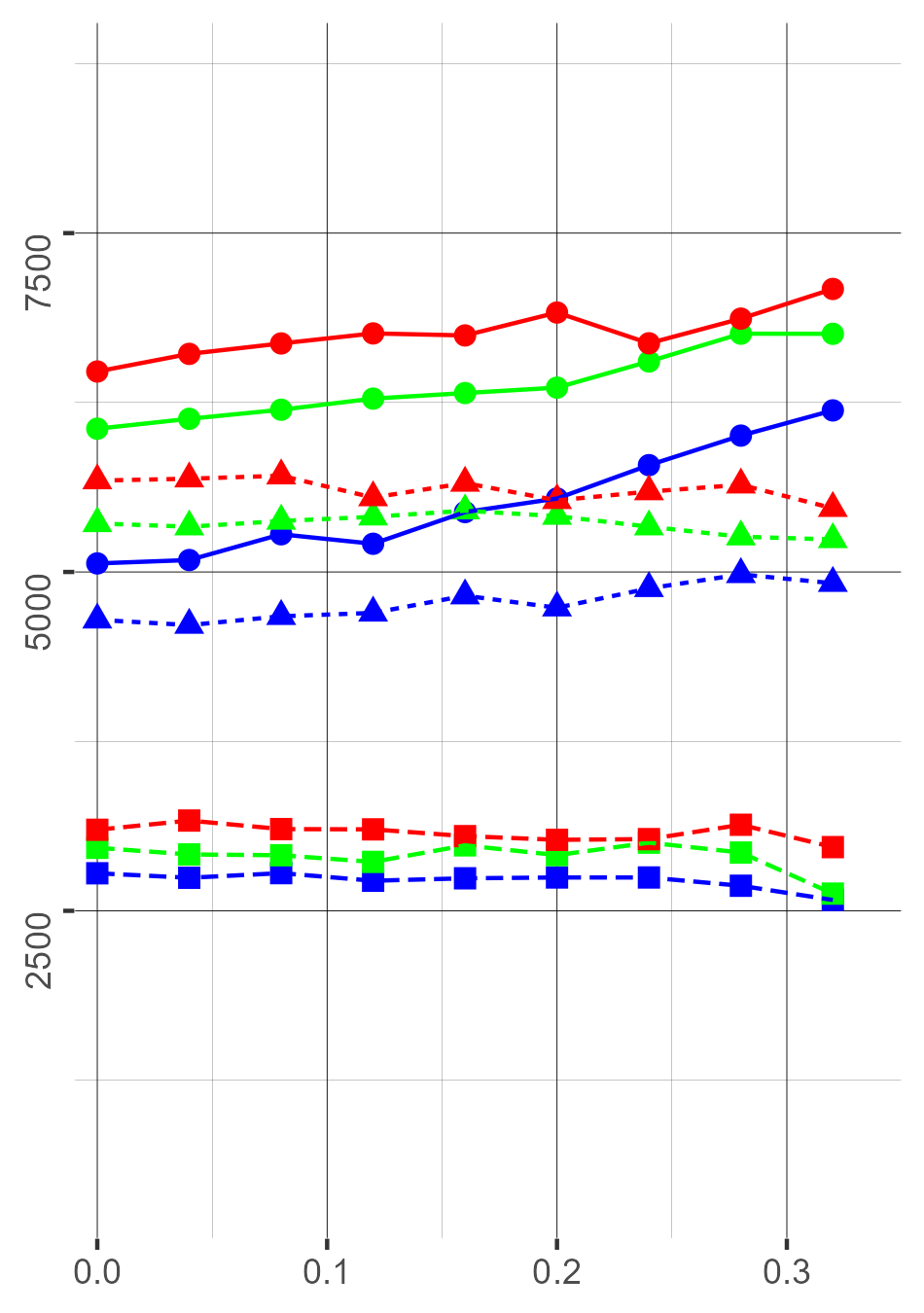}
&
\scalebox{.7}{
\begin{tikzpicture}
\node[align=center] (leg1) at (0,0) {\textcolor{blue}{$\blacksquare$} {\footnotesize Small delays 
 \textcolor{blue}{\faClockO}}}; 
\node[below=.25cm of leg1.south west,anchor=west,align=center] (leg2) {\textcolor{green}{$\blacksquare$} {\footnotesize Medium delays 
 \textcolor{green}{\faClockO}}};
\node[below=.25cm of leg2.south west,anchor=west,align=center] (leg3) {\textcolor{red}{$\blacksquare$} {\footnotesize Large delays 
 \textcolor{red}{\faClockO}}};
\node[below right=.3cm and -2.5cm of leg3] (leg4) {\footnotesize $\%$ of infected peers :};
\node[below right=0cm and -1.5cm of leg4] (leg5) {\footnotesize $0\%$};
\node[draw=black,fill=black,left=.2cm of leg5,circle,inner sep=2pt] (leg5s) {};
\draw[thick] ($(leg5s) + (-.3,0) $) -- ($ (leg5s) + (.3,0) $);
\node[below=.25cm of leg5.south west,anchor=west,align=center] (leg6) {\footnotesize $24\%$};
\node[left=.2cm of leg6,inner sep=0pt] (leg6s) {$\blacktriangle$};
\draw[thick,dotted] ($(leg6s) + (-.3,0) $) -- ($ (leg6s) + (.3,0) $);
\node[below=.25cm of leg6.south west,anchor=west,align=center] (leg7) {\footnotesize $48\%$};
\node[left=.2cm of leg7,inner sep=0pt] (leg7s) {$\blacksquare$};
\draw[thick,dashed] ($(leg7s) + (-.3,0) $) -- ($ (leg7s) + (.3,0) $);
\node[below=.25cm of leg7] {};
\end{tikzpicture}
}
&
\includegraphics[scale=.3]{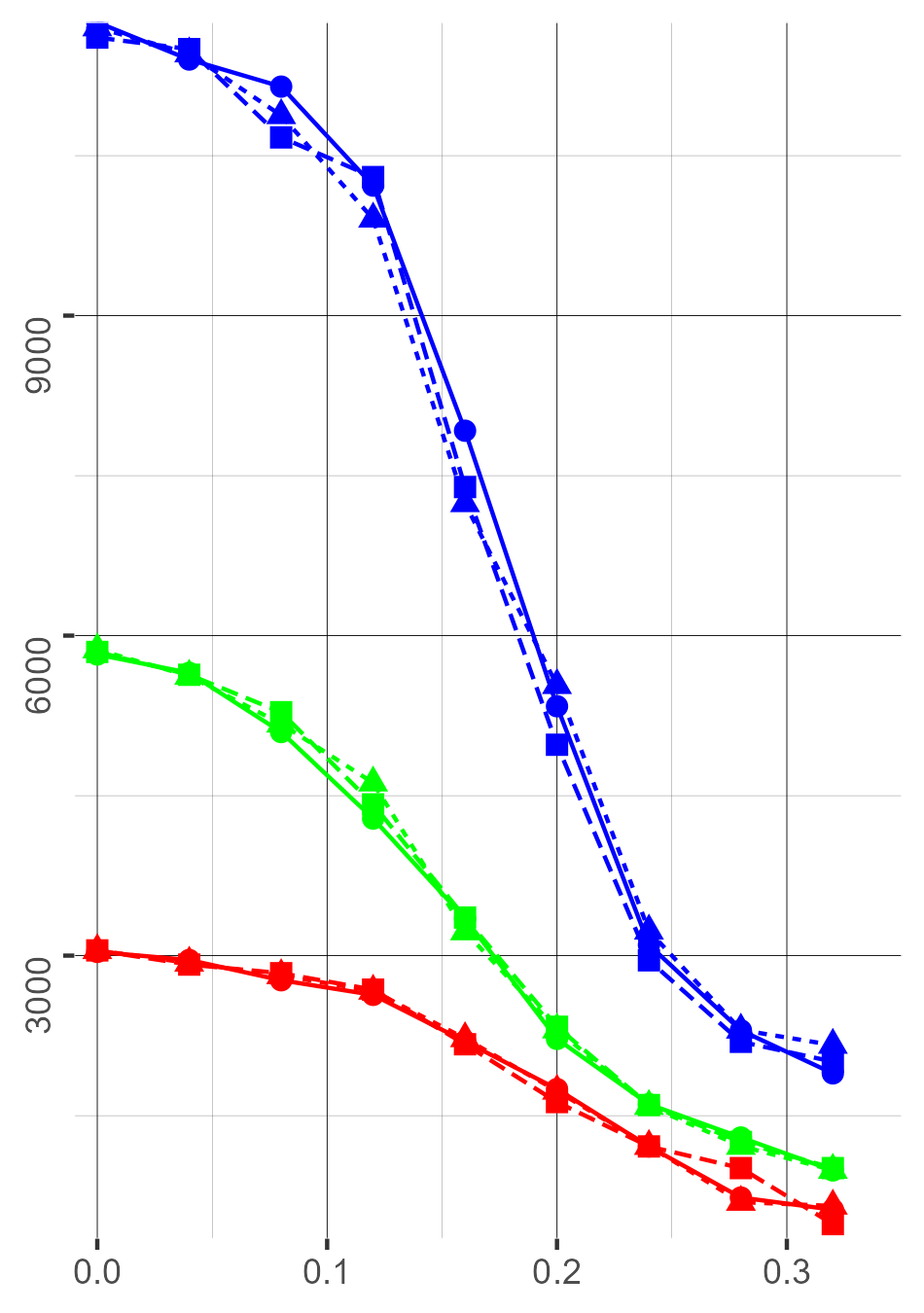}
\\
\hline 
\hline
{\scriptsize$OF_{S_{ND}}^{B_{LC}}$ violations}
&
{\scriptsize$OF_{E_{DS}}^{B_{LC}}$ violations}
&
{\scriptsize$OF_{O_{RD}}^{B_{LC}}$ violations}
&
{\scriptsize score}
&
{\scriptsize $3^{rd}$ quartile size of blocks}
\\
\hline 
\includegraphics[scale=.3]{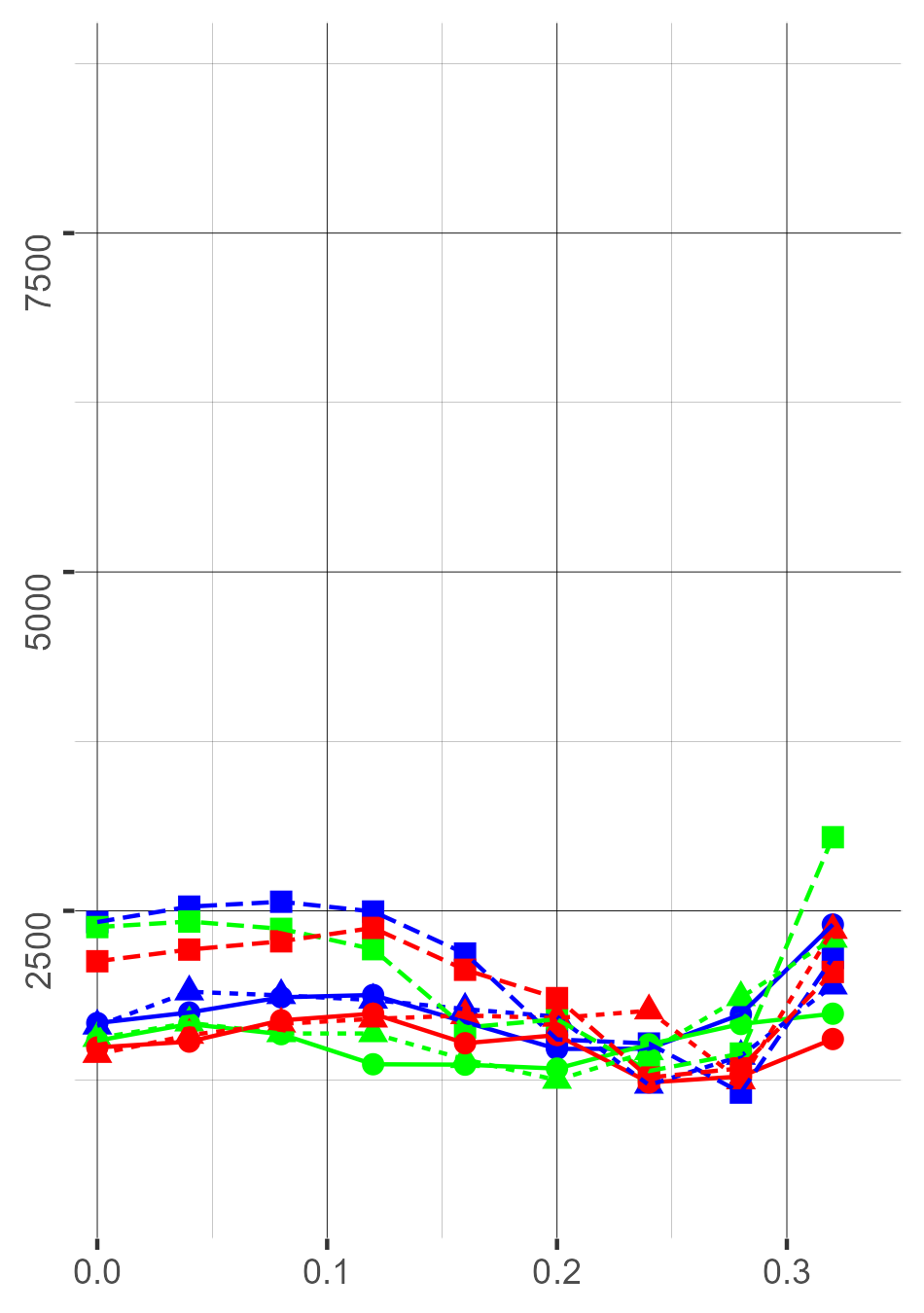}
&
\includegraphics[scale=.3]{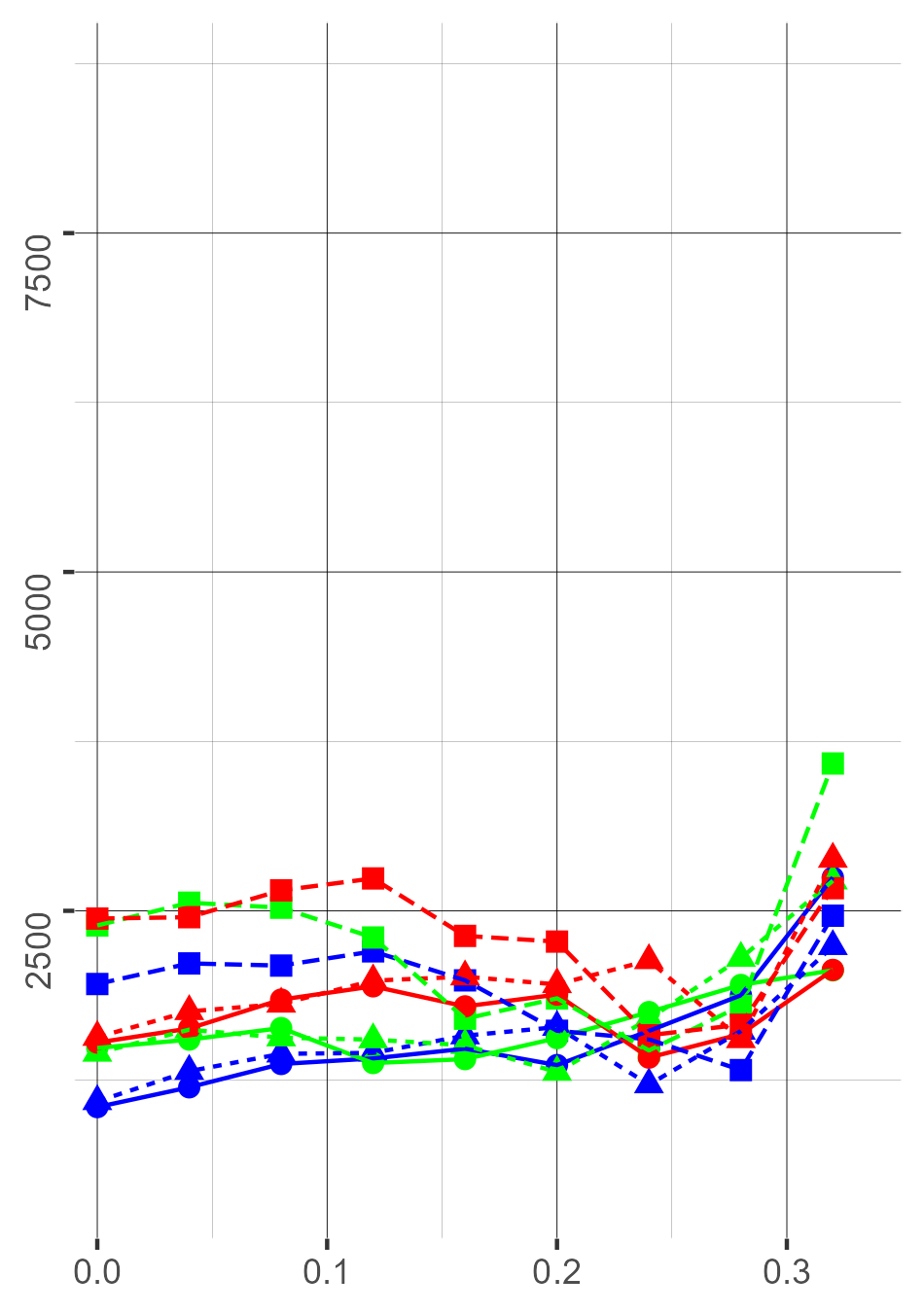}
&
\includegraphics[scale=.3]{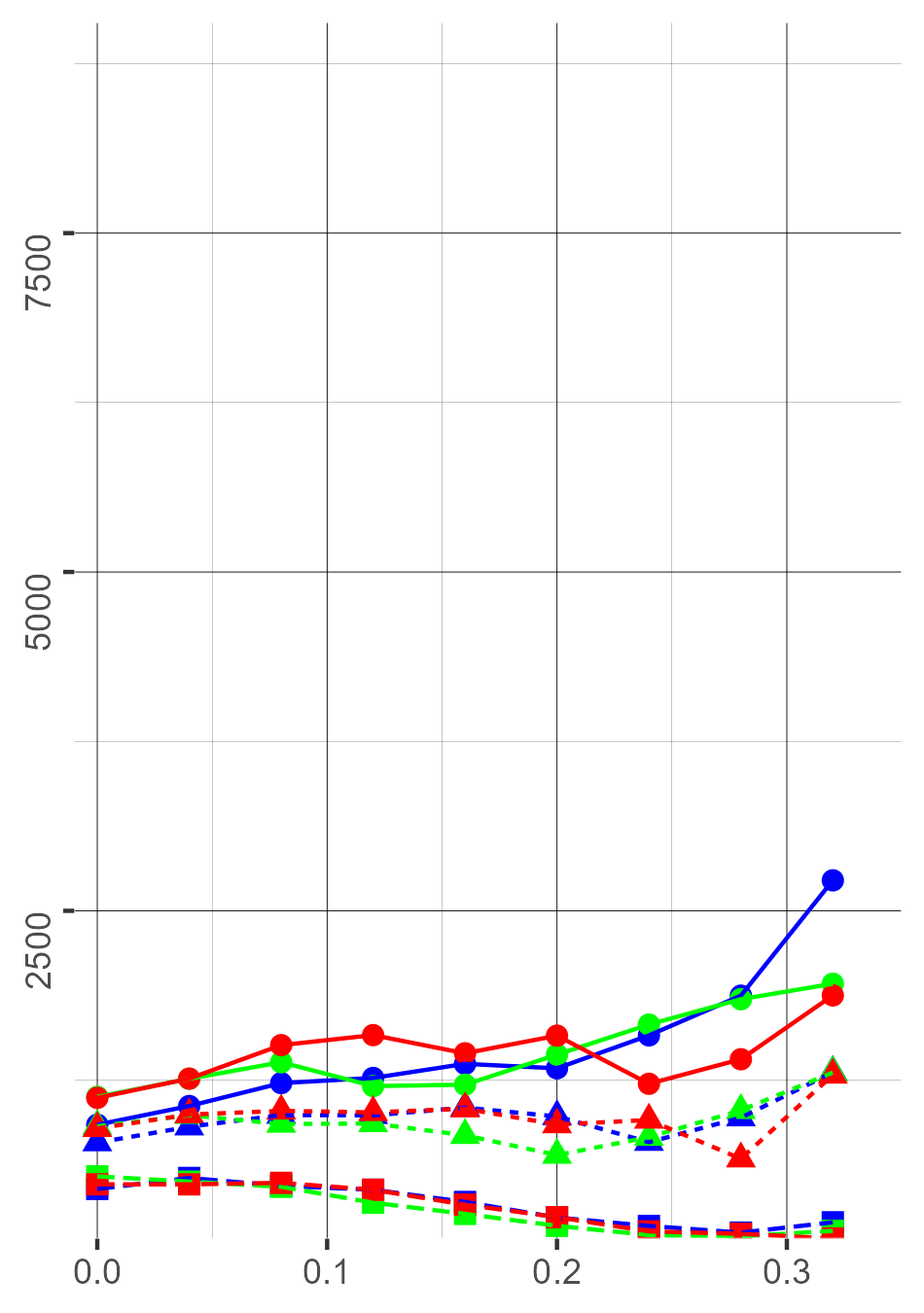}
&
\includegraphics[scale=.3]{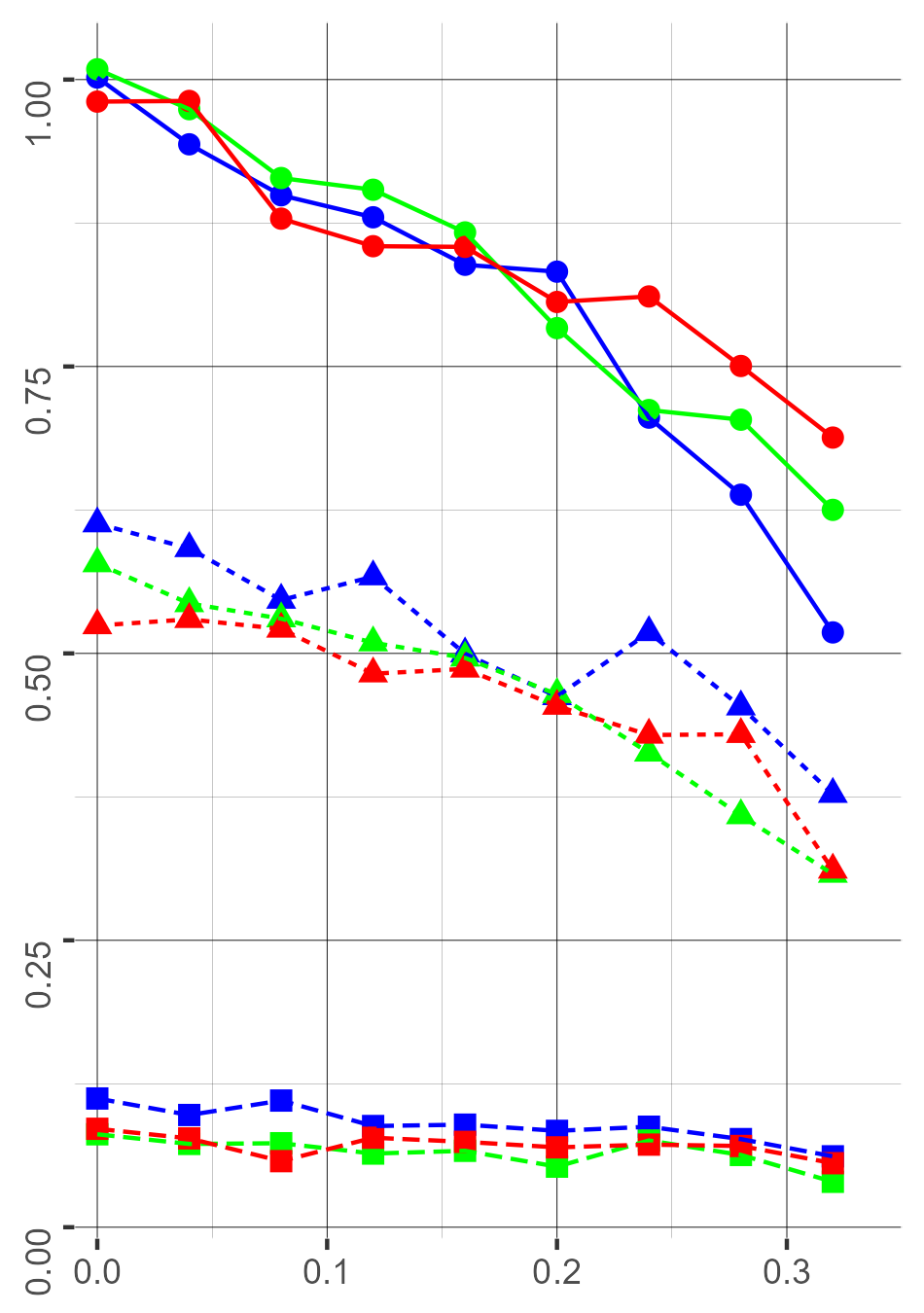}
&
\includegraphics[scale=.3]{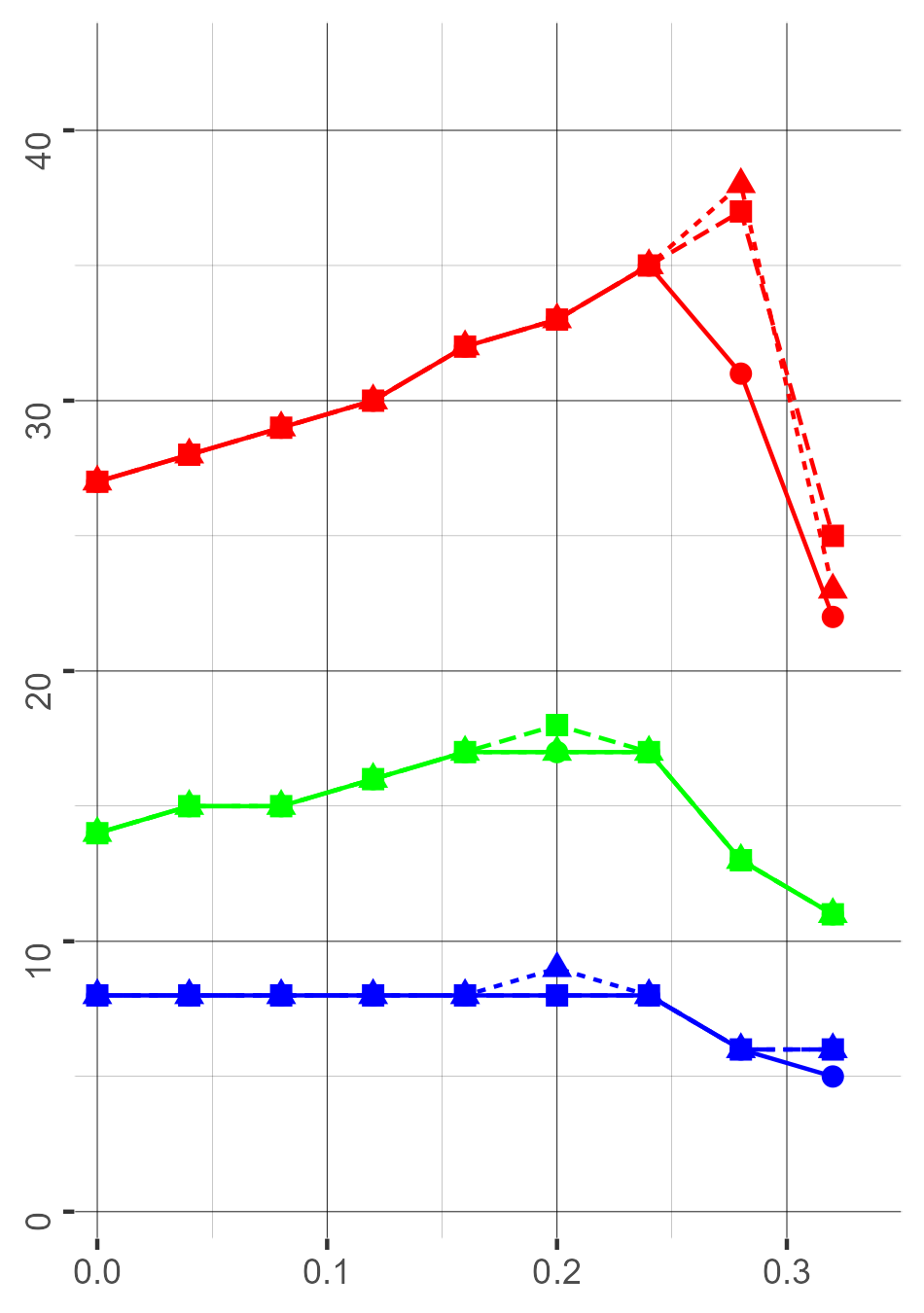}
\\
\hline 
\end{tabular}
\setlength\tabcolsep{6pt}
    
    \caption{Varying the number of infected Orderers}
    \label{fig:exp2_ords}
\vspace*{-.5cm}
\end{figure*}

\paragraph{Discussion on orderer sabotage.}
Fig.\ref{fig:exp2_ords} reports on $81$ simulations (having 9 curves, and 9 data-points per curve) that focus on the effect of infecting orderers. The diagrams of Fig.\ref{fig:exp2_ords} are similar than those of Fig.\ref{fig:exp1_peers}, the main difference being that the horizontal axes rather correspond to the proportion of infected orderers.
Also, the shape of points and style of the lines correspond here to proportions of infected peers ($0\%$, $24\%$ and $48\%$), highlighting the combined effect of both attacks.

Staying below the Byzantine threshold, we observe a moderate but still significant impact the more orderers are infected.
When no peers are infected (plain curves on Fig.\ref{fig:exp2_ords}), the diminution of the $\mathtt{score}$ is directly correlated to the proportion of infected orderers (e.g., $30\%$ of infected orderers yields a diminution of the score of around $30\%$). This is expected as it corresponds to the likelihood that an infected orderer is chosen as proposer.
In contrast to peer sabotage, we observe that the violations of both $OF_{E_{DS}}^{D_{LV}}$ and $OF_{O_{RD}}^{D_{LV}}$ increase with the number of infected orderers.
Indeed, this attack tamper with the delivery order but has no effect on the order of reception of transactions, w.r.t.~neither the peers nor the orderers.
Therefore its effect can be observed on violations of receive-order fairness w.r.t.~both the peers and the orderers.

We also remark that unlike peer sabotage, which has no noticeable effect on the number and size of blocks, this is not the case for orderer sabotage.
When there are more infected orderers, the number of blocks diminishes non-linearly in a S-shape sigmoid (top right diagram of Fig.\ref{fig:exp2_ords}). Also, the likelihood that some blocks are quite big increases before decreasing again (bottom right diagrma of Fig.\ref{fig:exp2_ords}).
This can be explained as follows : when there is a small proportion of Byzantine orderers, the mean time required for consensus to be reached increases as these orderers will not send PREVOTE and PRECOMMIT messages for blocks that contain transactions from the target client. Because a majority of nodes are honest, a majority of these blocks are likely to contain such transactions. As a result, the overall rate at which blocks are delivered diminishes. In turn, this lead to bigger blocks as transactions accumulate in each orderer's local mempool. In turn, these bigger blocks are more likely to contain transactions from the target client (simply because they contain more transactions, collected during a larger span of time). Thus the proportion of blocks which agreement are slowed by the Byzantine nodes increases, which further increases the overall diminution in the rate at which blocks are delivered.
This explains the sharp decrease part of the non-linear effect (center of the sigmoid).
As the proportion of infected orderers further increases, this increase in the likelihood that new block proposals contain transactions from the target client is counterbalanced by the fact that proposers are more likely to be Byzantine themselves and therefore that their proposal do not include such transactions.
This then explains the last part of the non-linear effect (end of the sigmoid).
This discussion also explains what is observed w.r.t.~ the $3^{rd}$ quartile in the size of blocks.
Indeed, when most block proposals include transactions from the target client, the rate at which blocks are delivered diminishes, which results in bigger blocks, more transactions accumulating in the orderers' mempools. Once the proportion of infected orderers is past a certain point, the likelihood that block proposals contain transactions from the target client diminishes again, leading to smaller blocks.

Now that we have understood the side effect orderer sabotage has on the number and size of blocks, we can take a look at the number of $OF_{S_{ND}}^{B_{LC}}$, $OF_{E_{DS}}^{B_{LC}}$ and $OF_{O_{RD}}^{B_{LC}}$ violations.
Byzantine orderers do not include transactions from the target client in their block proposals.
This trivially causes an increase in the number of $OF_{S_{ND}}^{B_{LC}}$, $OF_{E_{DS}}^{B_{LC}}$ and $OF_{O_{RD}}^{B_{LC}}$ violations. On Fig.\ref{fig:exp2_ords}, this is particularly noticeable for $OF_{O_{RD}}^{B_{LC}}$ (see plain curves, with no infected peers).
However, the aforementioned side effect on the size of blocks creates an inflection point in the curves (this is especially visible for $OF_{S_{ND}}^{B_{LC}}$).
As the proportion of infected orderers increases, the average size of the blocks increases before decreasing again.
With bigger blocks, the likelihood of violating OF defined w.r.t.~the partial order in which transactions are included into blocks decreases.
This explains why both $OF_{S_{ND}}^{B_{LC}}$ and $OF_{E_{DS}}^{B_{LC}}$ decrease before increasing again.

%% file: 6_4.tex
\paragraph{Combining peer and orderer sabotage.}
On both Fig.\ref{fig:exp1_peers} and Fig.\ref{fig:exp2_ords}, we have information related to the combined effect of peer and orderer sabotage in the form of the dotted and dashed curves.
We observe that the more there are infected peers and orderers, the more the $\mathtt{score}$ decreases.
However, there are diminishing returns between the number of infected peers and orderers.
Indeed, on Fig.\ref{fig:exp1_peers}, when the proportion of infected peers is low, the infection of orderers has a significant effect on the score.
However, closer to $50\%$ of infected peers, we can see that there is few to no advantage in infecting additional orderers below the BFT threshold.
Yet, overall, it seems that infecting peers is more efficient than infecting orderers, especially when network delays are large.
This is particularly visible on the $\mathtt{score}$ diagram of Fig.\ref{fig:exp2_ords}.

%% file: 7_0_mitig.tex
\section{A mitigation mechanism\label{sec:mitig_def}}

\input{7_1}

\input{7_2}

\input{7_3}

%% file: 7_1.tex
The experiments presented in Sec.\ref{sec:sim_attacks} suggest that HF is particularly vulnerable to transaction reordering attacks.
In this section, we propose a simple mitigation mechanism to make HF more robust to such attacks.
In this context, improving robustness amounts to modifying the protocol so as to reduce the ability of the adversary to increase the number of OF violations (for our specific use-case, this would also limit the ability of the adversary to reduce the $\mathtt{score}$).

In the following, we focus on reducing the number of $OF_{E_{DS}}^{D_{LV}}$ violations.
Indeed, it is not actually possible to consider a total order of send events across distant clients (as we have seen in Sec.\ref{sec:prelim}, upholding send-order fairness remains an open problem \cite{order_fairness_for_byzantine_consensus}) and reducing $OF_{E_{DS}}^{D_{LV}}$ violations would be more efficient than reducing $OF_{O_{RD}}^{D_{LV}}$ (as the reception orders on peers are likely closer to the send order than the reception orders on the orderers are).

Let us recall that receive-order fairness (which is the strongest OF defined w.r.t. receptions) is impossible to uphold \cite{order_fairness_for_byzantine_consensus}.
As a result, our mitigation mechanism may, at most, reduce the number of violations (it cannot prevent all pairs of transactions to violate $OF_{E_{DS}}^{D_{LV}}$, as HF must output a total order on transactions).

Let us also recall that batch-order fairness \cite{themis_fast_strong_order_fairness_in_byzantine_consensus} (block-order fairness in \cite{order_fairness_for_byzantine_consensus}), ignores part of the problem by excluding pairs of transactions that are in the same Condorcet cycle. 
Because we are considering $OF_{E_{DS}}^{D_{LV}}$, and because a total order must be agreed upon, mechanisms adapted from algorithms that uphold a form of batch-order fairness (see \cite{order_fairness_for_byzantine_consensus,themis_fast_strong_order_fairness_in_byzantine_consensus,quick_order_fairness}) may not be adapted.

Our goal of ``reducing the number of $OF_{E_{DS}}^{D_{LV}}$ violations'' is rather related to the notion of ``bounded unfairness'' from \cite{ordering_transactions_with_bounded_unfairness}.
However, we have seen in Sec.\ref{ssec:related_works_order_fairness} that finding a total order that minimizes the unfairness amounts to a NP-Hard problem.
Practical solutions to that problem can only be approximate heuristics.
Our mitigation mechanism is one such practical solution.

%% file: 7_2.tex
\subsection{Description of the mechanism}

Let us recall that, as described on Fig.\ref{fig:hf_tm_sd}, the endorsement of a transaction corresponds to collecting a quorum of signatures (denotes as $\{s_{k_1},\cdots,s_{k_j}\}$ on Fig.\ref{fig:hf_tm_sd}) from the peers.
Upon receiving a sufficient number of signatures, a client may forward its transaction to the orderers.
Concurrently, the orderers participate in repeated instances of consensus.
Upon receiving an endorsed transactions, an orderer puts it in its local mempool.
As a result, transactions can be progressively delivered.

\begin{figure}[h!]
    \centering
    \includegraphics[width=.45\textwidth]{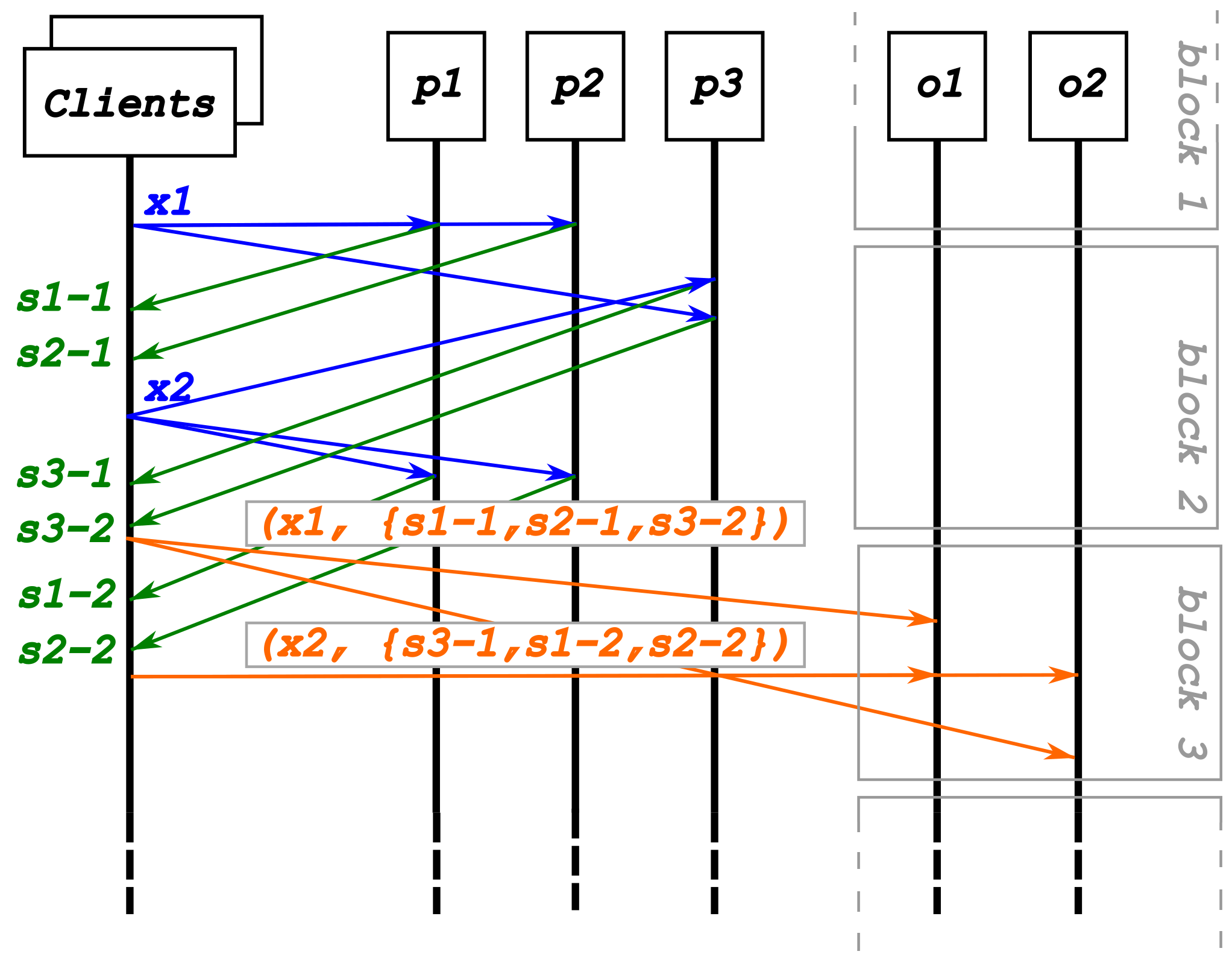}
    \caption{Example scenario}
    \label{fig:scenario_hf_rw}
\end{figure}

Fig.\ref{fig:scenario_hf_rw} describes a simple scenario in which we focus on the lifecycle of two transactions $x_1$ and $x_2$. In this simplified system we consider that there are three peers and that the endorsement policy is such that all of them must endorse a transaction for it to be ordered.
In our scenario, $x_1$ is sent before $x_2$. However, while $p_1$ and $p_2$ do receive $x_1$ before $x_2$, $p_3$ receives $x_2$ before $x_1$.
Once the corresponding has received the 3 endorsements (represented by the signatures $s_{1\text{-}1}$, $s_{2\text{-}1}$ and $s_{3\text{-}2}$) for $x_1$, it forwards it to the orderer.
The same thing occurs for $x_2$.
There are a number of orderers that participate in repeated instances of (Tendermint) consensus. On Fig.\ref{fig:scenario_hf_rw} only two of them : $o_1$ and $o_2$ are represented.
The Tendermint protocol is continuously executed concurrently w.r.t.~the endorsing process.
On the right of Fig.\ref{fig:scenario_hf_rw}, the grey squares roughly represents the spans of time taken to resolve an instance of consensus.
On Fig.\ref{fig:scenario_hf_rw}, we can see that both $o_1$ and $o_2$ (and we suppose this is the case for all the other orderers) receive the endorsed $x_1$ and $x_2$ during the vote of block 3. Supposing the vote is resolved at round 0, neither $x_1$ not $x_2$ will be included in block 3.
Later on, when the consensus instance for block 4 starts, the proposer selected for the first round of that instance may then include $x_1$ and $x_2$ in its proposal for block 4.

Let us then suppose that the proposer for the first round of the consensus instance for block 4 is either $o_1$ or $o_2$. Let us also suppose that both of them are honest and, in particular, that they have no interest in manipulating the relative delivery order of $x_1$ and $x_2$.
The default mempool implementation in Tendermint is a thread-safe FIFO queue\footnote{see {\tiny\url{https://github.com/tendermint/tendermint/discussions/6295}}}.
As a result, if $o_1$ is the proposer, it will propose a block in which $x_1$ is ordered before $x_2$, having received the endorsed $x_1$ before $x_2$.
However, if $o_2$ is the proposer, its block will order $x_2$ first (for the same reason).
Thus, if $o_2$ is the proposer, $OF_{E_{DS}}^{D_{LV}}$ will be violated for the pair $(x_1,x_2)$, a majority of
peers having received $x_1$ before $x_2$.

Our mitigation mechanism prevents such scenarios from occurring without any communication overhead and a slight computation overhead.
We propose that each peer maintains a local counter that keeps track of the total number of transactions it has endorsed. 
When endorsing a transaction, a peer piggybacks the current value of its local counter in its endorsement message (the signature authenticating this value) and increments the counter.
Then, when the client has received sufficiently many endorsements, the set of signatures ($\{s_{k_1},\cdots,s_{k_j}\}$ on Fig.\ref{fig:hf_tm_sd}) that it sends to the orderers includes information about the local orders in which individual peers endorsed it.

As a result, whenever an orderer is selected to propose a block and has to propose a new value for that block, it may use this information as a guide in order to determine the order of the transactions within that block.
In our case, let us consider that $o_2$ is the proposer and has exactly two transactions : $x_1$ and $x_2$ in its mempool at the moment it is selected to propose a block.
$o_1$ knows that $x_1$ has $\{s_{1\text{-}1},s_{2\text{-}1},s_{3\text{-}2}\}$ as set of endorsements and also knows that $s_{1\text{-}1}$ (resp.~$,s_{2\text{-}1}$ and $s_{3\text{-}2}$) implies that $x_1$ is the first (resp.~first and second) transaction endorsed by $p_1$ (resp.~$p_2$ and $p_3$).
Similarly, $o_2$ infers that $x_2$ is the second (resp.~second and first) transaction endorsed by $p_1$ (resp.~$p_2$ and $p_3$).
As a result, the orderer can infer a set of $3$ ballots, one for each peer, that indicate their ranked preferences as for the ordering of the transactions.

\begin{algorithm}
\caption{Extracting ballots from order metadata}\label{alg:extract_ballots}
\KwCustomType{$Key$ \Comment*[r]{\scriptsize public key}}
\KwCustomType{$Tx$ \Comment*[r]{\scriptsize endorsed transaction}}
\KwCustomInput{$peers : Set\langle Key\rangle $\Comment*[r]{\scriptsize peers' keys set}} 
\KwCustomInput{$pool : Set\langle Tx\rangle $\Comment*[r]{\scriptsize orderer's local mempool}} 
$localOrders : Map\langle Key,Map\langle Int,Tx\rangle \rangle  \gets \{\}$\;
\For{$x \in pool$}{
    \For{$s \in sigs(x)$}{
        $localOrders.addOrInsert(s.key,s.idx,x)$\;
    }
}
$ballots : Set\langle List\langle Tx\rangle \rangle  \gets \{\}$\;
\For{$(p,loc) \in localOrders$}{
    $ballot \gets loc.values().sortByKeys()$\;
    $ballots.add(ballot)$\;
}
\Return{$ballots$};
\end{algorithm}

Alg.\ref{alg:extract_ballots} details the process of extracting ranked preferences in the ordering of transactions from the orderer's mempool for each peer.
These preferences are extracted from the order-related metadata that is piggybacked on the endorsements of each transaction.
This yields a set of ballots (one per peer) that may then be used to determine a total order between the transactions that are currently present in the orderers' mempool.
Each ballot consists of a list of transactions in the order in which the corresponding peer has endorsed them.
In our example from Fig.\ref{fig:scenario_hf_rw} the ballots for $p_1$ and $p_2$ are both $\lbrack x_1,x_2 \rbrack$ and the one for $p_3$ is $\lbrack x_2,x_1 \rbrack$.
However, let us remark that, depending on the endorsing policy, each transaction may only be endorsed by a fraction of the nodes.
As a result, ballots may be of different lengths. In our example, if we had a fourth peer $p_4$, its ballot would be the empty list as it endorsed neither $x_1$ nor $x_2$.

In any case, this set of ballots may then be used by the orderer to compute a total order.
A wide variety of voting algorithms may be applied to that end.
Our only two requirements for choosing one such algorithm is that it is compatible with ranked ballots \cite{handbook_of_computational_social_choice} and that voting allows establishing an ordered list of winners.

%% file: 7_3.tex
\subsection{Selection of the voting algorithm}

Algorithms that satisfy our two requirements include the following three families: positional voting (e.g., Borda count, Dowdall system), Condorcet voting (e.g., Copeland's method, Tideman's Ranked Pairs, Kemeny-Young method) and runoff voting (e.g., Nanson's and Baldwin's methods) \cite{handbook_of_computational_social_choice}.

Positional voting would consist here in computing scores for each transaction based on their indices in each ballot and ordering them according to these scores.
Positional voting algorithms generally have a $\mathcal{O}(n*X)$ time complexity where $n$ is the number of peers (and therefore of ballots) and $X$ is the number of transactions (and therefore the size of the block).
These algorithms generally uphold monotonicity (increasing the rank of a transaction in some ballots should not cause a decrease in its overall ranking).


\input{9_fig_mitig}

Condorcet voting would consist here in reasoning on each transaction's pairwise wins and losses, a transaction winning against another if a majority of peers ranks it higher. 
Such algorithms have at least a $\mathcal{O}(n^2*X)$ time complexity.
They also uphold the Condorcet criterion, which, in our case, is particularly interesting to minimize $OF_{E_{DS}}^{D_{LV}}$ violations.

Runoff voting consists in determining a single winner or loser, removing it from the set of candidates and repeating the process until no candidate is left.
Some runoff voting algorithms satisfy the Condorcet criterion (e.g., Nanson's and Baldwin's methods which eliminate candidates below the average Borda score) and others do not (e.g., Bucklin, which reasons on the highest median ranking).
Their time complexity also varies.

There are many parameters one ought to consider when choosing a specific voting algorithm for the orderers.
Voting criteria such as the Condorcet criterion or the Majority criterion might be considered so as to reduce the probability to violate $OF_{E_{DS}}^{D_{LV}}$.
However, this is not the only parameter to consider.
Indeed, proposing a new block is in the critical path of the HF process.
Choosing an algorithm with a high time complexity could negatively impact latency.
Given a block $h$ of a certain size, determining the order of transactions within the block takes more time with one such algorithm.
Moreover as the consensus for block $h$ takes a long time, the next block proposal for block $h+1$ may likely contain numerous transactions (which have accumulated in the next proposer's mempool during the time it took to reach consensus for block $h$).
As a result, the consensus for block $h+1$ may take even more time and so on.

Another important aspect to consider is that of the manipulability of the algorithm \cite{handbook_of_computational_social_choice,complexities_of_and_algorithms_for_the_manipulation_of_borda_nanson_baldwin_voting_rules}.
Indeed, an adversary may attempt to change the outcome of the election (and therefore the order of transactions within the next block) by coordinating the peers and clients it controls so as to manipulate the content of some ballots.
For instance, the peers it controls could assign high (resp.~low) indices to transactions that the adversary wants ordered first (resp.~last).
As for the clients it controls, they could carefully select the subsets of required endorsements that are forwarded to the ordering service (thus manipulating the ballots more directly).
However, to do so, the adversary must, in a timely manner, \textbf{(1)} predicts which transactions are likely to be part of the next block (i.e., predict the set of candidates), \textbf{(2)} compute a ``manipulation'' that would yield an order it wants to favor (the term ``manipulation'' having a formal meaning in e.g., \cite{complexities_of_and_algorithms_for_the_manipulation_of_borda_nanson_baldwin_voting_rules}), \textbf{(3)} predict how to coordinate the peers and clients it controls so as to perform this manipulation of the ballots and \textbf{(4)} communicate with these peers and clients in a timely manner to coordinate them.
Fortunately, as discussed in \cite{complexities_of_and_algorithms_for_the_manipulation_of_borda_nanson_baldwin_voting_rules}, the problem of computing a ``manipulation'' is NP-Hard for a variety of positional and runoff voting algorithms (Borda, Nanson, Baldwin). 
As a result, it is highly unlikely that an adversary may succeed except if there is very few ballots (i.e., the endorsing service has only a few peers and/or the endorsing policy only requires a few endorsements) and candidates (the block proposal contains few transactions) or if it has complete control over the network under an asynchronous communication model.

In this paper, we do not recommend any specific voting algorithm. Still, simple positional voting algorithms such as Borda and Dowdall are interesting as they have a low time complexity and are difficult to manipulate in a timely manner (as per \cite{complexities_of_and_algorithms_for_the_manipulation_of_borda_nanson_baldwin_voting_rules}).
Let us also remark that, in another context, \cite{condorcet_attack_against_fair_transaction_ordering} uses Tideman's Ranked Pairs to order transactions within Condorcet cycles. However, the $\mathcal{O}(n^3)$ time complexity of Ranked Pairs may be problematic for latency.

%% file: 9_fig_mitig.tex
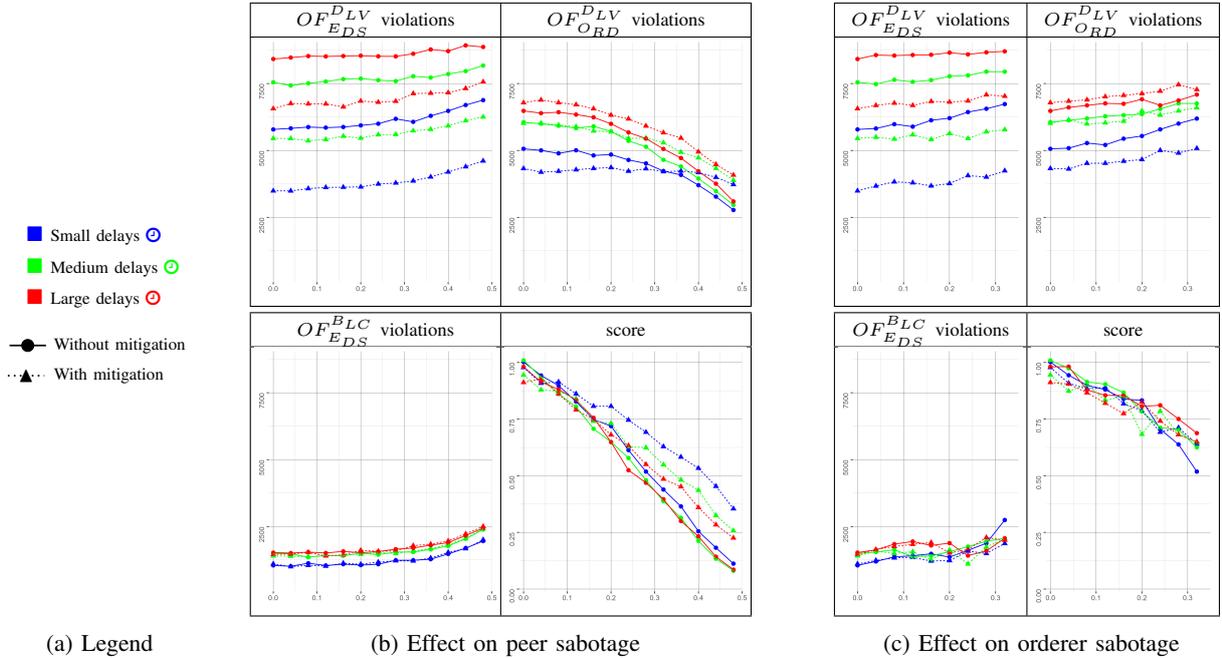
\begin{figure*}[h]
    \centering

\begin{subfigure}{.2\textwidth}
    \centering
    \scalebox{.8}{
\begin{tikzpicture}
\node[align=center] (leg1) at (0,0) {\textcolor{blue}{$\blacksquare$} {\footnotesize Small delays 
 \textcolor{blue}{\faClockO}}}; 
\node[below=.25cm of leg1.south west,anchor=west,align=center] (leg2) {\textcolor{green}{$\blacksquare$} {\footnotesize Medium delays 
 \textcolor{green}{\faClockO}}};
\node[below=.25cm of leg2.south west,anchor=west,align=center] (leg3) {\textcolor{red}{$\blacksquare$} {\footnotesize Large delays 
 \textcolor{red}{\faClockO}}};
\node[below right=.25cm and -2cm of leg3] (leg5) {\footnotesize Without mitigation};
\node[draw=black,fill=black,left=.2cm of leg5,circle,inner sep=2pt] (leg5s) {};
\draw[thick] ($(leg5s) + (-.3,0) $) -- ($ (leg5s) + (.3,0) $);
\node[below=.25cm of leg5.south west,anchor=west,align=center] (leg6) {\footnotesize With mitigation};
\node[left=.2cm of leg6,inner sep=0pt] (leg6s) {$\blacktriangle$};
\draw[thick,dotted] ($(leg6s) + (-.3,0) $) -- ($ (leg6s) + (.3,0) $);
\node[below=3.5cm of leg6] {};
\end{tikzpicture}
    }
    \caption{Legend}
    \label{fig:exp_mitig_legend}
\end{subfigure}
\begin{subfigure}{.375\textwidth}
    \centering
    \setlength\tabcolsep{1.5pt}
\begin{tabular}{|c|c|}
\hline
{\scriptsize$OF_{E_{DS}}^{D_{LV}}$ violations}
&
{\scriptsize$OF_{O_{RD}}^{D_{LV}}$ violations}
\\
\hline 
\includegraphics[scale=.3]{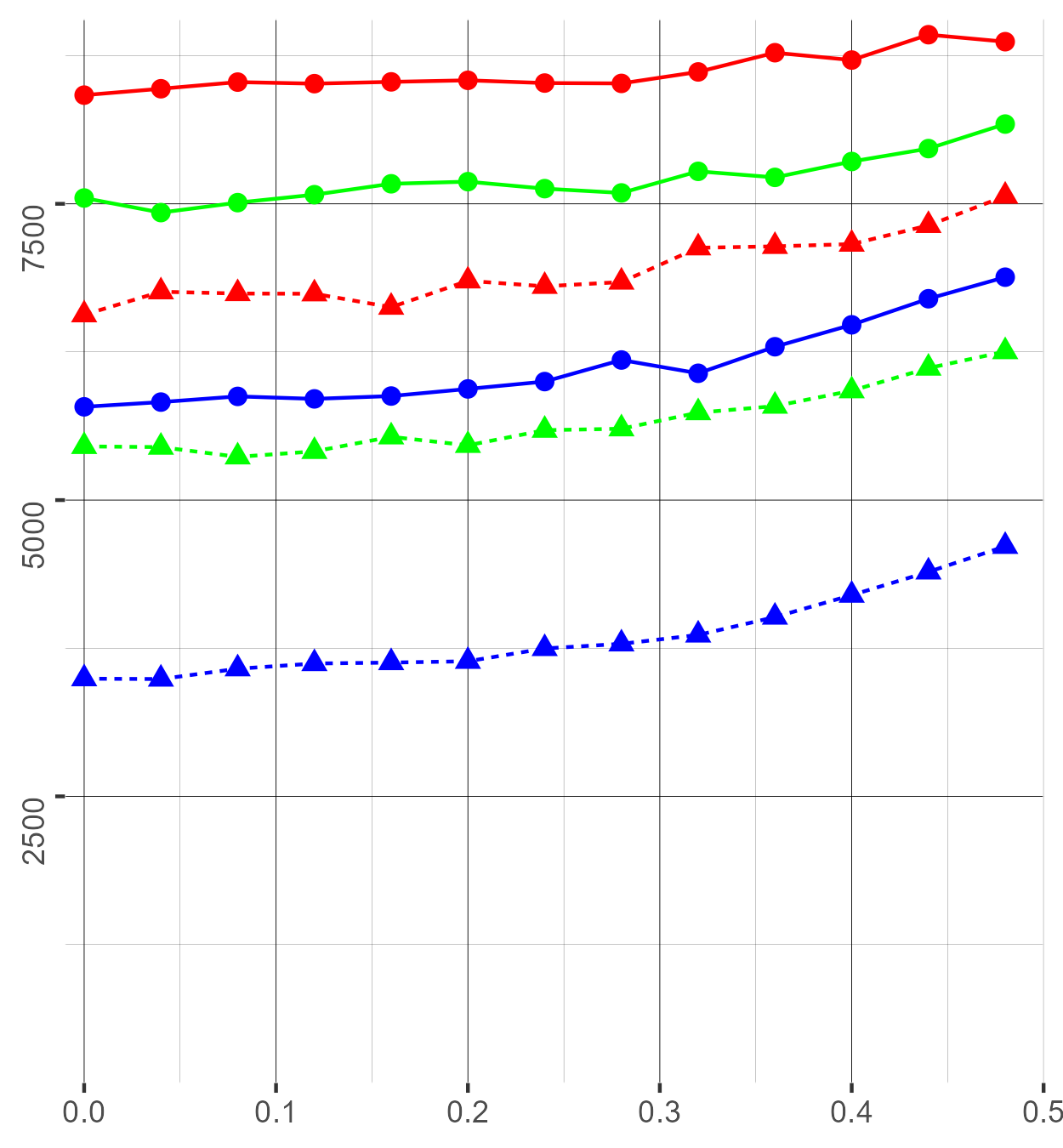}
&
\includegraphics[scale=.3]{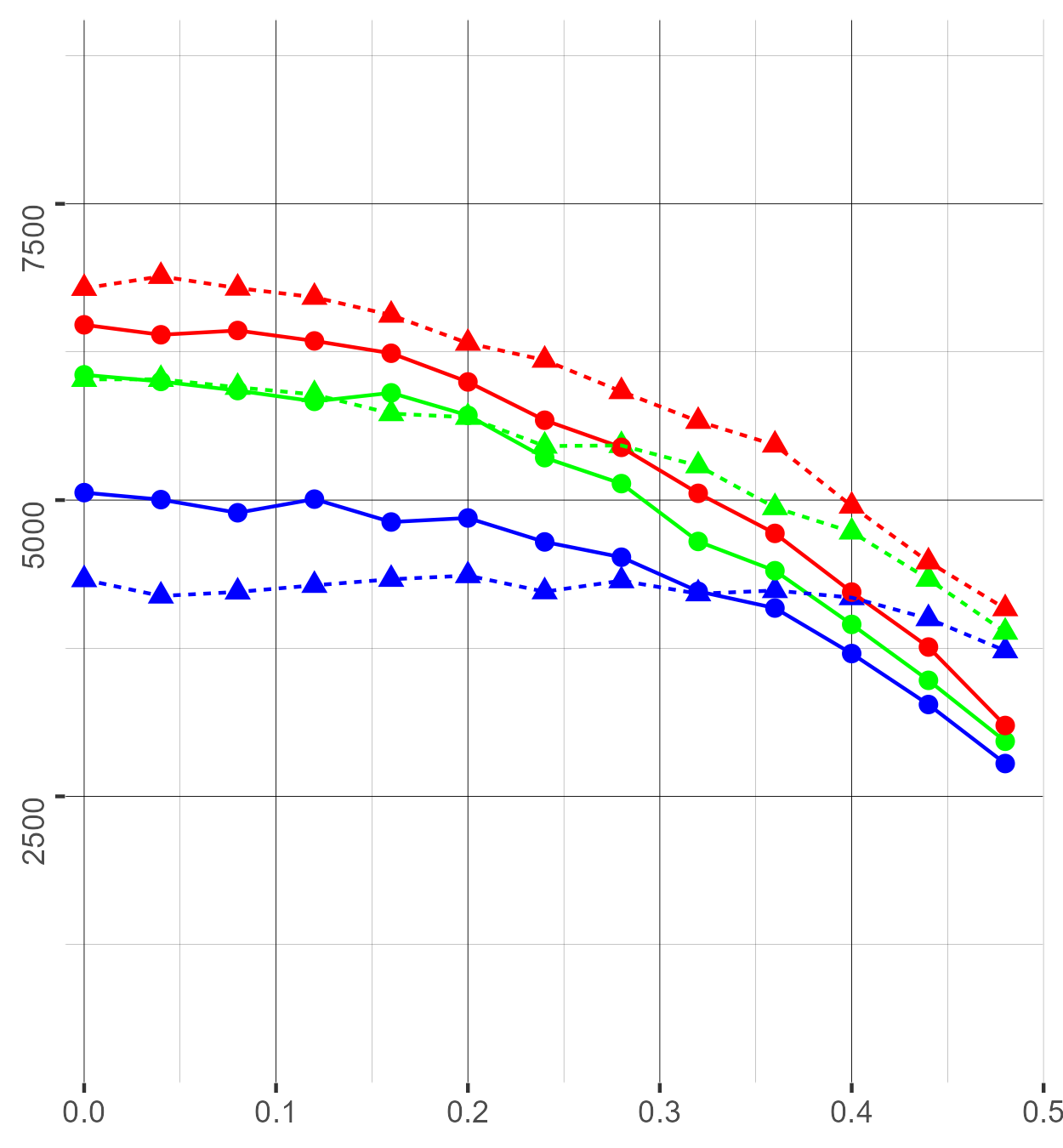}
\\
\hline 
\hline
{\scriptsize$OF_{E_{DS}}^{B_{LC}}$ violations}
&
{\scriptsize score}
\\
\hline 
\includegraphics[scale=.3]{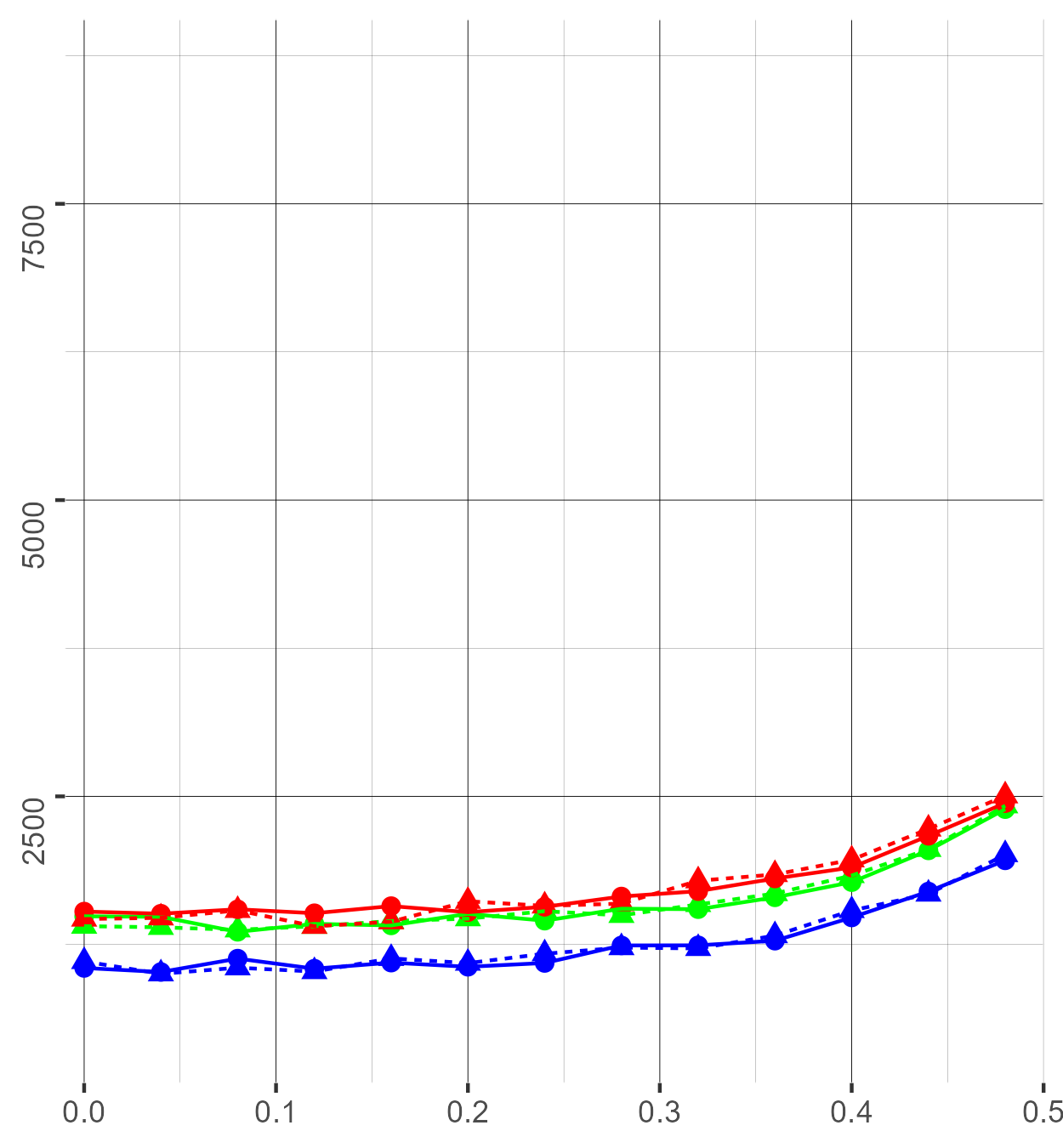}
&
\includegraphics[scale=.3]{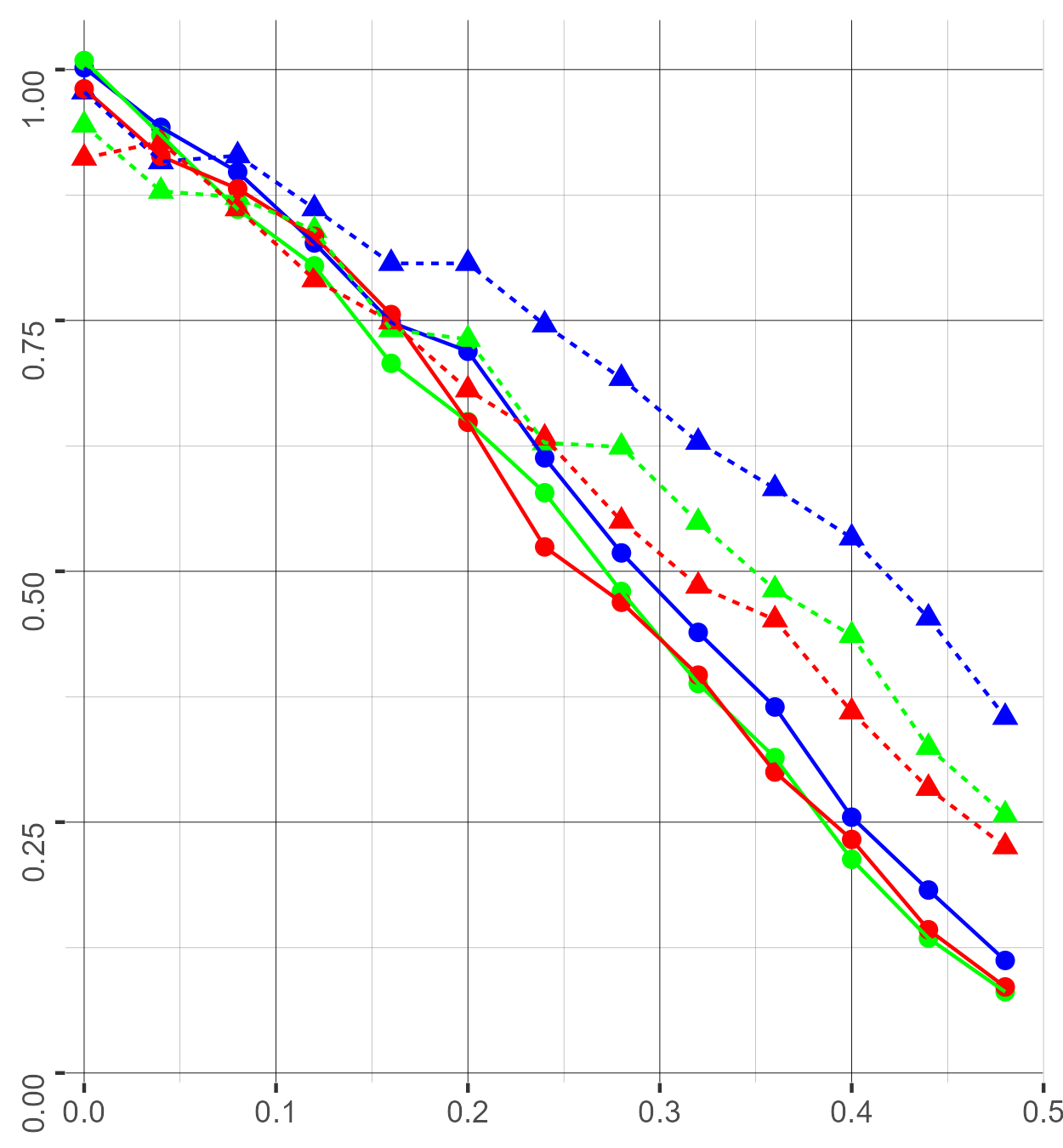}
\\
\hline 
\end{tabular}
\setlength\tabcolsep{6pt}
    \caption{Effect on peer sabotage}
    \label{fig:exp_mitig_peers}
\end{subfigure}
\begin{subfigure}{.375\textwidth}
    \centering

\setlength\tabcolsep{1.5pt}
\begin{tabular}{|c|c|}
\hline
{\scriptsize$OF_{E_{DS}}^{D_{LV}}$ violations}
&
{\scriptsize$OF_{O_{RD}}^{D_{LV}}$ violations}
\\
\hline 
\includegraphics[scale=.3]{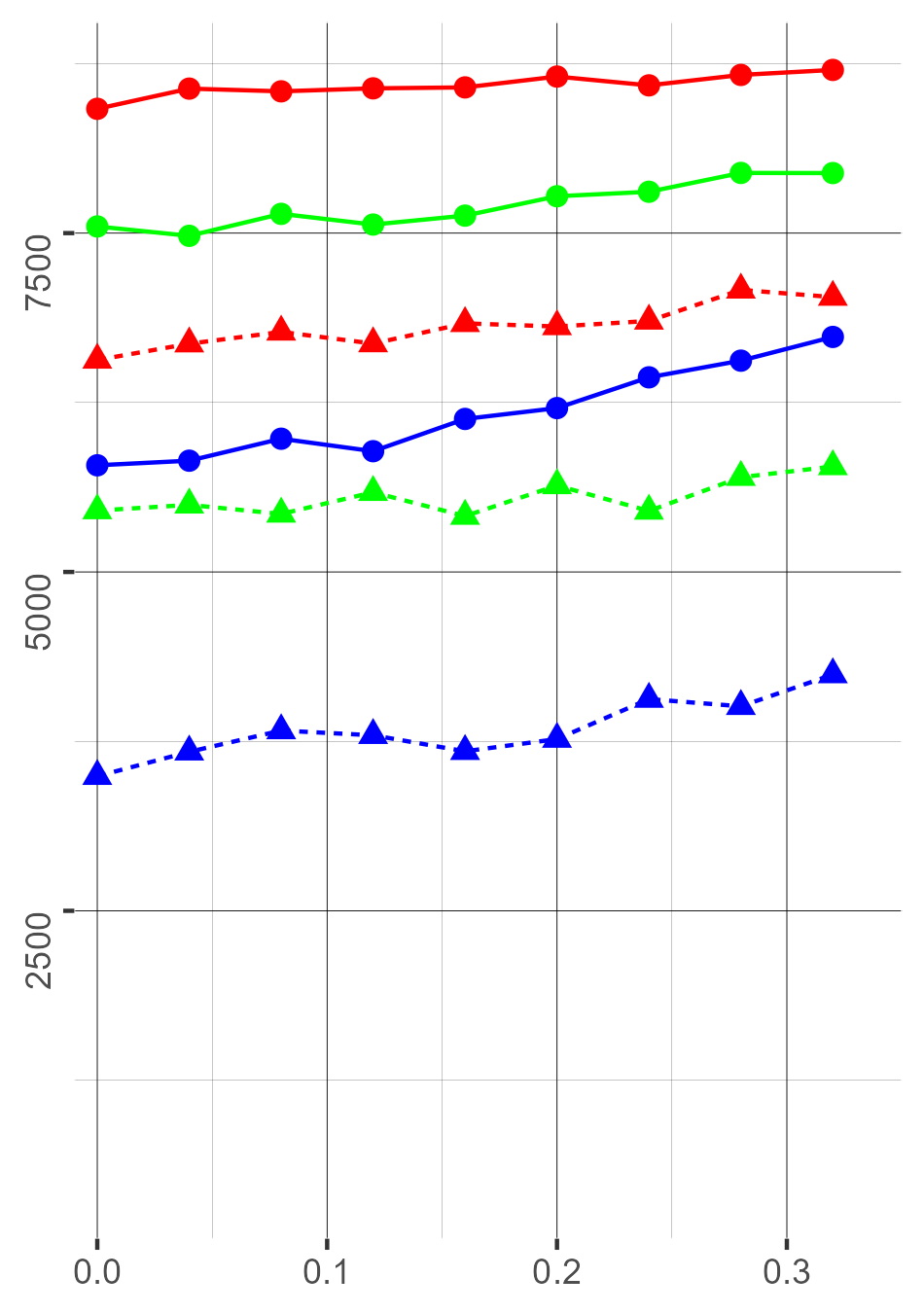}
&
\includegraphics[scale=.3]{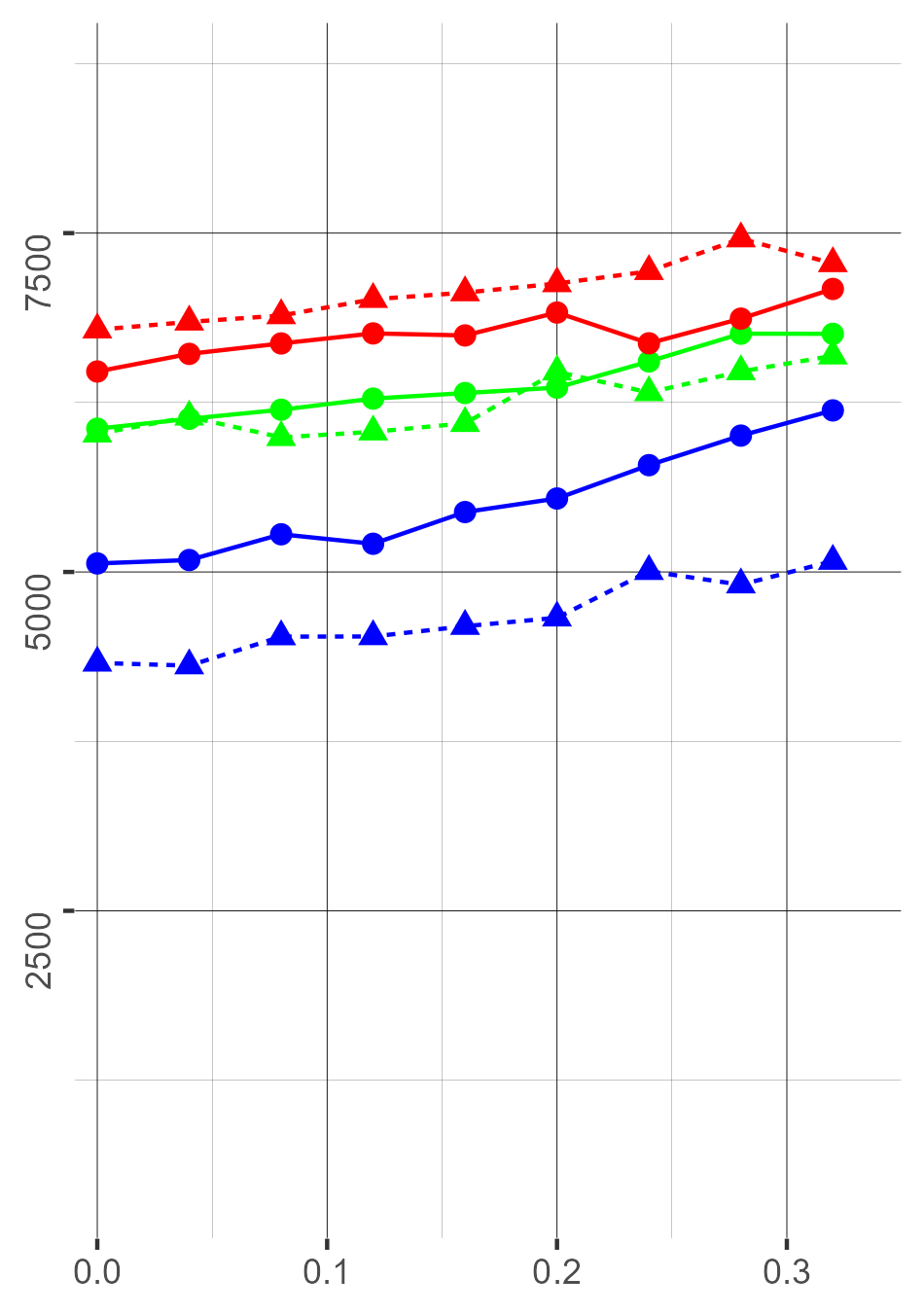}
\\
\hline 
\hline
{\scriptsize$OF_{E_{DS}}^{B_{LC}}$ violations}
&
{\scriptsize score}
\\
\hline 
\includegraphics[scale=.3]{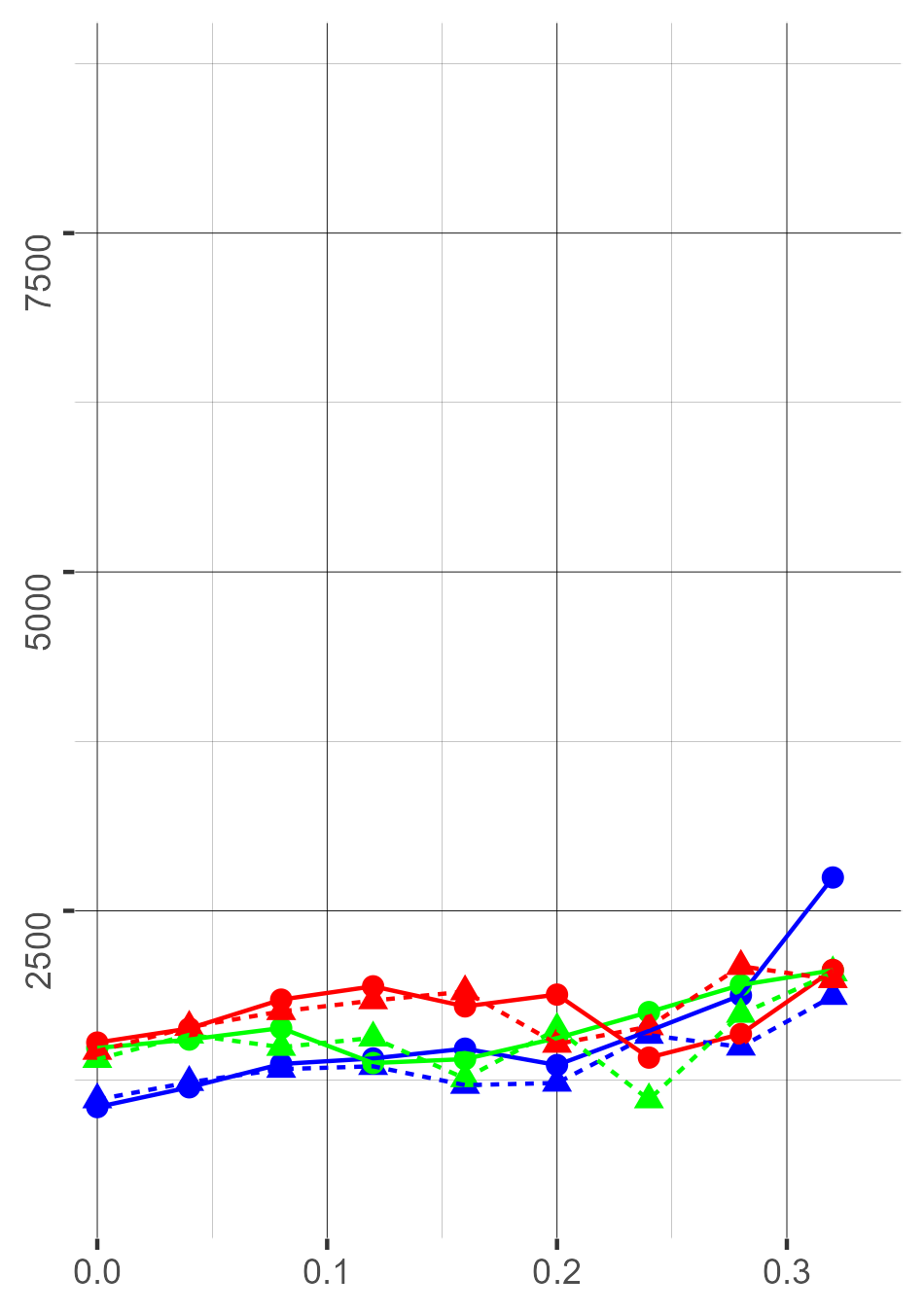}
&
\includegraphics[scale=.3]{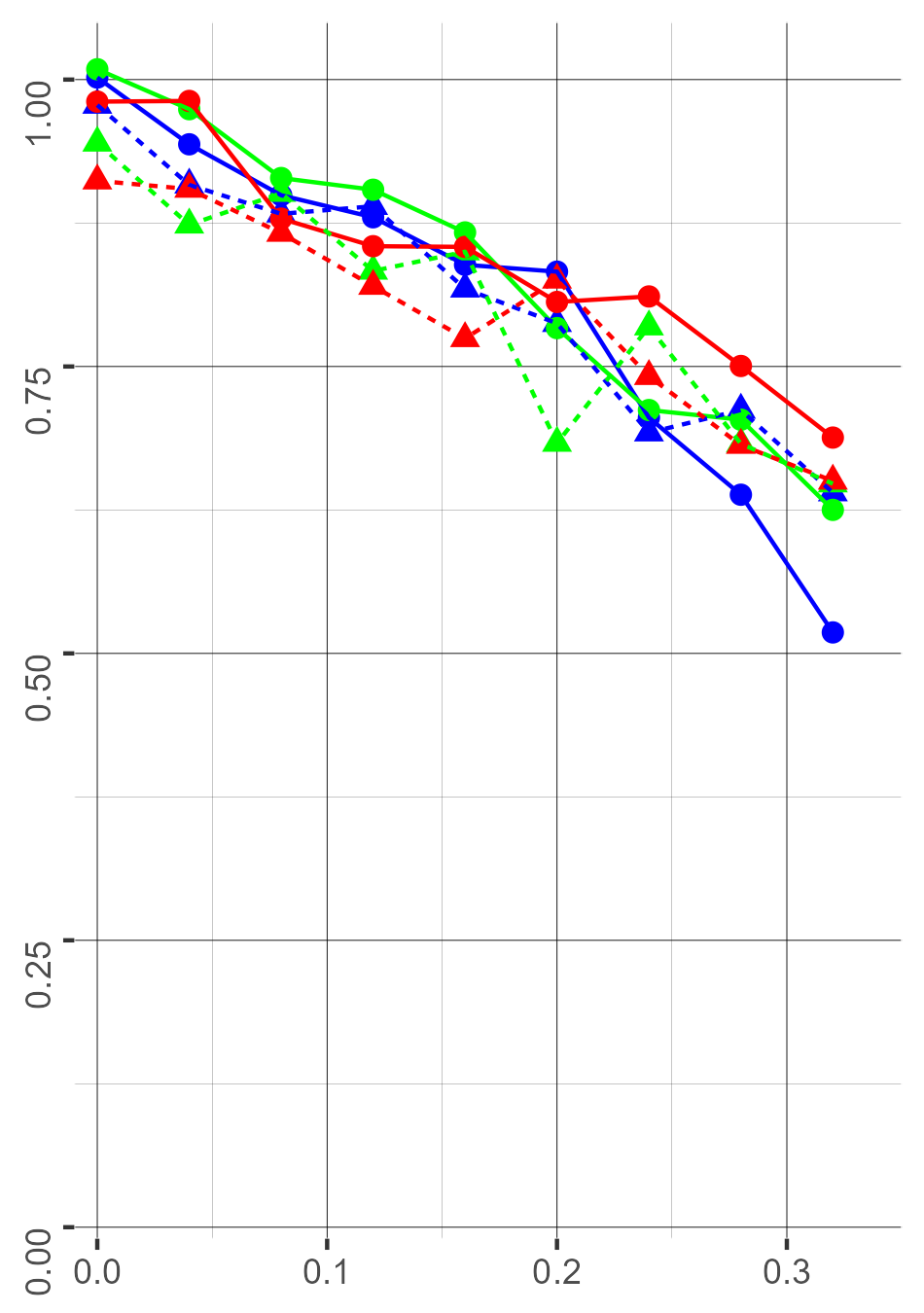}
\\
\hline 
\end{tabular}
\setlength\tabcolsep{6pt}
    
    \caption{Effect on orderer sabotage}
    \label{fig:exp_mitig_ords}
\end{subfigure}
    
    \caption{Effects of the mitigation mechanism}
    \label{fig:exp_mitig}
\vspace*{-.5cm}
\end{figure*}

%% file: 8_0_sims.tex
\section{Additional simulations\label{sec:sim_additional}}

\subsection{Effect of the mitigation mechanism\label{sec:sim_mitig}}

\input{9_fig_vary}

\input{8_1}

\subsection{Varying the numbers of sub-systems\label{sec:sim_params}}

\input{8_2}

\input{8_3}

%% file: 9_fig_vary.tex
\definecolor{varyNumClients3}{RGB}{255,110,180}
\definecolor{varyNumClients4}{RGB}{139,0,139}
\definecolor{varyNumClients5}{RGB}{132,112,255}

\definecolor{varyNumPeers25}{RGB}{0,139,0}
\definecolor{varyNumPeers35}{RGB}{127,255,0}
\definecolor{varyNumPeers45}{RGB}{238,201,0}

\definecolor{varyNumOrds25}{RGB}{107,142,35}
\definecolor{varyNumOrds31}{RGB}{205,170,125}
\definecolor{varyNumOrds37}{RGB}{255,69,0}

\begin{figure*}[h]
    \centering

    \setlength\tabcolsep{1.5pt}
\begin{tabular}{|c||c|c|c||c|}
\hline
{\scriptsize legend}
&
{\scriptsize$OF_{E_{DS}}^{D_{LV}}$ violations}
&
{\scriptsize$OF_{O_{RD}}^{D_{LV}}$ violations}
&
{\scriptsize score w.r.t.~peer sabotage}
&
{\scriptsize score w.r.t.~orderer sabotage}
\\
\hline 
\scalebox{.7}{
\begin{tikzpicture}
\node[align=center] (leg1) at (0,0) {\textcolor{varyNumClients3}{$\blacksquare$} {\footnotesize $m=3$}};
\node[above left =0cm and -1.5cm of leg1] (leg0) {\footnotesize \# of clients $m$:};
\node[below=.175cm of leg1.south west,anchor=west,align=center] (leg2) {\textcolor{varyNumClients4}{$\blacksquare$} {\footnotesize $m=4$}};
\node[below=.175cm of leg2.south west,anchor=west,align=center] (leg3) {\textcolor{varyNumClients5}{$\blacksquare$} {\footnotesize $m=5$}};
\node[below right=.25cm and -1.5cm of leg3] (leg5) {\footnotesize no mitigation};
\node[draw=black,fill=black,left=.2cm of leg5,circle,inner sep=2pt] (leg5s) {};
\node[below=.2cm of leg5.south west,anchor=west,align=center] (leg6) {\footnotesize mitigation};
\node[left=.2cm of leg6,inner sep=0pt] (leg6s) {$\blacktriangle$};
\node[below=.5cm of leg6.south west,anchor=west,align=center] (leg7) {\footnotesize Small Delays};
\node[left=.2cm of leg7,inner sep=0pt] (leg7s) {$~$};
\draw[thick] ($(leg7s) + (-.3,0) $) -- ($ (leg7s) + (.3,0) $);
\node[below=.25cm of leg7.south west,anchor=west,align=center] (leg8) {\footnotesize Large Delays};
\node[left=.2cm of leg8,inner sep=0pt] (leg8s) {$~$};
\draw[thick,dotted] ($(leg8s) + (-.3,0) $) -- ($ (leg8s) + (.3,0) $);
\end{tikzpicture}
}
&
\includegraphics[scale=.3]{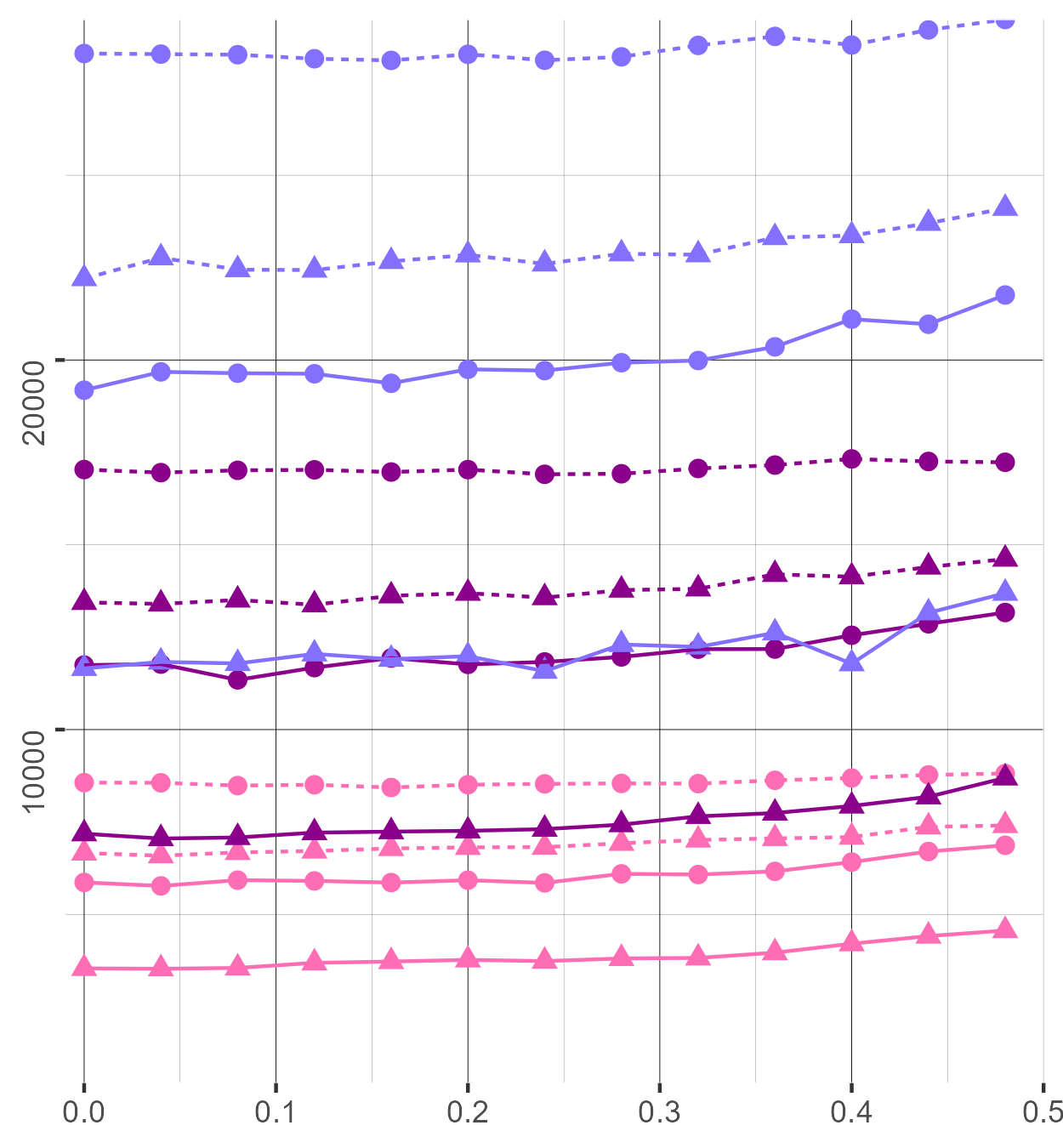}
&
\includegraphics[scale=.3]{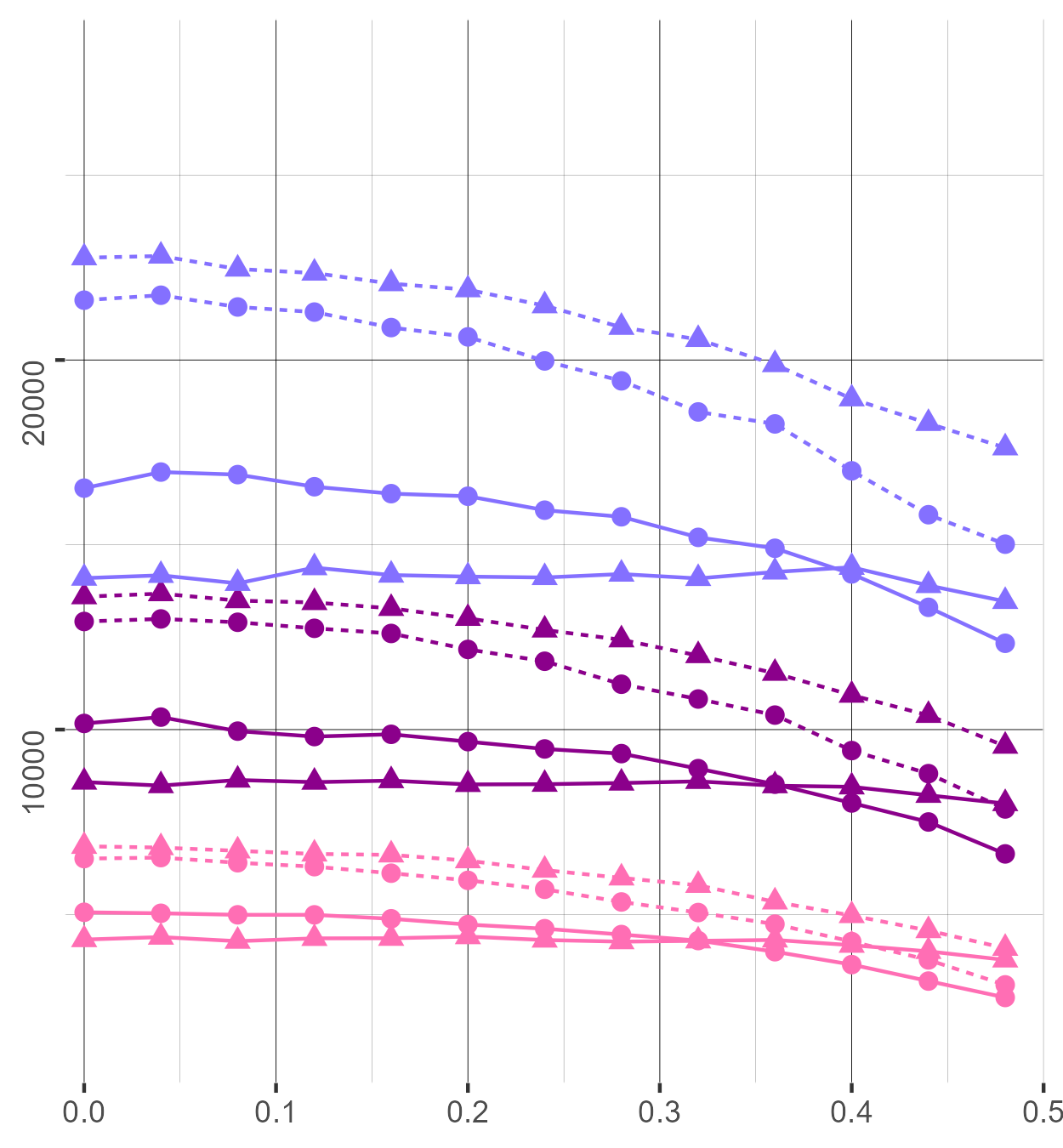}
&
\includegraphics[scale=.3]{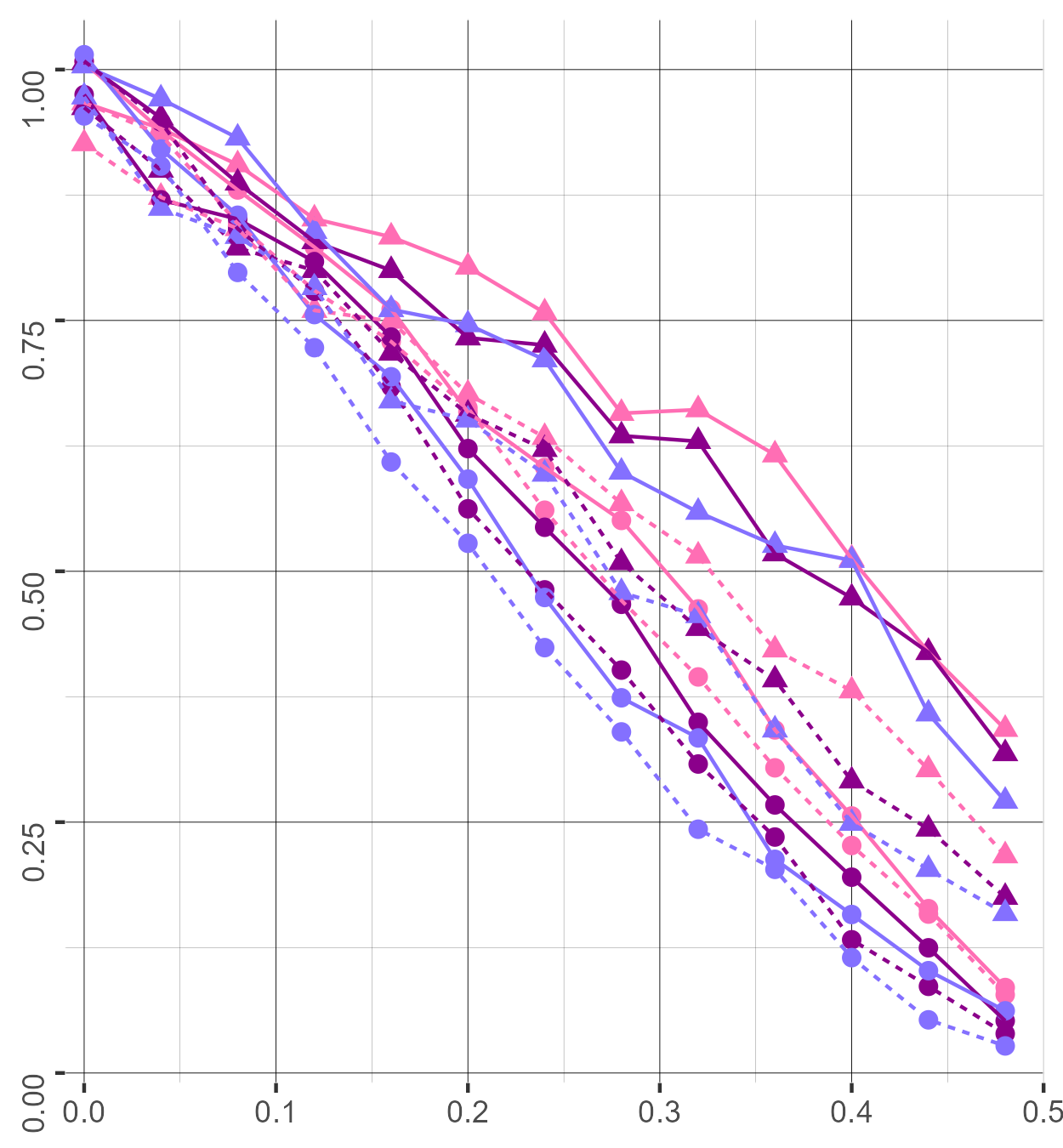}
&
\includegraphics[scale=.3]{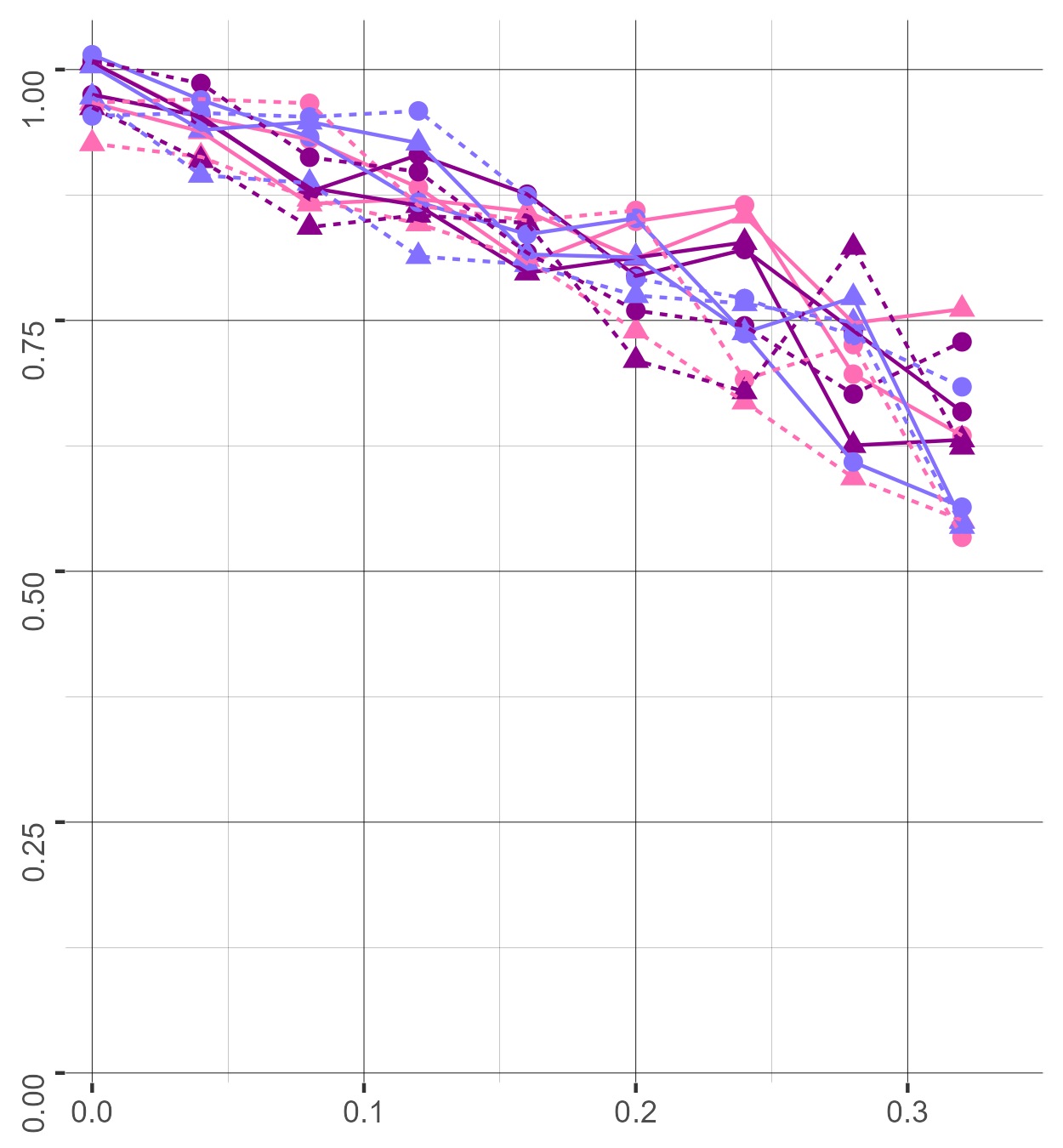}
\\
\hline 
    \scalebox{.7}{
\begin{tikzpicture}
\node[align=center] (leg1) at (0,0) {\textcolor{varyNumPeers25}{$\blacksquare$} {\footnotesize $n=25$}};
\node[above left =0cm and -1.5cm of leg1] (leg0) {\footnotesize \# of peers $n$:};
\node[below=.175cm of leg1.south west,anchor=west,align=center] (leg2) {\textcolor{varyNumPeers35}{$\blacksquare$} {\footnotesize $n=35$}};
\node[below=.175cm of leg2.south west,anchor=west,align=center] (leg3) {\textcolor{varyNumPeers45}{$\blacksquare$} {\footnotesize $n=45$}};
\node[below right=.25cm and -1.5cm of leg3] (leg5) {\footnotesize no mitigation};
\node[draw=black,fill=black,left=.2cm of leg5,circle,inner sep=2pt] (leg5s) {};
\node[below=.2cm of leg5.south west,anchor=west,align=center] (leg6) {\footnotesize mitigation};
\node[left=.2cm of leg6,inner sep=0pt] (leg6s) {$\blacktriangle$};
\node[below=.5cm of leg6.south west,anchor=west,align=center] (leg7) {\footnotesize Small Delays};
\node[left=.2cm of leg7,inner sep=0pt] (leg7s) {$~$};
\draw[thick] ($(leg7s) + (-.3,0) $) -- ($ (leg7s) + (.3,0) $);
\node[below=.25cm of leg7.south west,anchor=west,align=center] (leg8) {\footnotesize Large Delays};
\node[left=.2cm of leg8,inner sep=0pt] (leg8s) {$~$};
\draw[thick,dotted] ($(leg8s) + (-.3,0) $) -- ($ (leg8s) + (.3,0) $);
\end{tikzpicture}
}
&
\includegraphics[scale=.3]{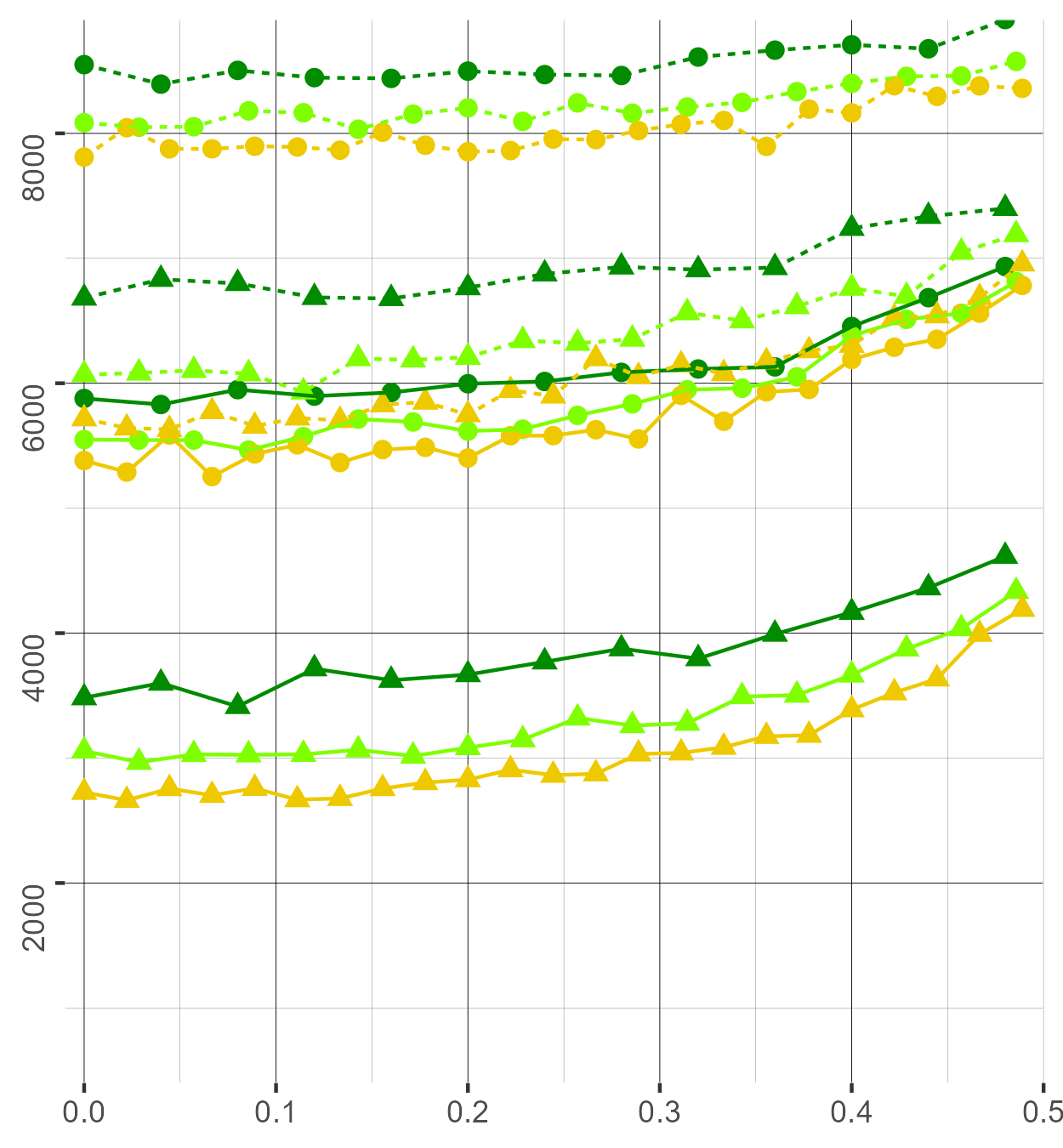}
&
\includegraphics[scale=.3]{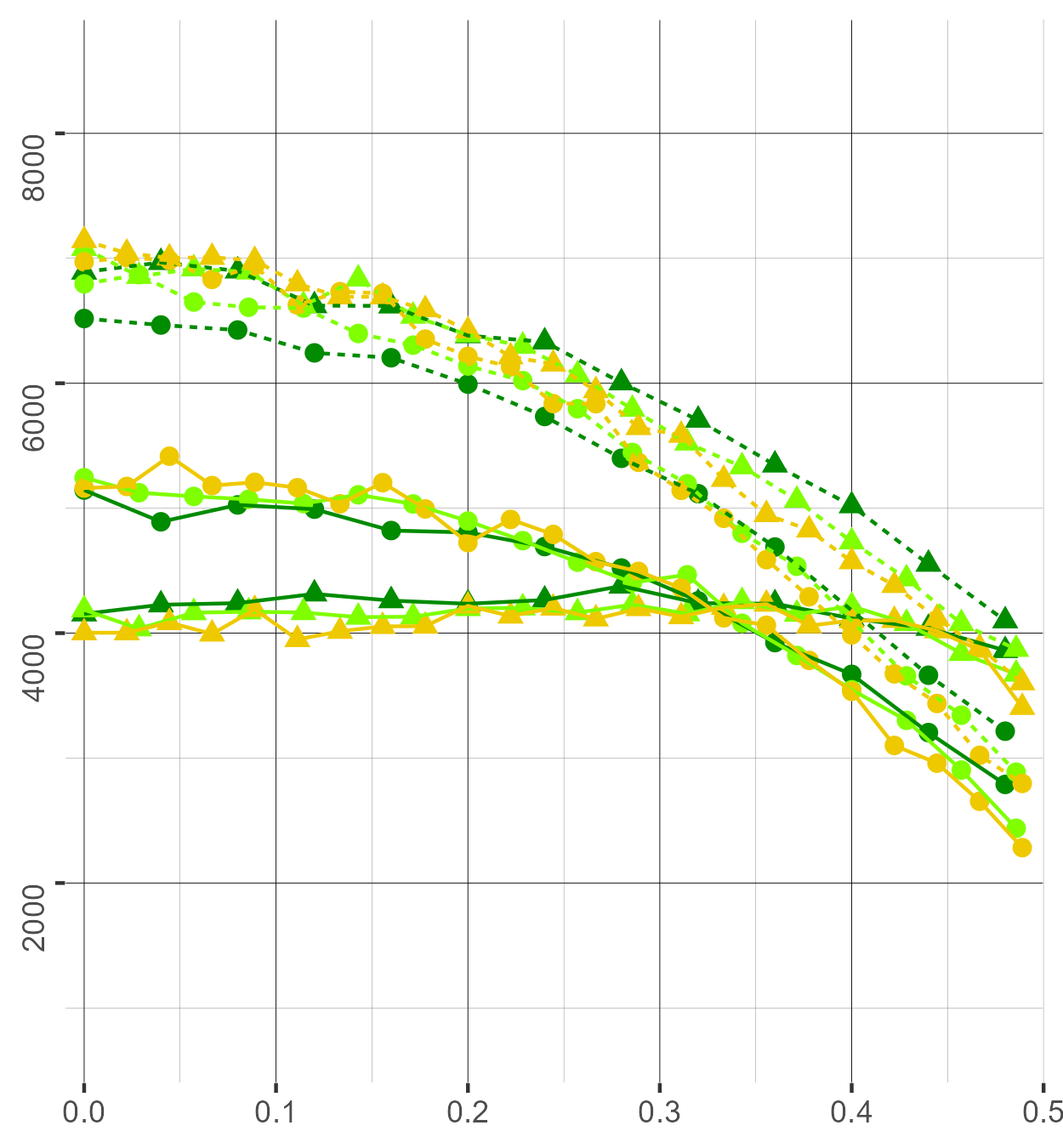}
&
\includegraphics[scale=.3]{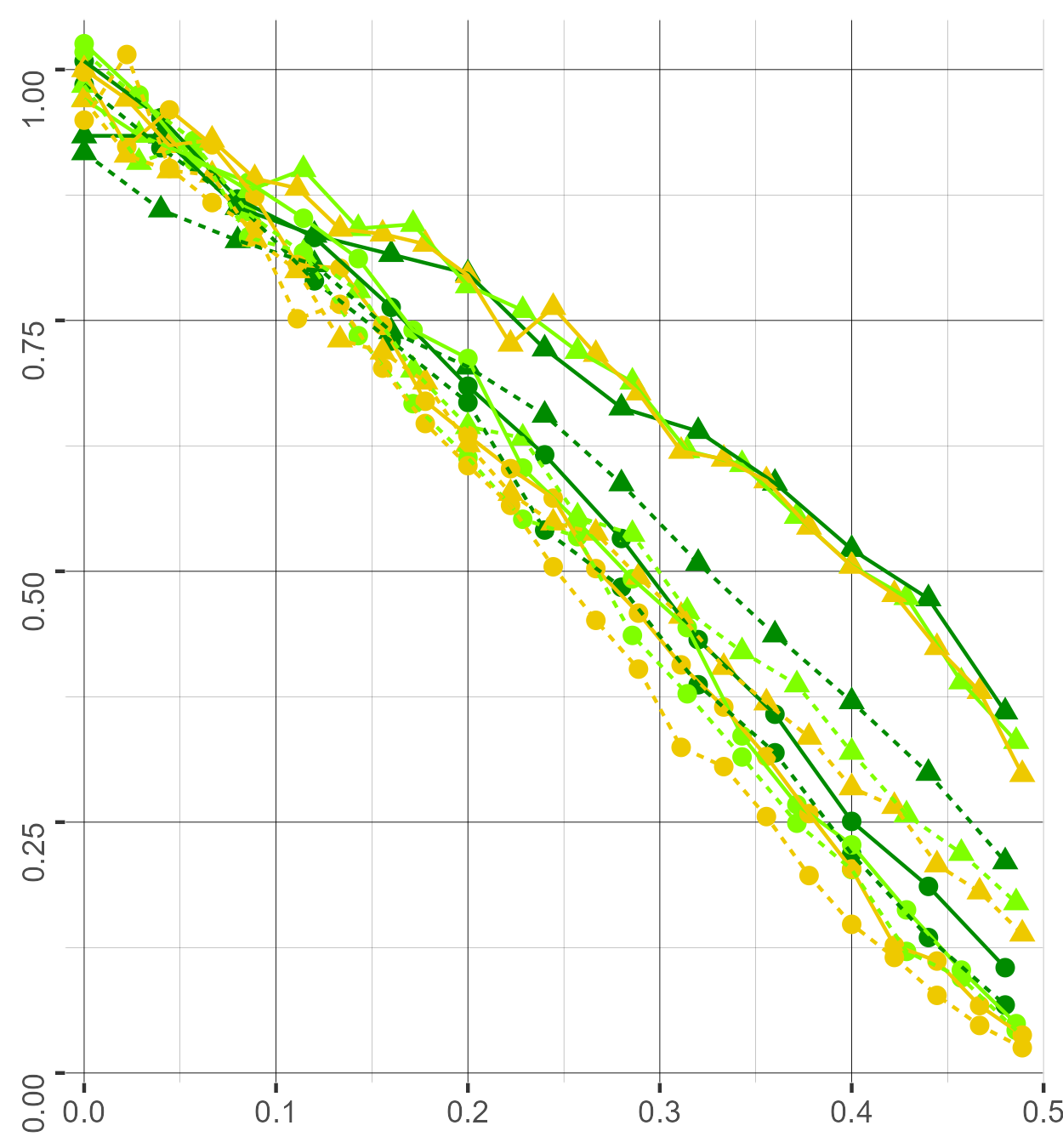}
&
\includegraphics[scale=.3]{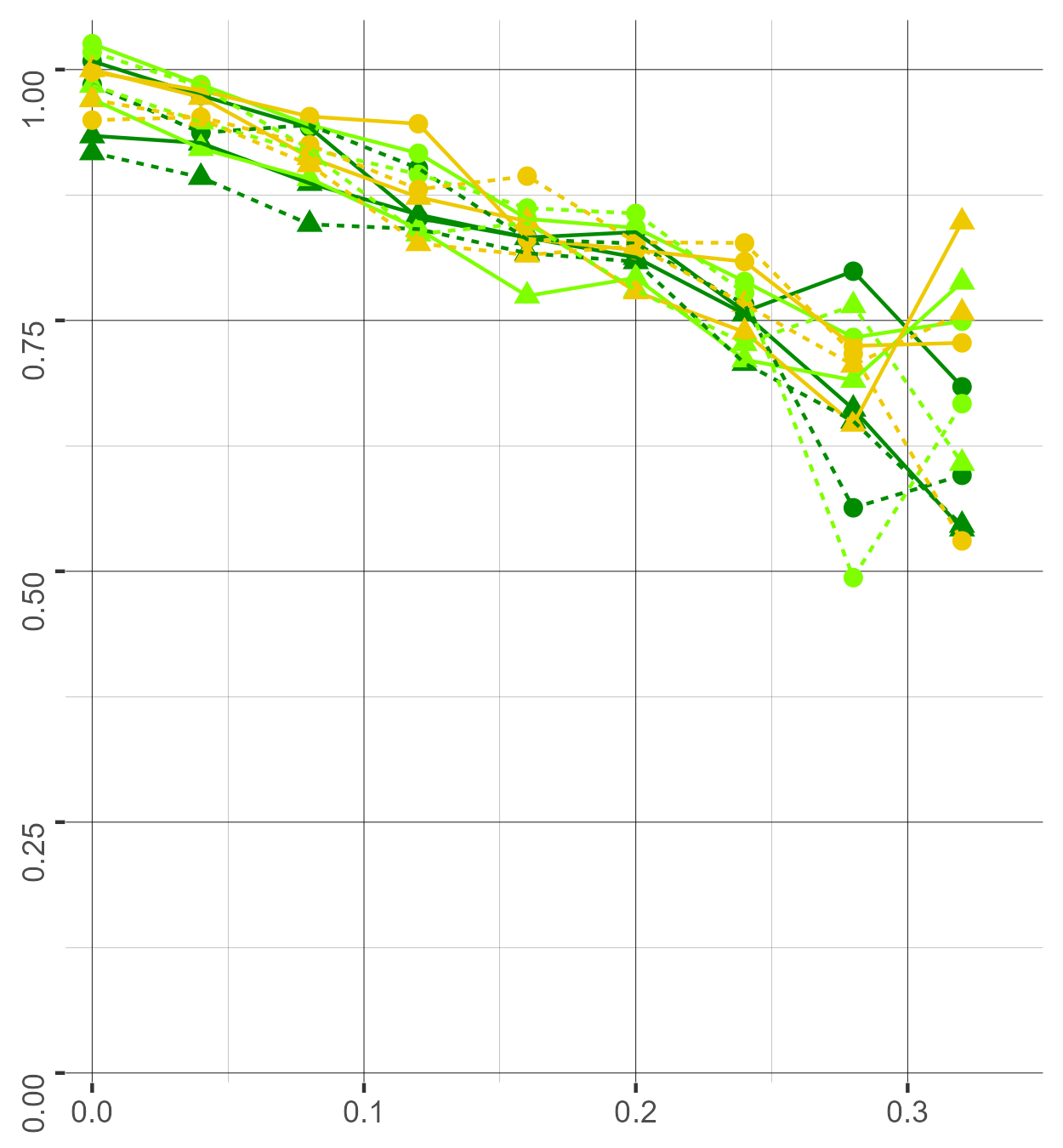}
\\
\hline 
\scalebox{.7}{
\begin{tikzpicture}
\node[align=center] (leg1) at (0,0) {\textcolor{varyNumOrds25}{$\blacksquare$} {\footnotesize $n'=25$}};
\node[above left =0cm and -1.75cm of leg1] (leg0) {\footnotesize \# of orderers $n'$:};
\node[below=.175cm of leg1.south west,anchor=west,align=center] (leg2) {\textcolor{varyNumOrds31}{$\blacksquare$} {\footnotesize $n'=31$}};
\node[below=.175cm of leg2.south west,anchor=west,align=center] (leg3) {\textcolor{varyNumOrds37}{$\blacksquare$} {\footnotesize $n'=37$}};
\node[below right=.25cm and -1.5cm of leg3] (leg5) {\footnotesize no mitigation};
\node[draw=black,fill=black,left=.2cm of leg5,circle,inner sep=2pt] (leg5s) {};
\node[below=.2cm of leg5.south west,anchor=west,align=center] (leg6) {\footnotesize mitigation};
\node[left=.2cm of leg6,inner sep=0pt] (leg6s) {$\blacktriangle$};
\node[below=.5cm of leg6.south west,anchor=west,align=center] (leg7) {\footnotesize Small Delays};
\node[left=.2cm of leg7,inner sep=0pt] (leg7s) {$~$};
\draw[thick] ($(leg7s) + (-.3,0) $) -- ($ (leg7s) + (.3,0) $);
\node[below=.25cm of leg7.south west,anchor=west,align=center] (leg8) {\footnotesize Large Delays};
\node[left=.2cm of leg8,inner sep=0pt] (leg8s) {$~$};
\draw[thick,dotted] ($(leg8s) + (-.3,0) $) -- ($ (leg8s) + (.3,0) $);
\end{tikzpicture}
}
&
\includegraphics[scale=.3]{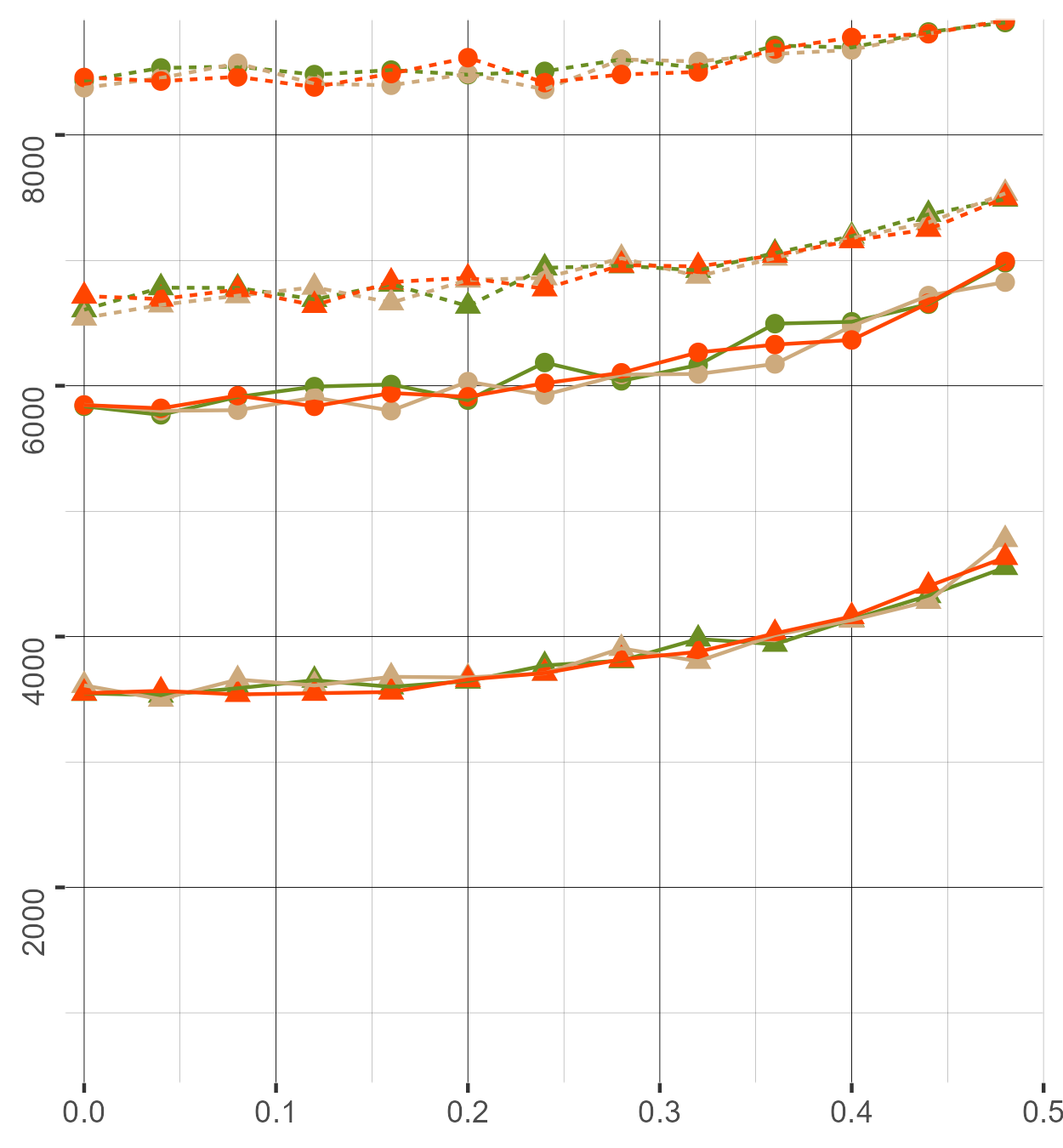}
&
\includegraphics[scale=.3]{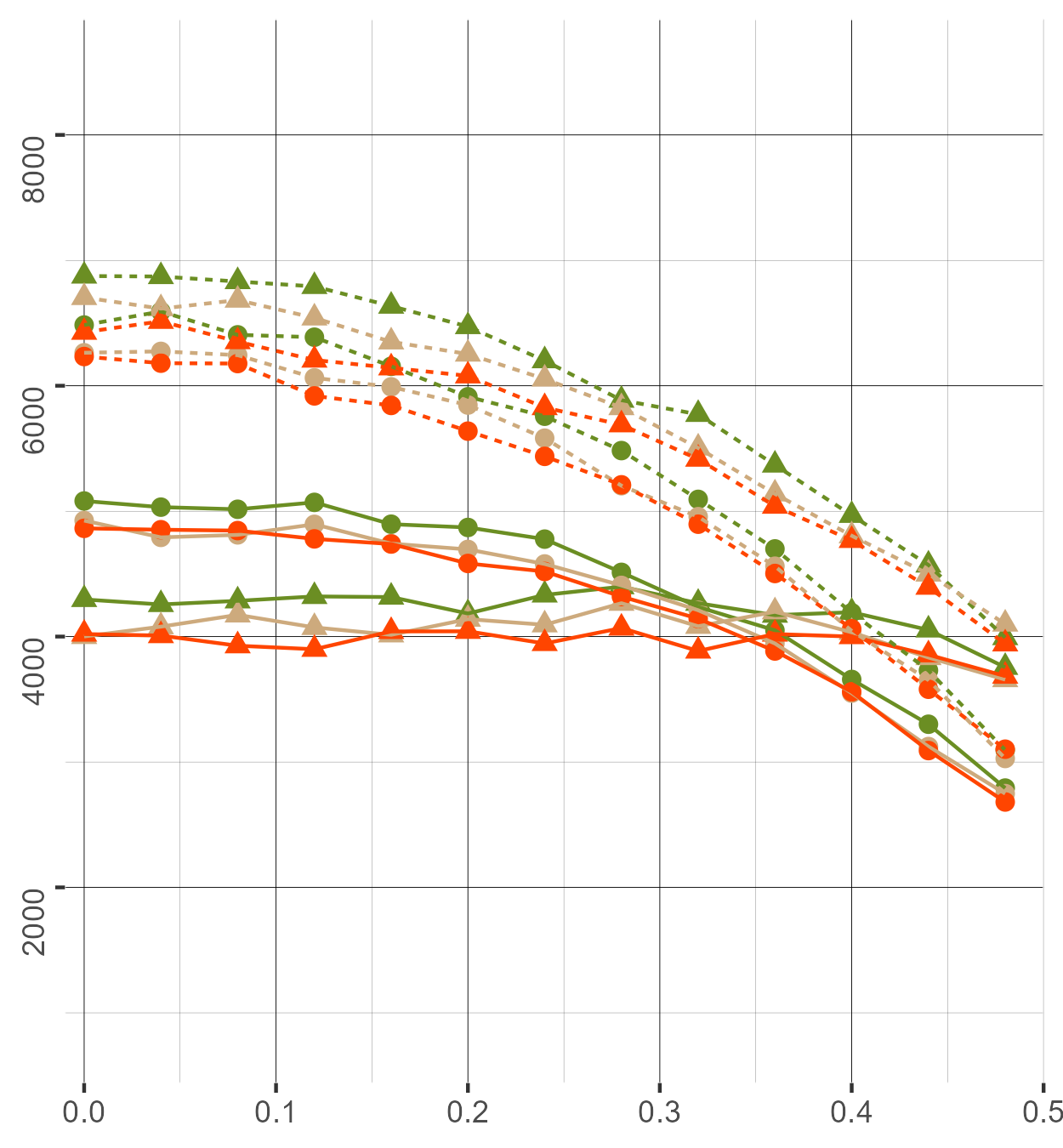}
&
\includegraphics[scale=.3]{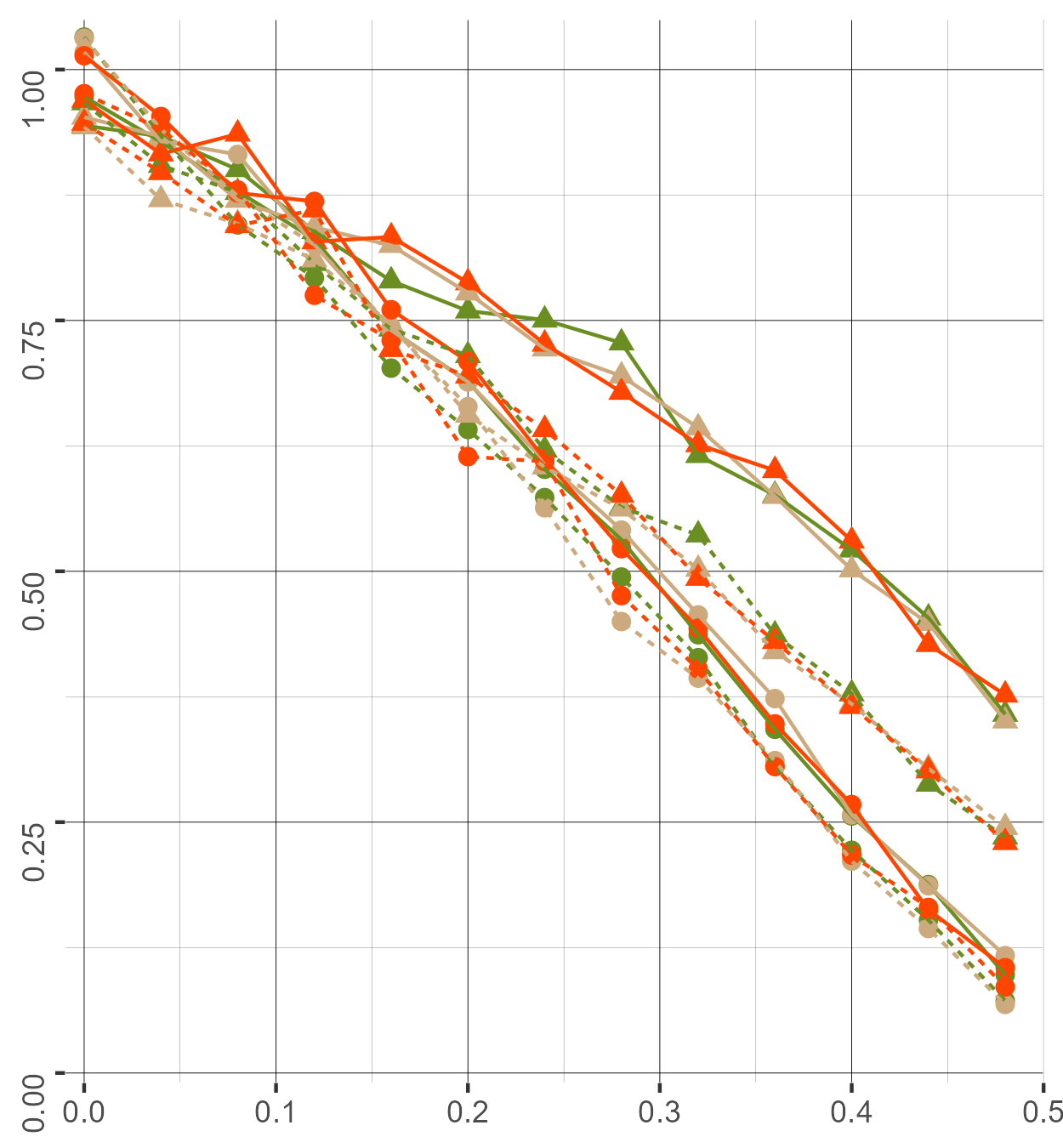}
&
\includegraphics[scale=.3]{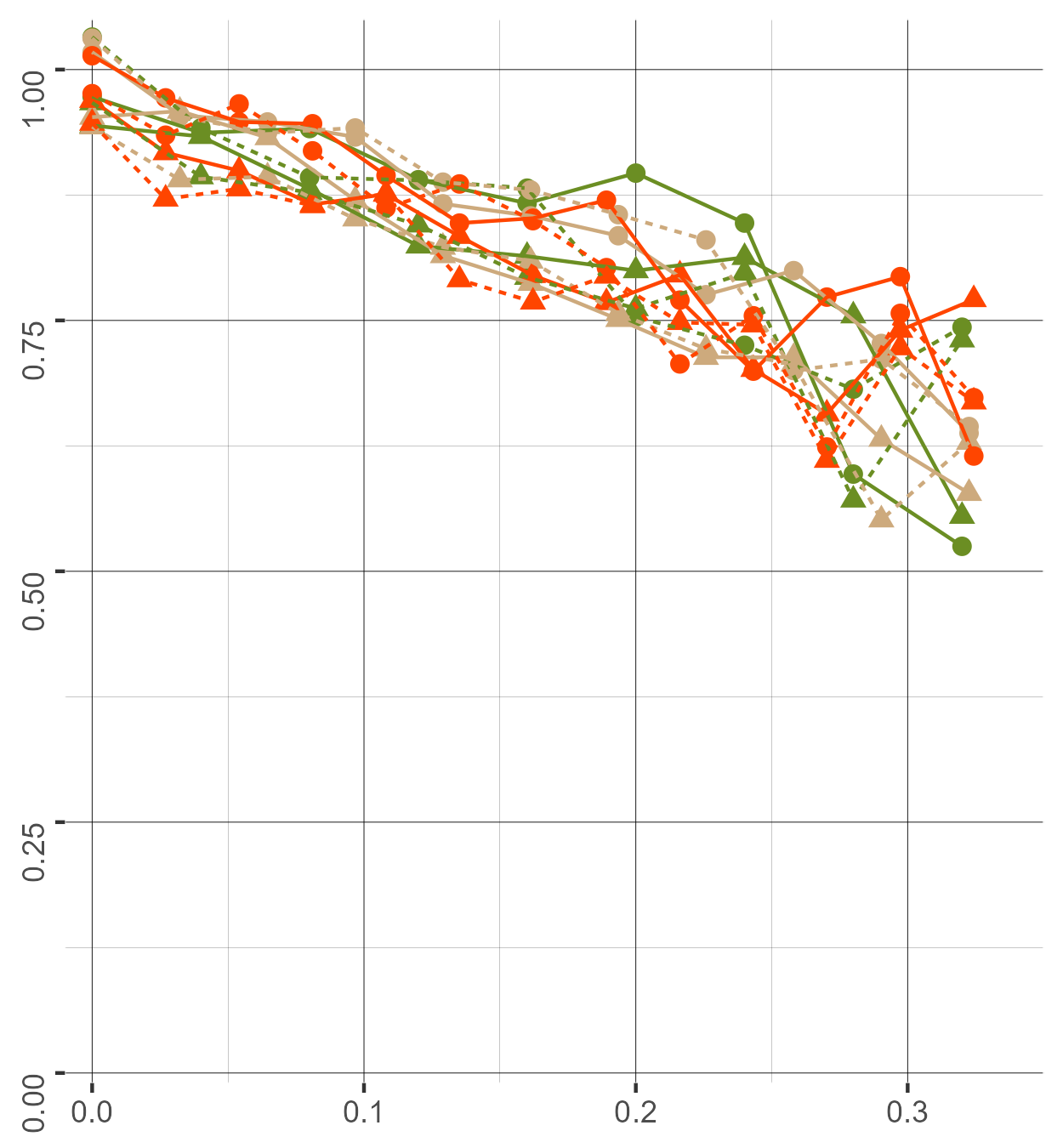}
\\
\hline 
\end{tabular}
\setlength\tabcolsep{6pt}
    
    \caption{Effects of varying the number of clients, peers, and orderers}
    \label{fig:exp_vary_participants}
\vspace*{-.5cm}
\end{figure*}
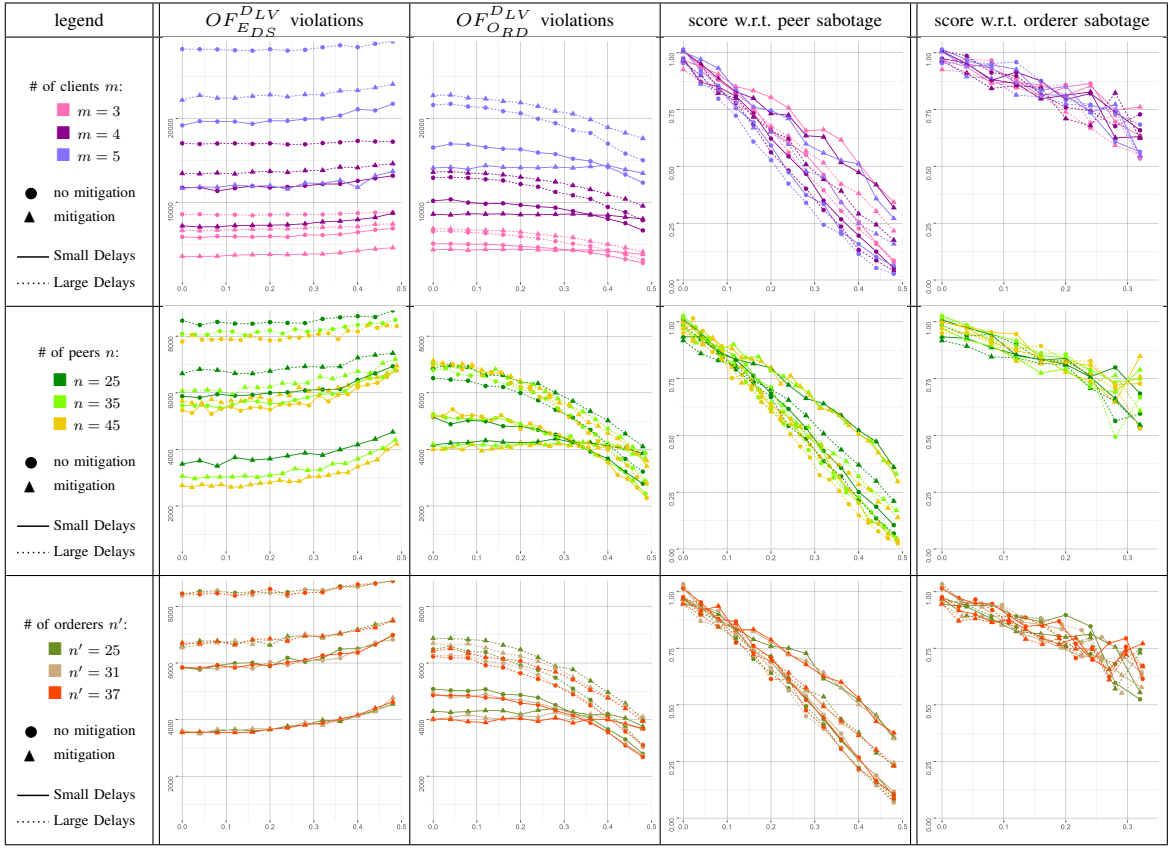

%% file: 8_1.tex
We use MAX \cite{max_tool} to evaluate the effectiveness of the mitigation mechanism described in Sec.\ref{sec:mitig_def} in making our HF system more robust to the attacks performed by the adversary.
For our experiments, the orderers use the Dowdall positional voting algorithm to order transactions.
We otherwise keep the same experimental setting as that of Sec.\ref{sec:sim_attacks}.

Fig.\ref{fig:exp_mitig} summarizes our experimental results.
In each diagram, the plain (resp.~dotted) curves correspond to simulations without using (resp.~that use) the mitigation mechanism.
Overall, the number of $OF_{E_{DS}}^{D_{LV}}$ violations is smaller when using the mechanism (at any proportion of infected peers and orderers and under any network condition), which was its purported goal (it being defined as a vote on the ordering preferences of the peers).

However, the mechanism might increase the number of $OF_{O_{RD}}^{D_{LV}}$ violations in certain cases.
On Fig.\ref{fig:exp2_ords}, using the mechanism causes a decrease under the smaller delays \textcolor{blue}{\faClockO} in blue but an increase under the larger delays \textcolor{red}{\faClockO} in red.
Indeed, when network delays are particularly high and random, the order with which orderers receive endorsed transactions might not be correlated to the order with which peers receive transactions.
Using the mechanism replaces the default FIFO order on the orderers' local mempools by an order that is computed according to the preferences of the peers.
If we recall our example from Fig.\ref{fig:scenario_hf_rw}, under high network delays it may be so that a majority of orderers receive $x_2$ before $x_1$, even though a majority of peers received $x_1$ before $x_2$. As a result, using the default FIFO order would result in a violation of $OF_{E_{DS}}^{D_{LV}}$ and complying with $OF_{O_{RD}}^{D_{LV}}$ while using the mitigation mechanism would result in complying with $OF_{E_{DS}}^{D_{LV}}$ and a violation of $OF_{O_{RD}}^{D_{LV}}$.

On Fig.\ref{fig:exp_mitig_peers}, we remark that the mitigation mechanism has a protective effect against the attack on the endorsing service, especially under smaller network delays.
Indeed, when using the mitigation, the decrease of the $\mathtt{score}$ with the number of infected peers is less pronounced. 
This protection is less visible against the ordering service attack except under smaller delays.
Still, as peer sabotage is more efficient than orderer sabotage, overall, using the mechanism should prove advantageous.

%% file: 8_2.tex
In Sec.\ref{sec:sim_attacks} and Sec.\ref{sec:sim_mitig}, we considered $n = 25$ peers, $n'=25$ orderers and $m=3$ clients and we varied the number of Byzantine peers (resp.~orderers) between $0$ and $12$ (resp.~$0$ and $8$) so that their proportion varies between $0\%$ and $50\%$ (resp.~$0\%$ and $33\%$).
As we need long simulations to make the $\mathtt{score}$ converge, considering higher numbers of participants involves costly simulations.
These arbitrary values for $n$, $n'$ and $m$ were chosen so as to make the experiments reproducible without a prohibitive computational cost.
In the following, we show that our results and remarks still hold when varying $n$, $n'$ and $m$.

%% file: 8_3.tex
The first line of Fig.\ref{fig:exp_vary_participants} summarizes results from 
$264$ 
distinct simulations, varying the number of clients $m$ from $3$ to $5$.
While the three diagrams on the left correspond to $156$ simulations, varying the number of infected peers between $0$ and $12$ on the horizontal axis, the one on the right corresponds to $108$ simulation, varying the number of infected orderers between $0$ and $8$ on the horizontal axis.
We observe that the number of violations increases with $m$, as there are more pairs of non commutative transactions to consider.
Still, we observe the same effect as in Sec.\ref{sec:sim_attacks} w.r.t.~the power of the adversary : increasing the proportion of infected peers increases the number of $OF_{E_{DS}}^{D_{LV}}$ violations and decreases the number of $OF_{O_{RD}}^{D_{LV}}$ violations.
Also, the effectiveness of the attack increases w.r.t.~the number of clients $m$, which is expected, as slowing down the delivery of the transactions from the target clients is more likely to make that client loose if it has more competitors.
As previously, we also remark that larger delays make the system more vulnerable while the mitigation mechanism makes it more robust.
Similarly, modifying $m$ does not significantly change the effect of orderer sabotage.

The second line of Fig.\ref{fig:exp_vary_participants} summarizes results from 
$324$
distinct simulations with the number of peers $n$ being either $25$, $35$ or $45$.
We remark that the number of $OF_{E_{DS}}^{D_{LV}}$ violations decreases with $n$. Indeed, increasing $n$ reduces the variance of the time taken to obtain $n/2$ endorsements.
However, increasing the proportion of infected peers still lead to an increase in $OF_{E_{DS}}^{D_{LV}}$ violations.
Overall the effect of peer sabotage and orderer sabotage is not impacted by modifying $n$.
Also, the effect of the delay distribution and the mitigation system does not depend on $n$.

Finally, the third line of Fig.\ref{fig:exp_vary_participants} reports results of 
$288$
simulations with the number of orderers $n'$ being either $25$, $31$ or $37$.
Likewise, modifying $n'$ does not change the effect of peers sabotage, orderer sabotage, of the distribution of delays or of the mitigation system.

%% file: 10_conc.tex
\section{Conclusion\label{sec:conclusion}}

In \cite{adversary_augmented_simulation_to_evaluate_order_fairness_on_HF}, we have introduced a novel adversary model tailored to distributed systems and blockchain technologies.
The adversary operates by performing actions that are bound by a failure and a communication model.
Chaining such actions, it executes attacks within predefined fault-tolerance thresholds.
This approach facilitates a more direct and fine-grained integration of adversarial behavior in practical scenarios.
Furthermore, by integrating this adversary model into a multi-agent-based simulator, we enable the simulation of realistic adversarial attacks.
We applied our approach on an HF-based blockchain system, simulating several attacks on a client-fairness property while evaluating their side effects on several order-fairness properties.

This present paper extends the study from \cite{adversary_augmented_simulation_to_evaluate_order_fairness_on_HF} via considering more realistic network assumptions (e.g., hypoexponential delay distributions \cite{models_of_network_delay}) and a wider range of metrics. We also propose a novel mitigation mechanism (with no overhead) to make HF more robust to transaction reordering attacks \cite{sok_preventing_transaction_reordering_manipulations_in_decentralized_finance}.
we extend our simulation study to this mechanisms to demonstrate its protective effect, in particular against peer sabotage.
Finally, we vary additional simulation parameters to comfort our results.
Varying the number of clients, peers and orderers do not call into question the effect of the attacks from \cite{adversary_augmented_simulation_to_evaluate_order_fairness_on_HF} and of our mitigation mechanism.

Our study thus highlights the vulnerability of HF to transaction reordering attacks.
Although achieving order fairness is not realistic (as discussed in Sec.\ref{ssec:related_works_order_fairness} and in \cite{ordering_transactions_with_bounded_unfairness}),
simple mechanisms can be put in place to mitigate this vulnerability.